\documentclass[11pt]{article}
\pdfoutput=1

\usepackage{jheppub}
\usepackage{graphicx,epsfig}
\usepackage{float}
\usepackage{subcaption}
\usepackage[utf8]{inputenc}
\usepackage{hyperref}
\usepackage{caption}
\usepackage{xcolor}
\usepackage{physics,xparse,comment}
\usepackage[mathscr]{euscript}

\newcommand{\rmd}{\mathrm d}

\definecolor{darkpastelgreen}{rgb}{0.01, 0.75, 0.24}

\newcommand{\beq}{\begin{equation}}

\newcommand{\eeq}{\end{equation}}

 \newcommand{\be}{\begin{equation}}
 \newcommand{\ee}{\end{equation}}
 \newcommand{\bea}{\begin{eqnarray}}
 \newcommand{\eea}{\end{eqnarray}}

\usepackage{tikz}
\usetikzlibrary{positioning}
\usetikzlibrary{intersections}
\usetikzlibrary{fadings} 
\usetikzlibrary{arrows.meta} 
\usetikzlibrary{arrows}

\tikzfading[name=fade out,
inner color=transparent!0,
outer color=transparent!100]

\definecolor{cherryblossompink}{rgb}{1.0, 0.72, 0.77}
\definecolor{lightblue}{rgb}{0.68, 0.85, 0.9}

\usetikzlibrary{decorations.pathmorphing}
\usetikzlibrary{decorations.pathreplacing,decorations.markings}

\usetikzlibrary{backgrounds,automata}

\title{$\text{T}\overline{\text{T}}$ deformations from AdS$_2$ to dS$_2$}
\author[a,b]{Sergio E. Aguilar-Gutierrez,}
\author[c]{Andrew Svesko,}
\author[d]{and Manus R. Visser}
\affiliation[a]{Okinawa Institute of Science and Technology Graduate University, Onna, Okinawa 904 0495, Japan}
\affiliation[b]{Institute for Theoretical Physics, KU Leuven, Celestijnenlaan 200D, B-3001 Leuven, Belgium}
\affiliation[c]{Department of Mathematics, King’s College London, Strand, London WC2R 2LS, UK}
\affiliation[d]{Institute for Mathematics, Astrophysics and Particle Physics, and Radboud Center for Natural Philosophy, Radboud University, 6525 AJ Nijmegen, The Netherlands}
\emailAdd{sergio.ernesto.aguilar@gmail.com,  andrew.svesko@kcl.ac.uk, manus.visser@ru.nl}

\abstract{We revisit the formalism of $\text{T}\overline{\text{T}}$ deformations for quantum theories that are holographically dual to two-dimensional dilaton-gravity theories with Dirichlet boundary conditions. To better understand the microscopics of de Sitter space, we focus on deformations for which the dual bulk geometry flows from  Anti-de Sitter  to de Sitter space. We explore two distinct ways to achieve this: either through so-called centaur geometries that interpolate between AdS$_2$ and dS$_2$, or by a spherical dimensional reduction of $\text{T}\overline{\text{T}} + \Lambda_2$ theories that were proposed to give a microscopic interpretation of three-dimensional de Sitter entropy. We derive the microscopic energy spectrum, heat capacities, and deformed Cardy expressions for the thermodynamic entropy in the canonical and microcanonical ensembles for these two setups. In both setups a signature of the change from AdS to dS is that the heat capacity at a fixed deformation parameter of the boundary system changes sign, indicating the existence of a thermodynamically unstable de Sitter patch.  Our findings provide important consistency conditions for holographic models of the dS$_2$ static patch.}

\begin{document}

\maketitle

\section{Introduction} \label{sec:intro}

Like black hole horizons, the cosmological horizon of de Sitter (dS) space obeys an entropy-area law~\cite{Gibbons:1977mu,Gibbons:1976ue},
\beq S_{\text{GH}}=\frac{A}{4G}\;.\label{eq:GHentintro}\eeq
Here $S_{\text{GH}}$ denotes the thermal entropy associated to the dS static patch, where $A$ is the cross-sectional area of the horizon. The thermodynamic description of dS, however, is far more subtle than its asymptotically flat or anti-de Sitter (AdS) black hole counterparts. Indeed, unlike black holes, adding positive energy to the dS static patch at fixed cosmological constant leads to an entropy reduction, as follows from an additional minus sign appearing in the first law of cosmological horizons~\cite{Gibbons:1977mu}. Moreover, spatial sections of dS space are spheres, having no asymptotic boundary, such that the total gravitational energy (defined with respect to the time translation Killing vector of the static patch) vanishes.  Additionally,  the microscopic origin of Gibbons-Hawking entropy (\ref{eq:GHentintro}) is more mysterious than for black holes. Due to the apparent lack of unbroken supersymmetry in empty dS \cite{Banks:2000fe,Witten:2001kn,Goheer:2002vf},  it seems unlikely the string-theoretic explanation for the entropy of specific black holes \cite{Sen:1995in,Strominger:1996sh} applies to dS horizons. 

A promising way to address all of these puzzles is to introduce a finite timelike boundary satisfying Dirichlet boundary conditions inside the dS static patch, extending previous work by York and collaborators \cite{York:1986it,Whiting:1988qr,Braden:1990hw,Brown:1992br,Brown:1992bq,Brown:1994gs}. The Dirichlet wall delineates two distinct systems: the ``horizon patch'', the region between the wall and the horizon, and the ``pole patch'', the region between the wall and the pole. Thermal ensembles may then be consistently defined by fixing quasi-local thermodynamic data with respect to the Dirichlet wall \cite{Banihashemi:2022jys} (see also \cite{Carlip:2003ne,Hayward:1990zm,Svesko:2022txo,Draper:2022ofa,Anninos:2022hqo}). The quasi-local set-up also reveals the origin of the minus sign in the first law of cosmological horizons is a consequence of misinterpreting matter Killing energy as the internal energy of the system \cite{Banihashemi:2022htw}.

Timelike Dirichlet walls, moreover, provide a natural home for holographic quantum theories dual to a finite region of bulk gravity. Indeed, this is one approach to provide a holographic description of the de Sitter static patch. Specifically, in \cite{Anninos:2011af,Anninos:2017hhn} it was proposed that the non-gravitating dual theory resides on a timelike boundary near the pole of de Sitter space, whereas in \cite{Susskind:2021omt} the dual description lives on the (stretched) cosmological horizon. Our perspective is that a Dirichlet wall can interpolate between these two proposals since it can sit at any location in de Sitter static patch \cite{Leuven:2018ejp,Svesko:2022txo}.
Holography of timelike Dirichlet boundaries thus offers a way to explore the microscopic origins of de Sitter thermodynamics.


Indeed, Dirichlet walls play a critical role in providing a microstate accounting of the Gibbons-Hawking entropy, including its 1-loop logarithmic correction \cite{Shyam:2021ciy,Coleman:2021nor,Silverstein:2022dfj,Batra:2024kjl}. Broadly, the idea is to start with an AdS$_{3}$ black hole near the Hawking-Page (HP) transition and push its asymptotic timelike boundary inward until positioned just outside of the black hole horizon, where it becomes indistinguishable from the dS$_{3}$ cosmological horizon. The timelike boundary is then pushed outward in a specific way to recover the dS static patch geometry. This procedure, in three bulk spacetime dimensions,  
has a dual description in terms of solvable deformations of two-dimensional conformal field theories (CFT) living on the Dirichlet wall. Precisely, pushing the $\text{AdS}_{3}$ boundary inward corresponds to an  irrelevant $\text{T}\overline{\text{T}}$ deformation of a holographic $\text{CFT}_{2}$ \cite{Zamolodchikov:2004ce,Smirnov:2016lqw,Dubovsky:2012wk,Cavaglia:2016oda,McGough:2016lol,Kraus:2018xrn}, while a $\text{T}\overline{\text{T}}+\Lambda_{2}$ deformation builds the $\text{dS}_{3}$ static patch \cite{Gorbenko:2018oov,Shyam:2021ciy,Coleman:2021nor,Silverstein:2022dfj,Batra:2024kjl}. With this patch-wise prescription, ``dressed'' microstates of the $\text{AdS}_{3}$ black hole are identified with the microstates associated with the dS$_3$ static patch, such that the $\text{dS}_{3}$ entropy is exactly equal to the Cardy entropy of the dual $\text{CFT}_{2}$. Thence, analogous to $\text{AdS}_{3}$ black holes \cite{Strominger:1997eq,Carlip:2000nv},\footnote{A similar conclusion is reached for Kerr-$\text{dS}_{3}$ \cite{Bousso:2001mw}, where, via $\text{dS}_{3}/\text{CFT}_{2}$ duality, the cosmological horizon entropy equals the Cardy entropy \cite{Cardy:1986ie} of $\text{CFT}_{2}$ deformed by a marginal operator located at past infinity.} the Gibbons-Hawking entropy of $\text{dS}_{3}$ follows from a counting of dual CFT microstates, consistent with a different approach using the long-string phenomenon \cite{Leuven:2018ejp}.

In this article we develop a dual holographic description for the quasi-local thermodynamics of two-dimensional de Sitter space. We do so by adapting the $\text{T}\overline{\text{T}}$ formalism dual to bulk geometric flows from AdS$_2$ to~dS$_2$. In particular, we study the properties of the microscopic theory dual to the ``centaur'' model~\cite{Anninos:2017hhn}, a specific two-dimensional model of dilaton gravity, that describes an interpolation between AdS$_2$ and dS$_2$. Further, we holographically characterize a similar type of geometric flow via a one-dimensional analog of $\text{T}\overline{\text{T}} + \Lambda_2$-deformations.

More precisely, we first revisit a conjectured microscopic description of the quasi-local energy for a general class of two-dimensional dilaton theories of gravity \cite{Gross:2019ach}. The correspondence essentially follows from a dimensional reduction of $\text{T}\overline{\text{T}}$-deformations to a two-dimensional holographic CFT. 
Within this framework, we provide the first dual holographic description of the quasi-local thermodynamics of Jackiw-Teitelboim (JT) gravity \cite{Jackiw:1984je,Teitelboim:1983ux} in de Sitter space \cite{Svesko:2022txo}, and the centaur model \cite{Anninos:2022hqo}. 
In particular, we find that the microcanonical entropy of the dual deformed theory of the dS$_{2}$ geometry has a Cardy-like form (cf. (\ref{eq:SmicrocanonicaldS})),
\beq S (\lambda, \mathcal E) =S_0 \pm 2 \pi \sqrt{\frac{c}{6}\left ( \mathcal E (\lambda\mathcal E  -2) + \frac{2}{\lambda} \right)} \quad \Longleftrightarrow \quad S_{\text{BH}} =  \frac{\Phi_0}{4 G_2} + \frac{\Phi_h}{4 G_2  }\,, \eeq
where `$+$' refers to states dual to the   patch in dS$_2$  between the  Dirichlet wall and the cosmological horizon, and `$-$' to states dual to  the   patch between the Dirichlet boundary and the black hole horizon. Further, $\mathcal{E}$ is a (dimensionless) microscopic energy, $\lambda$ is a dimensionless parameter characterizing the $\text{T}\overline{\text{T}}$-flow, and $c$ is the one-dimensional analog of a central charge of the seed ($\lambda=0$) conformal quantum theory.  Note that the microcanonical entropy reduces to the Cardy formula for $\lambda = 1/ \mathcal E = 1/ (2 \mathcal E_0),$ for energy $\mathcal{E}_{0}$ of the undeformed theory (where the Dirichlet wall is located at the horizon).  Applying the appropriate holographic dictionary, we recover the classical gravitational entropy of   dS$_{2}$  \eqref{BHentropy}, where $\Phi_{0}$ is proportional to $S_{0}$ and $\Phi_{h}$ is the value of the dilaton at either the black hole or cosmological horizon. Further, we compute the microscopic energy spectrum and heat capacity, finding that the boundary dual to the dS$_{2}$ cosmic patch is thermally unstable (see also \cite{Banihashemi:2022jys,Svesko:2022txo,Anninos:2022hqo}). 

 We further consider a spherical dimensional reduction of a generalized $\text{T}\overline{\text{T}}+\Lambda_2$ flow connecting a BTZ black hole of arbitrary mass to a locally conical dS$_{3}$ spacetime.  This leads us to a new type of flow connecting AdS$_{2}$ and dS$_{2}$ along their respective horizons, in contrast with the centaur model. Moreover,      the AdS$_2$ region  in this case corresponds to an eternal AdS$_2$ black hole (i.e.,  Rindler-AdS$_2$), whereas the AdS$_2$ region in the centaur geometry corresponds to global AdS. Consequently, the thermal entropy of states dual to AdS$_{2}$ regions differ from those in the centaur model. The quasi-local heat capacities of  the interpolating system are qualitatively the same as the centaur model, however. In particular,  the boundary dual to the AdS$_2$ black hole that is smoothly connected to the seed theory is thermally stable, while the boundary dual to the cosmic patch dS$_2$ solution is unstable.

The remainder of this article is as follows. In Section \ref{sec:dilaton gravity TTbar} we study $\text{T}\overline{\text{T}}$ deformed theories dual to general dilaton-gravity models; reviewing the procedure developed in \cite{Gross:2019ach} and adapting it to (A)dS JT gravity. We derive the microscopic energy spectrum, thermodynamic entropy and heat capacity, finding precise agreement with the bulk quasi-local thermodynamics. 
We provide a dual holographic description of the centaur model in terms $\text{T}\overline{\text{T}}$-flows in Section~\ref{sec:centaur TTbar}. 
In Section \ref{Sec:TTbar+Lambda2}, we first generalize $\text{T}\overline{\text{T}}+\Lambda_2$ deformations connecting a BTZ black hole of arbitrary mass to a locally conical dS$_{3}$ spacetime of the same mass.\footnote{A version of this more general flow was considered in \cite{Coleman:2021nor}. Specifically, flows dual to gluing the BTZ geometry to  dS$_3$ with a conical \emph{excess}.  } We then dimensionally reduce this flow, leading to   another type of geometric flow between (A)dS$_{2}$ patches, and analyze the quasi-local thermodynamics. 
We conclude in Section \ref{Sec:Discussion} with a summary and discussion of future work. Appendix~\ref{app:TTdefsgen2D} provides explicit details deriving the flow equation and microscopic energy spectrum for deformed theories dual to general 2D dilaton theories of gravity.


\section{Quasi-local thermodynamics in general dilaton-gravity}\label{sec:dilaton gravity TTbar}

In this section, we summarize essential elements of the thermodynamics of general 2D dilaton-gravity theories with Dirichlet walls and the one-dimensional analog of $\text{T}\overline{\text{T}}$ deformations. We follow the previous work     \cite{Gross:2019ach,Gross:2019uxi,Svesko:2022txo,Anninos:2022hqo} in this section, highlighting aspects that have not been analyzed in the literature.

\subsection{General 2D dilaton-gravity with Dirichlet walls}

Pure general relativity in two dimensions is topological, having trivial dynamics. However, 2D dilaton-gravity theories have nontrivial dynamics and are characterized by actions of the type (in Euclidean signature)
\begin{equation}\label{eq:I centaur}
    I_{\text{E}}=I_{\text{top}}-\frac{1}{16\pi G_2}\int_{\mathcal{M}} \rmd^2x\sqrt{g}(R\Phi+V(\Phi))-\frac{1}{8\pi G_2}\int_{\partial\mathcal{M}}\rmd \tau\sqrt{h}\,\Phi K+I_{\text{ct}}~,
\end{equation}
with the topological term
\begin{align}
    I_{\text{top}}&=-\frac{\Phi_{0}}{16\pi G_{2}}\int_{\mathcal{M}}\rmd^{2}x\sqrt{g}R-\frac{\Phi_{0}}{8\pi G_{2}}\int_{\partial\mathcal{M}}\rmd\tau\sqrt{h}K=-\frac{2\pi}{8\pi G_{2}}\Phi_{0}\chi\;,\label{eq:topterm}
\end{align}
and $I_{\rm ct}$ is a counterterm action.
The theory consists of a contribution over bulk spacetime $\mathcal{M}$ endowed with metric $g_{\mu\nu}$, and a (timelike) boundary $\partial\mathcal{M}$ term, with induced metric $h_{\mu\nu}$ and (trace of) extrinsic curvature $K$. In Euclidean signature, we take $\mathcal{M}$ to have the topology of a disk, such that $\partial\mathcal{M}=S^{1}$. The dilaton $\Phi$,\footnote{Newton's constant $G_{2}$ is dimensionless in two spacetime dimensions, but we keep it for bookkeeping purposes. In fact, the prefactor $(\Phi_{0}+\Phi)$ plays
the natural role of a gravitational constant, where a diverging dilaton indicates regions of weak gravity.} resides on all of $\mathcal{M}$ including $\partial\mathcal{M}$, and has general potential $V(\Phi)$. 
The contribution (\ref{eq:topterm}), with positive, dimensionless constant $\Phi_{0}\gg\Phi$ and Euler character $\chi$ of $\mathcal{M}$, describes a topological contribution, having no effect on the gravitational dynamics, though it will contribute to the thermodynamics. All interesting dynamics arise from the remaining terms in (\ref{eq:I centaur}), with the third being the Gibbons-Hawking-York contribution, while the fourth $I_{\text{ct}}$ is a local counterterm rendering the action finite. For our purposes, since we are interested in 2D geometries with an AdS$_2$ boundary, the counterterm is
\beq I_{\text{ct}}^{\text{JT}}=\frac{1}{8\pi G_{2}}\int_{\partial\mathcal{M}}\rmd \tau\sqrt{h}\frac{\Phi}{\ell}\;.\label{eq:IctJT}\eeq
with $\ell$ being a bulk curvature scale (determined by the dilaton potential $V(\Phi)$, as in (\ref{eq:EOM dilaton})).

Models of the type (\ref{eq:I centaur}) have a higher-dimensional pedigree for specific dilaton potentials and asymptotic structure. In particular, when $V(\Phi)=+2\Phi/\ell^{2}$, the action is that of $\text{AdS}_{2}$ Jackiw-Teitelboim (JT) gravity \cite{Jackiw:1984je,Teitelboim:1983ux}, which follows from a spherical-dimensional reduction of black hole spacetimes whose extremal limit takes the form $\text{AdS}_{2}\times X$ for compact space~$X$. In this context, the dilaton $\Phi$ controls the size of $X$, and thus represents (small) deviations away from extremality. The resulting 2D geometry describes an $\text{AdS}_{2}$ black hole, with horizon entropy $S_{\text{BH}}=(\Phi_{0}+\Phi_{h})/4G_{2}$, where $\Phi_{0}$ corresponds to the extremal entropy of the higher-dimensional black hole, and $\Phi_{h}$ is the value of the dilaton evaluated at the $\text{AdS}_{2}$ black hole horizon. Analogously, the potential $V(\Phi)=-2\Phi/\ell^{2}$ yields $\text{dS}_{2}$ JT gravity, which follows from a spherical reduction of higher-dimensional de Sitter black holes in their near-horizon, near-Nariai limit, which has the product structure $\text{dS}_{2}\times X$, and $\Phi$ represents deviations from the Nariai limit, where the cosmological and black hole horizons coincide \cite{Maldacena:2019cbz,Svesko:2022txo}. The 2D spacetime can be interpreted as a $\text{dS}_{2}$ black hole, where $\Phi_{0}$ is proportional to the entropy of the higher-dimensional Nariai black hole. 
We will be interested in 2D spacetimes which interpolate between $\text{AdS}_{2}$ and $\text{dS}_{2}$, corresponding to an interpolation between the dilaton potentials characterizing $\text{(A)dS}_{2}$ JT gravity \cite{Anninos:2017hhn,Anninos:2022hqo}.

The gravitational and dilaton equations of motion of (\ref{eq:I centaur}) are, respectively,
\begin{align}
    \nabla_\mu\nabla_\nu\Phi-g_{\mu\nu}\nabla^2\Phi+\frac{1}{2}g_{\mu\nu}V(\Phi)=0&~,\label{eq:EOM gravity}\\
    R=-\partial_{\Phi}V(\Phi)&~.\label{eq:EOM dilaton}
\end{align}
Clearly, the gravity equation of motion governs the dynamics of the dilaton, while the dilaton equation of motion fixes the background metric. A general  solution to the equations of motion can be written in the form (cf. \cite{Cavaglia:1998xj})
\begin{equation}\label{eq:static patch sol}
    \rmd s^2=N(r)\rmd\tau^2+\frac{\rmd r^2}{N(r)}~,\quad \Phi=\Phi_r \frac{r}{\ell}~,
\end{equation}
for positive, dimensionless constant $\Phi_{r}$.  We further impose $\Phi_{0}\gg \Phi_{r}$; for theories that arise from a dimensional reduction, this condition is a direct consequence of the parent geometry being a nearly-extremal black hole.  
The Ricci scalar of the metric is $R=-\partial^{2}_{r}N(r)$ such that
\begin{equation}\label{eq:N(r) factor}
    N(r,r_{h})=\frac{\ell}{\Phi_{r}}\int_{r_h}^r\rmd r'\,V(r')~.
\end{equation}
These types of solutions describe a 2D Euclidean geometry with $r_h$ as the location of its horizon, where $N(r_{h},r_{h})=0$ and $N(r,r_{h})>0$ for all $r\neq r_{h}$ such that the geometry is Euclidean. Additionally, the period of Euclidean time $\tau$ is $\tau\sim \tau+\beta_{\text{H}}$, where \beq\beta_{\text{H}}=\frac{4\pi\Phi_{r}}{\ell|V(r_{h})|}\eeq upon expanding (\ref{eq:static patch sol}) near $r_{h}$ and removing the conical singularity at $r_h$.\footnote{\label{fn:beta def}The period $\beta_{\text{H}}$ coincides with the inverse of the Hawking temperature, $\beta_{\text{H}}=2\pi/\kappa=4\pi/|N'(r_{h})|$, for surface gravity $\kappa$, defined by $\xi^{\mu}\nabla_{\mu}\xi^{\nu}=\kappa \xi^{\nu}$ on the horizon with respect to the Killing field $\xi=\partial_{\tau}$.}

Some additional care is needed depending on whether we are describing a geometry with a black hole or cosmological horizon (both feature in de Sitter JT gravity).  In either case, being a Euclidean section requires $N(r,r_{h})>0$ for all $r\neq r_{h}$. As written, (\ref{eq:N(r) factor}) is non-negative for $r\geq r_{h}$, with $r_{h}$ denoting the black hole horizon. Alternatively, for systems with a cosmological horizon $r_{c}\geq r$, one instead has $N(r,r_{c})\propto \int_{r}^{r_{c}}dr' V(r')<0$ \cite{Anninos:2022hqo}.

\subsection*{Adding a Dirichlet wall and quasi-local thermodynamics}

Now we impose Dirichlet boundary conditions at the boundary $\partial \mathcal  M$ that fix   
the proper length $\beta_{\text{T}}$ of $\mathcal  M$, and the value of the dilaton at the boundary, $\Phi=\Phi_{B}$ \cite{Lemos:1996bq,Creighton:1995uj,Svesko:2022txo,Anninos:2022hqo}. On-shell, the value of the dilaton is proportional to the coordinate radius, $\Phi_B \propto r_B$. Hence, on-shell, fixing $\Phi_{B}$ amounts to fixing $r_B$. While the former boundary data properly defines the ensemble and is covariant, we often express thermodynamic quantities also in terms of $r_B$ for convenience when we relate the quasi-local thermodynamics to $\text{T}\overline{\text{T}}$ deformations. As pointed out above, the addition of a Dirichlet wall allows for a proper definition of thermal ensembles when the spacetime has a cosmological horizon. Indeed, Euclidean de Sitter space is a sphere, having no boundary, such that there is no location to fix thermodynamic data defining a thermal ensemble, as occurs in, for example, asymptotically AdS or flat black hole backgrounds.  The Dirichlet wall thus provides a location to fix the data defining a thermal ensemble. 

In 2D dilaton gravity, the solutions for the metric and dilaton with these boundary conditions are given by \eqref{eq:static patch sol}.  The periodicity of the Euclidean time is $\beta_{\text{H}}$, but the proper length of the time circle at $\partial \mathcal M$, which is fixed to be $\beta_{\text{T}}$, is equal on-shell to the inverse of the Tolman temperature $T=\beta_{\text{T}}^{-1}$,
\beq \beta_{\text{T}}(r_{B})=\int_{0}^{\beta_{\text{H}}}\rmd\tau\sqrt{N(r_{B},r_{h})}=\beta_{\text{H}}\sqrt{N(r_{B},r_{h})}\;.\eeq
Clearly, $T$ is the redshifted temperature locally measured at $\partial \mathcal M$. 

Following the standard Gibbons-Hawking analysis \cite{Gibbons:1976ue}, the canonical partition function $Z(\beta_{\text{T}})$ is represented as a Euclidean gravitational path integral, with integration measure over all the dynamical fields ($g_{\mu\nu}$ and $\Phi$) with the same Dirichlet boundary conditions defining the ensemble. In a semi-classical saddle-point approximation, the partition function is given by $Z(\beta_{\text{T}})\approx e^{-I_{\text{E}}}$, for on-shell action $I_{\text{E}}$. Specifically, for a system with a single black hole horizon (we describe the case with a cosmological horizon momentarily), the Euclidean action~(\ref{eq:I centaur}) on-shell is \cite{Svesko:2022txo,Anninos:2022hqo}
\beq \label{eq:action}
\begin{aligned}
I_{\text{E}}=&\frac{\beta_{\text{H}} \Phi_{0}}{16\pi G_{2}}\left(\int_{r_{h}}^{r_{B}}\rmd rN''(r)-N'(r_{B})\right)-\frac{\beta_{\text{H}}\Phi_{r}}{16\pi G_{2}\ell}\int_{r_{h}}^{r_{B}}\rmd r\left[-rN''(r)+N'(r)\right]\\
&-\frac{\beta_{\text{H}}\Phi_{B}}{16\pi G_{2}}N'(r_{B})+I_{\text{ct}}\;,    
\end{aligned}
\eeq
where we used $K=N'/2\sqrt{N}$, the trace of the extrinsic curvature of the boundary $\partial \mathcal M$,  and the prime denotes the derivative with respect to $r$. Evaluating and using integration by parts in the second term yields
\beq I_{\text{E}}=-\frac{1}{4G_{2}}(\Phi_{0}+\Phi_{h})-\beta_{\text{T}}\frac{\Phi_{r}\sqrt{N(r_{B})}}{8\pi G_{2}\ell}+\beta_{\text{T}}\frac{\Phi_{B}}{8\pi G_{2}\ell}\;,\label{eq:onshellactgenh}\eeq
with $\Phi_{h}=\Phi(r_{h})$ being the value of the dilaton at the black hole horizon. The third contribution is the  counterterm action evaluated on shell.
Using the identification $\log Z(\beta_{\text{T}})=-I_{\text{E}} $ in the saddle point approximation, the energy and entropy can be computed from the Euclidean on-shell action as follows
\begin{align}
E_{\text{BY}} &= \left ( \frac{\partial I_{\text{E}}}{\partial \beta_{\text{T}}}\right)_{\Phi_B} = \frac{\Phi_{r} }{8\pi G_{2}\ell} \left ( r_B /\ell- \sqrt{N(r_{B})} \right)\;,\label{eq:BYenbh} \\
S_{\text{BH}}&= \beta_{\text{T}} \left ( \frac{\partial I_{\text{E}}}{\partial \beta_{\text{T}}}\right)_{\Phi_B}  - I_{\text{E}}=\frac{\Phi_0 + \Phi_h}{4 G_2}\,, \label{BHentropy}
\end{align}
where we used $\partial_{r_{h}}N(r_{B},r_{h})=-\ell V(r_{h})/\Phi_{r}$ together with the chain rule $\partial_{\beta_{T}}=(\partial\beta_{\text{T}}/\partial r_{h})^{-1}\partial_{r_{h}}$.\footnote{It is useful to know $$\left(\frac{\partial\beta_{T}}{\partial r_{h}}\right)=-\frac{\beta_{\text{T}}}{N(r_{B})}\frac{[\ell V(r_{h})^{2}+2\Phi_{r} V'(r_{h})N(r_{B})]}{2\Phi_{r}V(r_{h})}\;.$$}
Here, $E_{\text{BY}}$ is the quasi-local thermodynamic energy, which can also be computed using the Brown-York stress-energy tensor  \cite{Brown:1992br}, see, e.g., equation (3.4) in \cite{Svesko:2022txo}. 
 The on-shell Euclidean action \eqref{eq:onshellactgenh} is thus proportional to the free energy 
\beq  I_{\text{E}}=\beta_{\text{T}} F =\beta_{\text{T}}E_{\text{BY}}-S_{\text{BH}}\;.\label{eq:log Z relation}\eeq
Further, the heat capacity at fixed $\Phi_B$ is
\beq C_{\Phi_{B}}\equiv-\beta_{\text{T}}^{2}\left(\frac{\partial E_{\text{BY}}}{\partial\beta_{\text{T}}}\right)_{\hspace{-1mm}\Phi_{B}}=\frac{\beta_{\text{T}}\Phi_{r}}{8\pi G_{2}\ell}\frac{\sqrt{N(r_{B})}V(r_{h})^{2}}{(V(r_{h})^{2}+2N(r_{B})V'(r_{h})\Phi_{r}/\ell)}\;,\label{eq:heatcapbh}\eeq
matching \cite{Grumiller:2007ju,Anninos:2022hqo}. We also define a ``surface pressure'' $\sigma$
\beq \label{eq:surface pressure}
\sigma\equiv-\left(\frac{\partial E_{\text{BY}}}{\partial\Phi_{B}}\right)_{\hspace{-1mm}S_{\text{BH}}}=\frac{1}{8\pi G_{2}}\left(\frac{1}{\ell}-\frac{N'(r_{B})}{2\sqrt{N(r_{B})}}\right)\;,\eeq
with fixed entropy $S_{\text{BH}}$ (equivalently, fixed $r_{h}$). 
The quasi-local energy, gravitational entropy, and surface pressure together obey the quasi-local first law for 2D dilaton-gravity \cite{Svesko:2022txo}
\beq\label{eq:1st law}
dE_{\text{BY}} = T d S_{\text{BH}} - \sigma d \Phi_B\;.
\eeq

\subsection*{\texorpdfstring{$\text{(A)dS}_{2}$}{} JT gravity}

For concreteness, let us apply the above formalism to two specific theories. 

\vspace{3mm}

\noindent \textbf{$\text{AdS}_{2}$:} First consider the case of $\text{AdS}_{2}$ JT gravity, characterized by Euclidean action (\ref{eq:I centaur}), with potential $V(\Phi)=2\Phi/\ell^{2}$. 
 The blackening factor is \beq N(r,r_{h})=(r^{2}-r_{h}^{2})/\ell^{2}\,,\eeq and the Ricci scalar is $R=-2/\ell^{2}$, such that the background is an eternal $\text{AdS}_{2}$ black hole. The (inverse) Tolman temperature is $\beta_{\text{T}}=\frac{\beta_{\text{H}}}{\ell}\sqrt{r_{B}^{2}-r_{h}^{2}}$, with $\beta_{\text{H}}=2\pi\ell^{2}/r_{h}$. Further, the quasi-local energy (\ref{eq:BYenbh}) takes the form
 \beq 
 \begin{split} 
 E_{\text{BY}}&=\frac{\Phi_{B}}{8\pi G_{2}\ell}\left(1-\frac{\beta_{\text{T}}}{\sqrt{4\pi^{2}\ell^{2}+\beta_{\text{T}}^{2}}}\right)=\frac{\Phi_{r}}{8\pi G_{2}\ell^{2}}\left(r_{B}-\sqrt{r_{B}^{2}-r_{h}^{2}}\right)\;.
 \end{split}
 \label{eq:AdS2BYen}\eeq
 The energy is 
 non-negative for all temperatures $\beta_{\text{T}} \in [0,\infty)$ (or for $r_{B}\geq r_{h}$). Notice the quasi-local energy multiplied by the lapse $\sqrt{N(r_{B})}$ and taking the $r_{B}\to\infty$ limit yields the black hole ADM mass \cite{Moitra:2019xoj,Pedraza:2021cvx}, 
 \begin{equation}\label{eq:M ADM}
     M_{\text{ADM}}=\frac{\Phi_{r}r_{h}^{2}}{16\pi G_{2}\ell^{3}}~.
 \end{equation} 
Notice also that, without multiplying by the lapse, for large $r_B$ the quasi-local energy   takes the form  $   M_{\text{ADM}} \ell / r_B$, which is according to the holographic dictionary \cite{Witten:1998zw,Savonije:2001nd,Visser:2021eqk} the dimensionful energy in the dual boundary theory. 

Meanwhile, the heat capacity (\ref{eq:heatcapbh}), 
 \beq\label{eq:Heat capacity rB}
 C_{\Phi_{B}}=\frac{4\pi^{2}\ell\Phi_{B}}{8\pi G_{2}}\frac{\beta_{\text{T}}^{2}}{(4\pi^{2}\ell^{2}+\beta_{\text{T}}^{2})^{3/2}}\,,\eeq
 is positive everywhere for $\beta_{\text{T}}>0$, except at $\beta_{\text{T}}=0$ where it vanishes. Hence, the AdS$_2$ black hole system with a Dirichlet boundary is thermodynamically stable.

\vspace{3mm}

\begin{figure}
    \centering
    \includegraphics[width=0.7\textwidth]{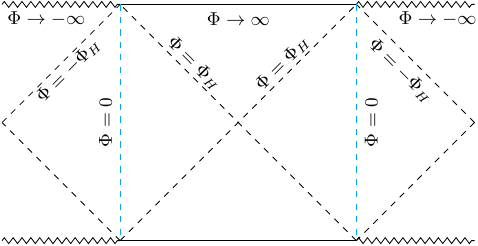}
    \caption{Full reduction model in dS$_2$ JT gravity. The black hole and cosmological horizons (black dashed lines) are located where $\Phi\rightarrow-\Phi_{H}$ and $\Phi\rightarrow+\Phi_{H}$. We have chosen the center of the static patches (cyan dashed lines) where $\Phi\rightarrow0$.}
    \label{fig:dS2Full}
\end{figure}

\noindent \textbf{$\text{dS}_{2}$:} Now consider the case of $\text{dS}_{2}$ JT gravity.  
The JT action (\ref{eq:I centaur}), with potential $V(\Phi)=-2\Phi/\ell^{2}$, follows from a spherical dimensional reduction of four or higher-dimensional de Sitter-Schwarzschild black hole in its near-horizon, near-Nariai limit (see, e.g., \cite{Maldacena:2019cbz,Cotler:2019nbi,Svesko:2022txo,Castro:2022cuo}). The solution to the equations of motion has Ricci scalar $R=+2/\ell^{2}$, such that the geometry is $\text{dS}_{2}$, namely, metric (\ref{eq:static patch sol}) with blackening factor \beq N(r_{H},r)=(r_{H}^{2}-r^{2})/\ell^{2}\,, \quad \text{with}\quad  -r_{H}\leq r\leq r_{H} \quad \text{for} \quad r_{H}>0\,. \eeq  The essential new feature with this set-up is that $\text{dS}_{2}$ has two horizons: a black hole horizon $r_{h}=-r_{H}$ and a cosmological horizon $r_{c}=r_{H}$ with $r_H >0$,\footnote{In \cite{Svesko:2022txo} $r_{H}=\ell$. Here we leave $r_{H}$ as an independent parameter.}
a herald of the higher-dimensional origin. A Penrose diagram for this  geometry is shown in Figure \ref{fig:dS2Full}.

Now introduce a timelike Dirichlet wall inside the static patch in between the two horizons, i.e., $r_{B}\in[r_{h},r_{c}]$.  
The quasi-local thermodynamics of this model was studied previously in \cite{Svesko:2022txo,Anninos:2022hqo}. The wall divides the static patch into two systems: the black hole system, $r\in[r_{h},r_{B}]$, and the cosmological system $r\in[r_{B},r_{c}]$. Treating each system separately gives\footnote{For the black hole system, the on-shell action is as in (\ref{eq:onshellactgenh}). Alternatively, to evaluate the on-shell action of the cosmological system, first note that the Gibbons-Hawking boundary term comes with a sign change since the outward pointing normal vector of the cosmological horizon has an opposite orientation to that of the black hole horizon. Additionally, the $r$-integration range is $r\in[r_{B},r_{c}]$.}
\beq \log Z_{c,h}(\beta_{\text{T}})=-\beta_{\text{T}}E^{c,h}_{\text{BY}}+S_{c,h}\;,\eeq
where the inverse Tolman temperature, horizon entropy   and quasi-local energies are, respectively, given on-shell by\footnote{To arrive at the second equality in \eqref{eq:EBY Nariai} we used $r_{H}=2\pi\ell |r_{B}|/\sqrt{4\pi^2\ell^{2}-\beta_{\text{T}}^{2}}$, which follows from inverting $\beta_{\text{T}}$ for $r_{H}$.}
\begin{align} \beta_{\text{T}}&=\frac{\beta_{\text{H}}}{\ell}\sqrt{r_{H}^{2}-r_{B}^{2}}  \\
S_{c,h}&=\frac{1}{4G_{2}}(\Phi_{0}\pm\Phi_{H})\\
E_{\text{BY}}^{c,h}&=\frac{1}{8\pi G_{2}\ell}\left(\Phi_{B}\pm\frac{|\Phi_{B}|\beta_{\text{T}}}{\sqrt{4\pi^{2}\ell^{2}-\beta^{2}_{\text{T}}}}\right) =\frac{\Phi_{r}}{8\pi G_{2}\ell^{2}}\left(r_{B}\pm\sqrt{r_{H}^{2}-r_{B}^{2}}\right)\;,\label{eq:EBY Nariai}\end{align}
where $\beta_{\text{H}}= 2\pi \ell^2 / r_H$ and $\Phi_{H}=r_{H}\Phi_{r}/\ell$. 
In both entropy and energy, the `$+$' sign refers to the cosmological system while the `$-$' sign corresponds to the black hole system. Geometrically, the `$\pm$' sign reflects the sign  change of the outward pointing normal vector $n^\alpha$ to $ \partial \mathcal M$, since the quasi-local energy (without counterterm) is defined in JT gravity as $E_{\text{BY}} = - \frac{1}{8\pi G_2} n^\alpha \partial_\alpha \Phi$~\cite{Svesko:2022txo}. Further, our expression for the quasi-local energy  (\ref{eq:EBY Nariai}) differs from  \cite{Svesko:2022txo,Anninos:2022hqo} as here we have included a local counterterm, since later we will embed $\text{dS}_{2}$ inside an asymptotically $\text{AdS}_{2}$ background.\footnote{Without the counterterm, the quasi-local energy of the cosmological system is always non-negative, while it is always non-positive for the black hole system.}
The quasi-local energy of the cosmological system is real and non-negative when
$\beta_{\text{T}}\in [0, 2\pi \ell]$.   
Similarly, for the black hole system the quasi-local energy is non-negative when $\beta_{\text{T}}\in [0,\sqrt{2}\pi \ell]$ and negative for $\beta_{\text{T}}> \sqrt{2}\pi \ell$. For $\beta_{\text{T}} > 2\pi \ell$, or equivalently for $r_{B}<-r_{H}$ and $r_{B}>r_{H}$, the energy of both systems  becomes complex, i.e., when the Dirichlet wall is placed in the black hole interior or inflating patch, respectively. 
Further, multiplying the energy by the lapse and taking the limit as the wall approaches either horizon yields $\lim_{r_{B}\to r_{c,h}}\sqrt{N(r_{B})}E_{\text{BY}}^{c,h}=0$. This is in accordance with the standard Gibbons-Hawking result of vanishing energy in the dS static patch. 

The heat capacity, meanwhile, is
\beq \label{eq:heat capacity BH CH}C^{c,h}_{\Phi_{B}}=\mp\frac{4\pi^{2}\ell|\Phi_{B}|}{8\pi G_{2}}\frac{\beta_{\text{T}}^{2}}{(4\pi^{2}\ell^{2}-\beta_{\text{T}}^{2})^{3/2}}\;,\eeq
where the minus sign refers to the cosmological system and plus sign to the black hole system. Thus, while the heat capacity of the black hole system is \emph{non-negative} everywhere in the range   $\beta_{\text{T}} \in [0, 2\pi \ell]$ (or $r_{B}\in[-r_{H},r_{H}]$), and hence thermally stable (at least locally),  the heat capacity for the cosmological system is everywhere \emph{non-positive} in the same range, indicating it is unstable against thermal fluctuations. Further, the black hole and cosmological systems represent two saddles to the Euclidean path integral, with the dominant one being the cosmological system as it has lower free energy, $F_{h}\geq F_{c}$, where $F_{c,h}=E^{c,h}_{\text{BY}}-T_{\text{T}}S_{c,h}$.

Note that $\text{dS}_{2}$ JT gravity also arises from a circular reduction of empty $\text{dS}_{3}$, leading to the so-called ``half-reduction'' model (opposed to the ``full-reduction'' model from near-Nariai de Sitter black holes considered here) \cite{Sybesma:2020fxg,Kames-King:2021etp}. The essential geometric difference is that now the $\text{dS}_{2}$ geometry has no black hole horizon, whereas the static patch radial coordinate has the range $r\in[0,\ell]$. Moreover, $\Phi_{0}=0$, as there is no Nariai black hole from whence the 2D geometry came. Consequently, the Dirichlet wall divides the static patch into a ``pole patch'' (the region between the boundary and the pole $r=0$) and ``cosmic horizon patch'' (the region between the cosmic horizon and the wall). The quasi-local thermodynamics of the cosmic horizon patch is identical to the cosmological system above, for the restricted range $r_{B}\in[0,r_{H}]$  \cite{Svesko:2022txo}. In this article, however, we focus on the full-reduction model.

For AdS$_{2}$-JT gravity, the black hole horizon $r_{h}$ is related to the ADM mass (\ref{eq:M ADM}). Analogously, the dS$_{2}$ horizon radius $r_{\text{H}}$ can be related to an ADM-like conserved charge as defined in \cite{Balasubramanian:2001nb}. This definition of mass for asymptotically dS$_{d+1}$ spacetimes directly  follows from constructing a quasi-local Brown-York stress tensor at early and late temporal infinity $\mathcal{I}^{\pm}$, i.e., outside the purview of a static patch observer.
For example, consider the Schwarzschild-dS$_{3}$ conical defect
\begin{equation}
    \rmd s^2=-N(r) \rmd\tau^2+\frac{\rmd r^2}{N(r)}+r^2 \rmd\phi^2\;, 
\end{equation}
with $N(r)=(8G_3M-r^2/\ell^{2})$, and cosmological horizon at $r_{c}=\ell\sqrt{8G_{3}M}$. The parameter $M$ is found to equal the ADM `mass', computed from the quasi-local stress tensor outside the cosmological horizon $(r>\ell\sqrt{8GM})$ near $\mathcal{I}^{\pm}$.

Motivated by \cite{Balasubramanian:2001nb}, which only considered pure Einstein-dS$_{d+1}$ gravity for $d\geq2$, we can  extend their proposal to dS-JT gravity and define an ADM-like mass for asymptotically dS$_{2}$ geometries. To wit, Lorentzian dS$_{2}$ has line element
\beq ds^{2}=-N(r) \rmd t^{2}+N^{-1}(r)\rmd r^{2} \quad \text{with} \quad N(r)=(r_{H}^{2}-r^{2})/\ell^{2}\,.\eeq Outside of the region accessible to a static patch observer, $r>r_{c}$, the coordinate $r$ becomes timelike while $t$ becomes a spatial coordinate, and  $\mathcal{I}^{\pm}$ are surfaces of large $r$. Using the definition of the quasi-local Brown-York stress-tensor at a constant $r=r_{B}>r_{c}$ slice, the quasi-local energy takes the same form as for AdS$_{2}$ (\ref{eq:AdS2BYen}), with ADM mass (\ref{eq:M ADM}), where $r_{h}$ has been replaced by $r_{H}$.

\subsection{\texorpdfstring{$\text{T}\overline{\text{T}}$}{} deformations and the microscopic energy spectrum}\label{SubSec:Energy spectrum}

\noindent \textbf{3D/2D correspondence.} Historically, $\text{T}\overline{\text{T}}$ deformations refer to the simplest example of solvable irrelevant deformations of a 2D CFT \cite{Smirnov:2016lqw,Cavaglia:2016oda}. This amounts to turning on a coupling $\mu$ resulting in a one-parameter family of local quantum field theories with Euclidean action 
\beq I_{\text{QFT}}=I_{\text{CFT}}+\mu\int \rmd^{2}x\sqrt{\gamma}\,\text{T}\overline{\text{T}}(x)\;,\label{eq:defQFTaction}\eeq
where $I_{\text{CFT}}$ is referred to as the `seed' CFT, and $\text{T}\overline{\text{T}}(x)$ is a specific irrelevant, local, composite operator, given in terms of the quadratic product of the components of the deformed field theory stress-energy tensor $T_{ij}$:\footnote{Technically $\text{T}\overline{\text{T}}$ is bilocal, $8\text{T}\overline{\text{T}}(x,y)=T^{ij}(x)T_{ij}(y)-T^{i}_{i}(x)T^{j}_{j}(y)$. Due to its OPE structure as $x\to y$, however, one typically works with the local operator $\text{T}\overline{\text{T}}(x)$ \cite{Zamolodchikov:2004ce}. Further, in complex Cartesian coordinates $z=x+i\tau$, $8\text{T}\overline{\text{T}}=T_{zz}T_{\bar{z}\bar{z}}-(T_{z\bar{z}})^{2}=\text{det}(T_{ij})$, with trace $T^{i}_{i}=4T_{z\bar{z}}$. Often one writes $T\equiv T_{zz}$, $\overline{T}\equiv T_{z\bar{z}}$, and $T_{z\bar{z}}\equiv\Theta$, such that $8\text{T}\overline{\text{T}}=T\bar{T}-\Theta^{2}$. When the deformed CFT is another CFT, $T^{i}_{i}=0$.}
\beq \text{T}\overline{\text{T}}(x)\equiv\frac{1}{8}\left[T^{ij}(x)T_{ij}(x)-(T^{i}_{i}(x))^{2}\right]\;.\label{eq:TTbardef}\eeq
Moreover, from $T_{ij}\equiv \frac{2}{\sqrt{\gamma}}\frac{\delta}{\delta\gamma^{ij}}I_{\text{QFT}}$, the deformed stress-tensor obeys $\gamma^{ij}T_{ij}=-2\mu\text{T}\overline{\text{T}}$. Incidentally, the deformation (\ref{eq:TTbardef}) preserves Lorentz invariance. 
The QFT generating function $Z_{\text{QFT}}$ obeys 
\beq \partial_{\mu}\log Z_{\text{QFT}}(\mu)=-\int \rmd^{2} x\sqrt{\gamma}\langle \text{T}\overline{\text{T}}\rangle\;,\eeq
where the initial condition of the flow is $Z_{\text{QFT}}(0)=Z_{\text{CFT}}$.

A key simplifying feature of $\text{T}\overline{\text{T}}$ deformations is that, for a finite $\mu$, the expectation value of the composite operator in any translation invariant, stationary state factorizes as \cite{Zamolodchikov:2004ce}
\beq \langle \text{T}\overline{\text{T}}\rangle= \langle T\rangle\langle \overline{T}\rangle-\langle \Theta\rangle^{2}\;,\label{eq:factorizationTT}\eeq
for $\Theta\equiv\frac{1}{4}T^{i}_{\;i}$. This factorization can be used to derive an exact renormalization group (RG) equation for the deformed theory, which coincides with a Hamilton-Jacobi equation whose classical limit corresponds to the Wheeler-DeWitt constraint characterizing the radial evolution of the three-dimensional gravity wavefunction \cite{McGough:2016lol}. Moreover, with respect to energy-momentum eigenstates $|n\rangle$ of the deformed Hamiltonian, the factorization (\ref{eq:factorizationTT}) allows for an exact computation of the deformed energy spectrum. Specifically, in complex Cartesian coordinates $z=x+i\tau$ with Euclidean time $\tau$, 
\beq \langle n|\text{T}\overline{\text{T}}|n\rangle=-\frac{1}{4}\left[\langle n|T_{xx}|n\rangle\langle n|T_{\tau\tau}|n\rangle-\langle n|T_{x\tau}|n\rangle^{2}\right]\;.\eeq
When the deformed CFT is on a cylinder with circumference $L$, one has
a quantized energy $E_{n}\equiv L\langle n|T_{\tau\tau}|n\rangle$, pressure $\partial_{L}E_{n}\equiv\langle n|T_{xx}|n\rangle$, and momentum $iP_{n}=L\langle n|T_{\tau x}|n\rangle$, where the factor of $i$ appears since $T_{zz}-T_{\bar{z}\bar{z}}=-iT_{x\tau}$. Moreover, from the (Euclidean) interaction Hamiltonian, one finds $\partial_{\mu}\langle H\rangle=L\langle \text{T}\overline{\text{T}}\rangle$, while $\partial_{\mu}P_{n}=0$, since the momentum is quantized in units of $2\pi/L$. 

Altogether, one finds the factorization (\ref{eq:factorizationTT}) becomes the (forced inviscid) Burgers equation describing turbulent fluids,
\beq\label{eq:Burges} 4\partial_{\mu}E_{n}+E_{n}\partial_{L}E_{n}+\frac{P_{n}^{2}}{L}=0\;.\eeq
Introducing the dimensionless parameter 
\begin{equation}\label{eq:def lambda}
\lambda\equiv\pi\mu/L^{2}~,    
\end{equation}
the Burgers equation becomes
\beq \label{eq:curly En}4\pi\partial_{\lambda}\mathcal{E}_{n}-2\lambda\mathcal{E}_{n}\partial_{\lambda}\mathcal{E}_{n}-\mathcal{E}_{n}^{2}+P_{n}^{2}L^{2}=0\;,\eeq
with $\mathcal{E}_{n}=E_{n}L$. The solution is found to be \cite{Smirnov:2016lqw,Cavaglia:2016oda} (see also \cite{Zamolodchikov:2004ce})
\beq\label{eq:En sols} E_{n}(\lambda)= \frac{2\pi}{\lambda L}\left(1-\sqrt{1-2\lambda M_{n}+\lambda^{2}J_{n}^{2}}\right)\;,\eeq
where the seed CFT gives the $\lambda=0$ initial conditions, i.e., $E_{n}(\lambda=0,L)=2\pi M_{n}/L$ and $P_{n}=2\pi J_{n}/L$. Using $M_{n}=\Delta_{n}+\bar{\Delta}_{n}-\frac{c}{12}$ and $J_{n}=\Delta_{n}-\bar{\Delta}_{n}$ for the undeformed CFT (where $(\Delta_{n},\bar{\Delta}_{n})$ are left and right scaling dimensions) the Casimir energy $E_{0}$ is 
\beq E_{0}(\lambda)=\frac{2\pi}{\lambda L}\left(1-\sqrt{1+\lambda\frac{c}{6}}\right)\;,\eeq
where $\Delta_{0}=\bar{\Delta}_{0}=0$, and $c$ is the central charge of the seed CFT.

Since the deformed energy is monotonic in $(\Delta_{n},\bar{\Delta}_{n})$, the $E_{n}$ levels do not intersect as $\mu$ or $L$ are changed. This implies the thermal entropy of the deformed theory is independent of $\mu$ and, at high energies, given by the Cardy entropy of the undeformed $\text{CFT}_{2}$, i.e., $S_{\text{CFT}}=2\pi\sqrt{\frac{c}{6}(\Delta_{n}-\frac{c}{24})}+2\pi\sqrt{\frac{c}{6}(\bar{\Delta}_{n}-\frac{c}{24})}$. 

Due to the square-root `shockwave singularity' \cite{Smirnov:2016lqw} that occurs for particular finite values of $\lambda$ (where the square-root vanishes), the energy spectrum of the deformed QFT truncates above some critical value of the scaling dimension $\Delta_{n}$, and places an upper bound on the energy and entropy. Thus, the deformed QFT on the cylinder has a finite number of quantum states \cite{McGough:2016lol}, bounding the energy and entropy. Lastly, observe, outside of the square-root singularity, the deformed energy spectrum (\ref{eq:En sols}) takes complex values when $1+\lambda^{2}J_{n}^{2}<2\lambda M_{n}$. 

When the deformed theory is taken to have a holographic dual,\footnote{The base assumption is to start with a holographic seed $\text{CFT}_{2}$ in a large$-c$ and large 't Hooft coupling limit, where one is able to restrict to the universal sector captured by three-dimensional Einstein gravity.} it has been proposed the deformation represents a geometric cutoff in a three-dimensional asymptotically AdS spacetime, where the QFT ends up residing on a Dirichlet wall at a fixed radial position $r_{B}$, such that tuning the parameter $\mu$ amounts to tuning the position of the wall \cite{McGough:2016lol}. This proposal is based on a number of observations. Specifically, directly compare the deformed spectrum (\ref{eq:En sols}) to the Brown-York quasi-local energy of the Ba\~{n}ados-Teitelboim-Zanelli (BTZ) black hole \cite{Banados:1992wn,Banados:1992gq} of mass $M_{\text{ADM}}$, spin $J$ and $\text{AdS}_{3}$ length scale $\ell_{3}$ (cf. \cite{Brown:1994gs})
\beq E_{\text{BY}}^{\text{BTZ}}=\frac{r_{B}}{4G_{3}\ell_{3}}\left(1-\sqrt{1-\frac{8G_{3}M_{\text{ADM}}\ell_{3}^{2}}{r_{B}^{2}}+\frac{16 G_{3}^{2}\ell^{2}_{3}J^{2}}{r_{B}^{4}}}\right)\;.\eeq
One finds precise agreement with the spectrum (\ref{eq:En sols}) upon invoking the standard $\text{AdS}_{3}/\text{CFT}_{2}$ dictionary to match the BTZ mass and angular momentum to the seed initial conditions, $\ell_{3}M_{\text{ADM}}=M_{n}$ and $J=J_{n}$, in addition to setting $L=2\pi r_{B}$ and identifying $\lambda\equiv \frac{4G_{3}\ell_{3}}{r_{B}^{2}}$. At $\lambda =0 $ the BY energy reduces to $E_{\text{BY}}^{\text{BTZ}} (r_B \to \infty) =  M_{\text{ADM}} \ell_3 / r_B$, which is dual according  to the holographic dictionary \cite{Witten:1998zw,Savonije:2001nd,Visser:2021eqk} to the (dimensionful) CFT$_2$ energy $2\pi M_n / L$. Thus, a $\text{T}\overline{\text{T}}$ flow starting at $\lambda=0$ and increasing toward some fixed $\lambda>0$ has a dual description as moving the $\text{AdS}_{3}$ conformal boundary inward from infinity to $r=r_{B}$.

Further, as demonstrated in the two-dimensional context, including a Dirichlet boundary does not alter the form of the horizon entropy: the thermal entropy of the black hole continues to follow the Bekenstein-Hawking area law, i.e., $S_{\text{BH}}=\frac{2\pi r_{h}}{4G_{3}}$ where $r_{h}$ is the black hole horizon radius. This is consistent with the dual picture, where the entropy of the deformed field theory is still given by the Cardy entropy. 
Moreover, gravitationally speaking, the square root singularity in the deformed energy spectrum occurs when the Dirichlet wall approaches the location of the BTZ event horizon, $r_{B}\to r_{h}$, where the upper bound on energy and entropy coincide with a BTZ black hole which entirely fills the $\text{AdS}_{3}$ region cutoff at the wall. 

Aside from matching the deformed energy spectrum to the quasi-local energy, additional non-trivial observations have been made relating bulk and boundary sides of the correspondence. These include computing propagation speeds of small perturbations about thermal states \cite{McGough:2016lol}, and the computation of stress-tensor correlation functions \cite{Kraus:2018xrn}.

\vspace{2mm}

\noindent \textbf{Dimensional reduction to 2D/1D system.} The goal now is to perform a similar analysis in one fewer dimension to provide a microscopic interpretation of the aforementioned quasi-local thermodynamics in $\text{A}\text{dS}_{2}$. This was accomplished by a dimensional reduction of $\text{T}\overline{\text{T}}$ deformations of two-dimensional CFTs \cite{Gross:2019ach,Gross:2019uxi}.\footnote{See also the dimensional reduction of massive gravity generalizations of stress-tensor deformations \cite{Tsolakidis:2024wut}.} To this end, it proves useful to first rewrite the composite operator $\text{T}\overline{\text{T}}$ in the following way. Introduce angular coordinate $\theta$ such that the QFT metric is diagonal. Using $T^{i}_{\;i}=-2\mu\text{T}\overline{\text{T}}$ to solve for $T^{\theta}_{\;\theta}$, the flow equation for the deformed QFT action (\ref{eq:defQFTaction}) becomes
\beq \partial_{\mu}I_{\text{QFT}}=\int \rmd^{2}x\sqrt{\gamma}\left(\frac{(T^{\tau}_{\;\tau})^{2}+T_{\tau\theta}T^{\tau\theta}}{4-2\mu T^{\tau}_{\;\tau}}\right)\;.\label{eq:flowv2}\eeq
A pole appears when $T^{\tau}_{\;\tau}=2/\mu$; with $E=\int d\theta T^{\tau}_{\;\tau}=LT^{\tau}_{\;\tau}$, this pole coincides with the square-root singularity of the deformed spectrum (\ref{eq:En sols}). Moreover, with $iP_{n}=T_{\tau\theta}$, one recovers the flow equation (\ref{eq:Burges}). The advantage of recasting the deformations in this way is one may now perform a simple dimensional reduction of the two-dimensional equation~(\ref{eq:flowv2})~\cite{Gross:2019ach}. This amounts to setting $T_{\tau\theta}=0$, or, equivalently, $P_{n}=J_{n}=0$ in the deformed energy spectrum~(\ref{eq:En sols}), yielding
\beq\partial_{\mu}(E/L)=\frac{(E/L)^{2}}{4-2\mu E/L}\;\;\Longrightarrow \;\; E(\lambda)=\frac{2\pi}{\lambda L}\qty(1-\sqrt{1-2\lambda\mathcal{E}_{0}})\;,\label{eq:defen1d}\eeq
where the undeformed CFT energy $M_{n}$ has been replaced by the $E(\lambda=0)\equiv 2\pi \mathcal{E}_{0}/L$ 
energy of the one-dimensional seed quantum mechanical theory, whatever it may be (more on this later). The square-root singularity appears when $1=2\lambda \mathcal{E}_{0}$, while the spectrum  goes complex if $2\lambda \mathcal{E}_{0}>1$. 

 \begin{figure}[t!]
    \centering
\includegraphics[width=9cm]{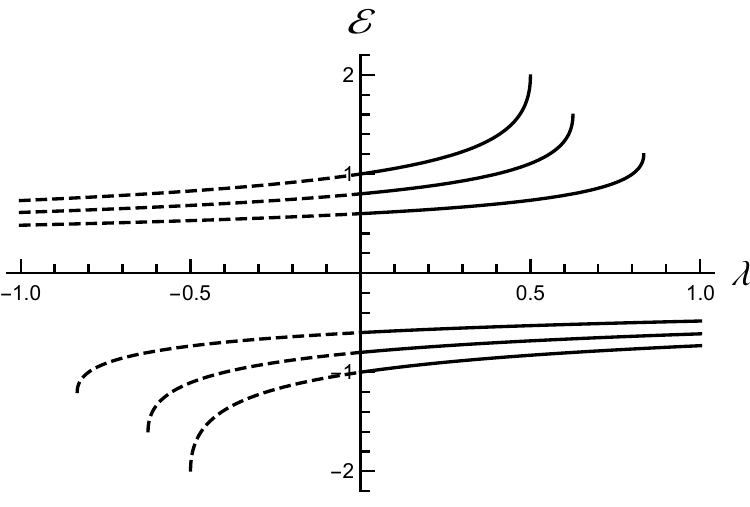}
    \caption{Energy levels $\mathcal{E}$ of the dual theory to AdS$_2$ JT gravity with a finite cutoff at fixed $\mathcal{E}_{0}$ (\ref{eq:EAdS2JT}) (from top to bottom: $\mathcal{E}_{0}=1,.8,.6,-.6,-.8,-1$). Solid lines indicate $\lambda>0$, while dashed lines correspond to $\lambda<0$. This plot is qualitatively identical to energy levels of the 2D effective theory dual to $\text{AdS}_{3}$ gravity with a finite cutoff \cite{McGough:2016lol}.}
    \label{fig:En lambda L}
\end{figure}

Notice, for example, the deformed energy spectrum (\ref{eq:defen1d}) precisely agrees with the quasi-local energy of the $\text{AdS}_{2}$ black hole (\ref{eq:AdS2BYen}) upon identifying $\mathcal{E}_{0}=\ell M_{\text{ADM}}$\footnote{We rescale the mass since $\Phi_{r}$ is dimensionless to keep $\lambda$ dimensionless.} together with $L=2\pi r_{B}$ and 
\beq\label{eq:lambda bulk}
\lambda\equiv\frac{8\pi G_{2}\ell^{2}}{\Phi_{r}r_{B}^{2}}\;.\eeq
As the Dirichlet wall is pushed to the $\text{AdS}_{2}$ black hole asymptotic boundary, the flow parameter vanishes. 
For a one-dimensional quantum mechanical theory, it is more natural to work with the dimensionless quantity\footnote{To arrive at the second equality it is useful to know $r_{h}=2\pi\ell r_{B}/\sqrt{4\pi^{2}\ell^{2}+\beta_{\text{T}}^{2}}$ and recall energy (\ref{eq:AdS2BYen}).} 
\beq
\mathcal{E}\equiv r_{B}E(\lambda)=\frac{1}{\lambda}\qty(1-\sqrt{1-2\lambda \mathcal{E}_{0}})= \frac{1}{\lambda} \left ( 1 - \frac{\tilde{\beta}_{\text{T}}}{\sqrt{\frac{2}{3}\pi^2 c \lambda+  \tilde{\beta}_{\text{T}}^{2}}}\right)\;,\label{eq:EAdS2JT}\eeq
  where $\tilde{\beta}_{\text{T}}\equiv \beta_{\text{T}}/ r_B  (=\tilde T^{-1})$ is the dimensionless (inverse) temperature of the deformed theory, and we defined a constant that is analogous to the central charge in 2D CFTs, 
  \beq \label{dictionaryc}
\frac{c}{3} \equiv \frac{\Phi_r}{4 \pi G_2}
  \,.  \eeq 
  At $\lambda=0$ the energy reduces to $\mathcal E (\lambda=0) = \frac{c}{3}\pi^2 \tilde T^2_{\text{H}}$, where $\tilde T_{\text{H}} = \tilde  T (\lambda =0) = T_{\text{H}} \ell $   is the dimensionful   boundary temperature of the seed theory. This is identical to the energy-temperature relation for 2D CFTs with $c$ is the central charge, showing the deformed theory in 1D can be understood as the dimensional reduction of the 2D $\text{T}\overline{\text{T}}$ deformed theory. 

   \begin{figure}[t!]
    \centering
\includegraphics[width=8cm]{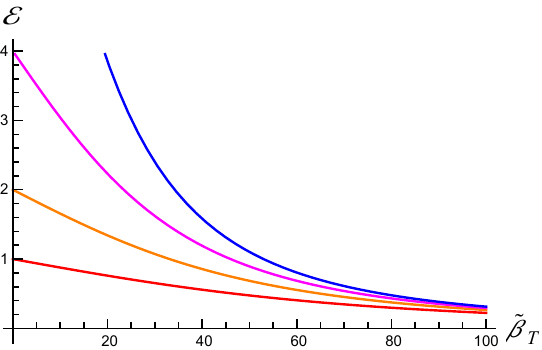}$\quad$\includegraphics[width=8cm]{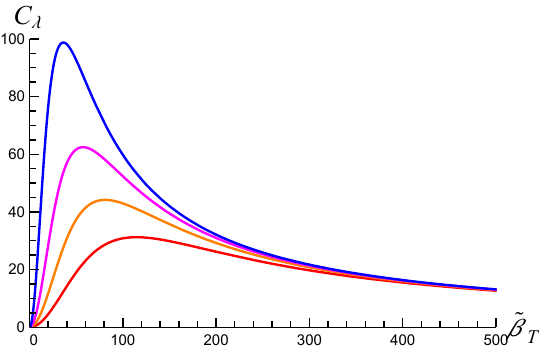}
    \caption{\textbf{Left:} Energy levels of the dual theory to AdS$_2$ JT gravity with a finite cutoff (\ref{eq:EAdS2JT}) at fixed $\lambda$ (from bottom to top: $\lambda=$ 1 (red), .5 (orange), .25 (magenta), .1 (blue)) and $c=1000$. \textbf{Right:} Heat capacity (\ref{eq:heatcapmic}) at fixed $\lambda$ for same parameter range.}
    \label{fig:EandCbeta}
\end{figure}
  
  Further, the heat capacity at fixed $\lambda$ is
\beq 
 C_\lambda\equiv - \tilde{\beta}_{\text{T}}^{2} \left   ( \frac{\partial \mathcal E (\lambda,~\tilde{\beta}_{\rm T})}{\partial \tilde{\beta}_{\text{T}}}\right )_{\hspace{-1mm}\lambda} = \frac{1}{\lambda}\frac{\frac{2}{3} \pi^2c \lambda \tilde{\beta}_{\text{T}}^2}{(\frac{2}{3} \pi^2 c \lambda  +  \tilde{\beta}_{\text{T}}^2)^{3/2}}= C_{\Phi_B} \,,
 \label{eq:heatcapmic}\eeq
 which is manifestly positive for all real values of $\tilde{\beta}_{\text{T}}$ and $\lambda>0$. We note that fixing $\Phi_B$ in the bulk corresponds to fixing $\lambda $ in the boundary theory, since $\lambda \propto 1/ \Phi_B$. Therefore, the proper data that   defines a canonical thermodynamic ensemble in the boundary theory is the (dimensionless) temperature $\tilde \beta_{\text T}$ and the (dimensionless) deformation parameter~$\lambda.$ Figure \ref{fig:En lambda L} displays the energy levels as a function of the flow parameter at fixed~$\mathcal E_0$, and is qualitatively identical to the energy levels of the 2D effective theory dual to $\text{AdS}_{3}$ gravity with a finite cutoff (see Figure 1 of \cite{McGough:2016lol}). However, in the canonical ensemble it is more appropriate to present  the microscopic energy levels and corresponding heat capacity as a function of inverse temperature $\tilde{\beta}_{\text{T}}$ at fixed $\lambda$ (see   Figure \ref{fig:EandCbeta}).

\subsection{\texorpdfstring{$\text{T}\overline{\text{T}}$}{} deformation for theories dual to general 2D dilaton theories} 

A more direct computation of the one-dimensional deformed energy spectrum (\ref{eq:defen1d}) -- one that is also valid for more general dilaton theories of gravity -- can be given by adapting a proposed generalization of higher-dimensional $\text{T}\overline{\text{T}}$ deformations of conformal field theories in a large-$N$ expansion \cite{Hartman:2018tkw} (see also \cite{Taylor:2018xcy}). The idea is the following. Begin by analyzing the Hamiltonian constraint of the bulk theory, e.g., the general dilaton gravity theory (\ref{eq:I centaur}), which is in part given in terms of components of the quasi-local Brown-York stress-tensor. From the Hamiltonian constraint, one defines a bulk flow equation for the on-shell dilaton gravity action. Assuming the standard holographic dictionary holds for a bulk gravity theory with Dirichlet boundary conditions at a finite timelike wall, the bulk flow equation then \emph{defines} the flow equation of the effective one-dimensional quantum mechanical theory, $I_{\text{EFT}}$. One is consequently led to the following flow equation \cite{Gross:2019ach} (see Appendix \ref{app:TTdefsgen2D} for details)
\beq \partial_{\lambda}I_{\text{EFT}}=\int\rmd\tau\sqrt{\gamma}\biggr[\frac{(T^{\tau}_{\;\tau})^{2}-(\ell\lambda)^{-2}\left(1-\frac{\ell^{2}}{2\Phi_{r}}\sqrt{\frac{c\lambda}{6}}V\qty(\Phi_{r}\sqrt{\frac{6}{c\lambda}})\right)}{2\ell^{-1}-2\lambda T^{\tau}_{\;\tau}}\biggr]\;,\label{eq:flowIFTgendil}\eeq
where we have introduced the parameter $c\equiv\frac{3\Phi_{r}}{4\pi G_{2}}$.
Writing $T^{\tau}_{\;\tau}=E(\lambda)$, the flow equation characterizing the deformed energy spectrum $E(\lambda)$ is
\beq \partial_{\lambda}E=\frac{E^{2}-(\ell\lambda)^{-2}\left(1-\frac{\ell^{2}}{2\Phi_{r}}\sqrt{\frac{c\lambda}{6}}V\qty(\Phi_{r}\sqrt{\frac{6}{c\lambda}})\right)}{2\ell^{-1}-2\lambda E}\;,\label{eq:Flow eq Dilaton gravity}\eeq
with a generic solution for $\mathcal{E}(\lambda)=r_BE(\lambda)$, 
\beq \mathcal{E}(\lambda)=\frac{1}{\lambda}\left(1\pm\sqrt{C_{1}\lambda+\lambda \frac{c}{6}~f\left(\Phi_{r}\sqrt{\frac{6}{c\lambda}}\right)}\right)\;,\label{eq:enspecgendila}\eeq
 for integration constant $C_{1}$ and where
    \beq  f(x)\equiv\frac{\ell^{2}}{\Phi^{2}_{r}}\int^{x}V(y)dy\;.\eeq
\begin{figure}[t!]
    \centering
\includegraphics[width=9cm]{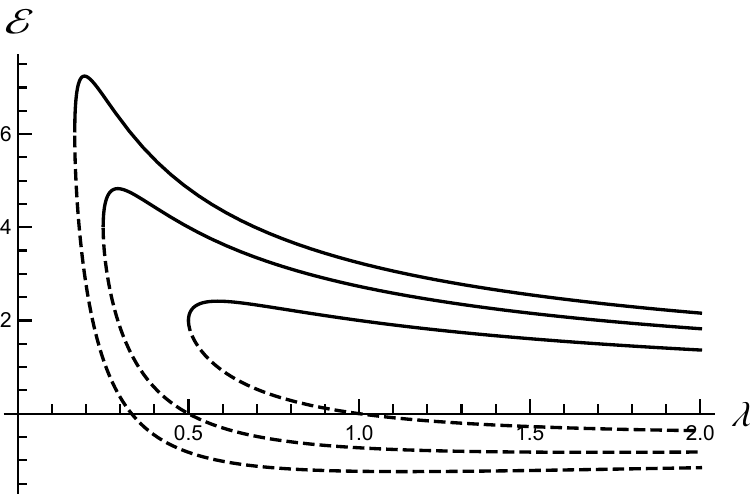}
    \caption{Energy levels $\mathcal{E}$ of the dual theory to dS$_2$ JT gravity with a timelike Dirichlet boundary as a function of $\lambda$ at fixed $\mathcal{E}_{H}$ (\ref{eq:micendS2}) (from left to right: $\mathcal{E}_{H}=3,2,1$). Solid (dashed) lines indicate energy levels of the cosmic (black hole) patch.}
    \label{fig:EndS2lam}
\end{figure} 
Notice the choice in `$\pm$' in front of the square root in (\ref{eq:enspecgendila}) leads to two different solutions. For example, consider the `$-$' branch. Then, imposing the initial data $\mathcal{E}(\lambda=0)\equiv \mathcal{E}_{0}$ fixes $C_{1}=-2\mathcal{E}_{0}$, from which we find agreement with the energy spectrum (\ref{eq:EAdS2JT}) dual to the $\text{AdS}_{2}$ JT black hole when $V(\Phi)=2\Phi/\ell^2$. Moreover, the choice of sign `$\pm$' also allows for matching to the quasi-local energies (\ref{eq:EBY Nariai}) of the cosmological ($+$) or black hole ($-$) patch in dS$_{2}$ JT gravity. Indeed, for $V(\Phi)=-2\Phi/\ell^{2}$, and setting $C_{1}\equiv 2\mathcal{E}_{H}$, Eq. (\ref{eq:enspecgendila}) yields
\beq \mathcal{E}^{c,h}(\lambda,\mathcal{E}_H)=\frac{1}{\lambda}\left(1\pm\sqrt{2\lambda\mathcal{E}_{H}-1}\right) = \frac{1}{\lambda}\left(1\pm\frac{\tilde{\beta}_{\text{T}}}{\sqrt{\frac{2}{3}\pi^2 c \lambda -\tilde{\beta}^{2}_{\text{T}}}}\right)\;,\label{eq:micendS2}\eeq
where `$+$' refers to the cosmological patch and `$-$' to the black hole patch. Using the $\lambda$ identification (\ref{eq:lambda bulk}) and $\mathcal{E}_{H}\equiv\ell M_{\text{ADM}}^{\text{dS}_{2}}=\frac{\Phi_{r}r_{H}^{2}}{16\pi G_{2}\ell^{2}}$, we recover the dS$_{2}$ quasi-local energy energies (\ref{eq:EBY Nariai}), with $\mathcal{E}^{c,h}(\lambda)=|r_{B}|E^{c,h}_{\text{BY}}$.
Notice that the range of $\lambda $ is restricted to $[\frac{1}{2 \mathcal E_H},\infty]$, where the lower bound corresponds to the situation in the bulk where the system boundary coincides with the horizon,  $r_B = r_h$.

\begin{figure}[t!]
    \centering
\includegraphics[width=8cm]{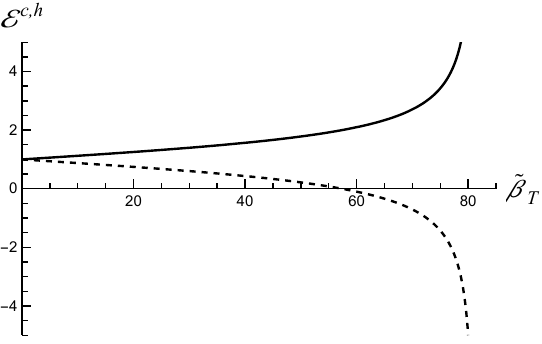}$\quad$\includegraphics[width=8cm]{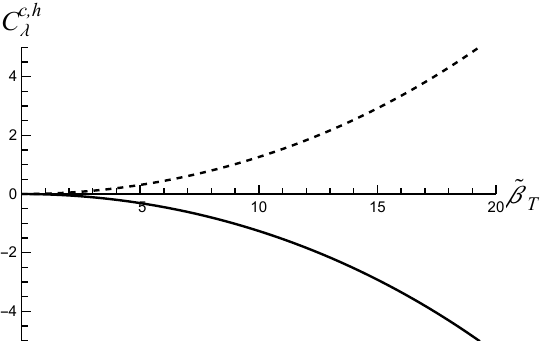}
    \caption{\textbf{Left:} Microscopic energy of the dual theory to dS$_2$ JT gravity with a timelike Dirichlet boundary (\ref{eq:micendS2}) at fixed $\lambda=1$, and $c=1000$. Solid (dashed) curve denotes cosmic (black hole) horizon patch. \textbf{Right:} Heat capacity (\ref{eq:C ch}) for same parameter range.}
    \label{fig:Energies TTbar dS2}
\end{figure}

The heat capacity (\ref{eq:heatcapmic}) for the quantum mechanical system dual to dS$_2$ JT gravity is
\beq \label{eq:C ch}
 C^{c,h}_\lambda 
 =\mp  \frac{1}{\lambda}\frac{\frac{2}{3} \pi^2c \lambda \tilde{\beta}_{\text{T}}^2}{\qty(\frac{2}{3} \pi^2 c \lambda  -  \tilde{\beta}_{\text{T}}^2)^{3/2}}\,,\eeq
with $-$ ($+$) for the cosmic (black hole) patch. Figure \ref{fig:EndS2lam} displays the microscopic energy spectrum (\ref{eq:micendS2}) as a function of $\lambda$ at fixed $\mathcal E_H$, and Figure \ref{fig:Energies TTbar dS2} portrays the microscopic energy and heat capacity as functions of $\tilde{\beta}_{\text{T}}=\beta_{\text{T}}/r_B$ at fixed $\lambda$. Crucially, the system that is dual to the cosmic patch is thermally unstable, while the system that is dual to the black hole patch is  stable against thermal fluctuations. As is well known from standard statistical mechanics, negative heat capacity implies the canonical ensemble partition function for the cosmological horizon patch (for bulk or boundary theory) is ill-defined. More precisely, positivity of heat capacity is linked to the convergence of the integral representation of the canonical partition function in a steepest descents approximation (see, e.g., \cite{Brown:1994su}).

\subsection{Cardy-like formulae for the thermal entropy of the dual deformed theory}
\label{subsec:cardy}

At fixed $\lambda$ the (dimensionless)  energy of the deformed boundary theory should satisfy the first law
\beq \label{firstlaw}
d \mathcal E = \tilde T d S \qquad \text{at fixed} \, \lambda\,,
\eeq
where $\tilde T = \tilde \beta_{\text{T}}^{-1}$ is the dimensionless boundary temperature. 
We can derive the thermal entropy $S$ (up to a constant) by integrating the first law. 
In the canonical ensemble (at fixed $\tilde T$) it follows from the final expressions in energies \eqref{eq:EAdS2JT} and \eqref{eq:micendS2} that the entropy takes the form
\begin{align}\label{eq:ScanonicalAdS}
&\text{for thermal states dual to AdS$_2$:} \qquad S (\lambda~\tilde{\beta}_{\rm T}) =S_0 + \frac{ \frac{2}{3} \pi^2 c }{\sqrt{ \frac{2}{3} \pi^2 c \lambda +\tilde{\beta}_{\rm T}^2}}\,, \\
&\text{for thermal states dual to  dS$_2$:} \qquad  \,\,\,\,S (\lambda,\tilde{\beta}_{\rm T}) =S_0 \pm \frac{ \frac{2}{3} \pi^2 c }{\sqrt{ \frac{2}{3} \pi^2 c \lambda-\tilde{\beta}_{\rm T}^2}}\,,\label{eq:ScanonicaldS}
\end{align}
where `$+$' refers to states dual to the cosmological patch, and `$-$' to states dual to  the black hole patch in dS$_2$.

For thermal states dual to AdS$_2$, the entropy of the undeformed theory is 
\beq S  (\lambda=0) =S_0 +  \frac{2}{3} \pi^2  c \tilde T_{\text{H}}\;,\label{eq:Slam0}\eeq
where $\tilde T_{\text{H}}= \tilde T (\lambda =0)=  T_{\text{H}}\ell$ is the dimensionless undeformed boundary temperature. This entropy agrees (up to a constant) with the Cardy formula for the canonical entropy in 2D CFTs with equal left- and right-moving energies and central charges. The formulae above are extensions of the canonical Cardy entropy to finite $\lambda$ and,  in fact, they also hold for $\text{T}\overline{\text{T}}$ deformations of 2D CFTs. Moreover, for thermal states dual to dS$_2$ there is a minimum value for $\lambda \tilde T^2$, given by $1/(\frac{2}{3}\pi^2 c )$. 

In the microcanonical ensemble (at fixed $\mathcal E$), on the other hand, the entropy for finite $\lambda$ is given by  
\begin{align}\label{eq:SmicrocanonicalAdS}
&\text{for thermal states dual to AdS$_2$:} \qquad S (\lambda, \mathcal E) =S_0 + 2 \pi \sqrt{\frac{c}{6} \mathcal E (2 -  \lambda \mathcal E)}\,, \\
&\text{for thermal states dual to  dS$_2$:} \qquad  \,\,\,\,S (\lambda, \mathcal E) =S_0 \pm 2 \pi \sqrt{\frac{c}{6}\left ( \mathcal E (\lambda\mathcal E  -2) + \frac{2}{\lambda} \right)} \,.\label{eq:SmicrocanonicaldS}
\end{align}
For states dual to AdS$_2$ the range of the deformation parameter is from $\lambda =0$ to $\lambda = 1/ \mathcal E$, where the lower bound corresponds to $r_B = \infty$ in the bulk and the upper bound  to $r_B = r_H$. For states dual to dS$_2$ the range is from $\lambda = 1/\mathcal E$ to $\lambda = \infty$, where the lower bound is dual to $r_B = r_h$ in the bulk and the upper bound to $r_B = 0$.
For the undeformed theory, the microcanonical entropy of thermal states dual to AdS$_2$ reduces to  
\beq\label{eq:EntropymicroAdS2}
S  (\lambda = 0, \mathcal E) = S_0+ 2 \pi \sqrt{\frac{c}{3} \mathcal E_0}~,
\eeq
which coincides (up to a constant) with the well-known Cardy formula for microcanonical entropy in 2D CFTs when the left- and right-moving energies are the same \cite{Cardy:1986ie,Bloete:1986qm}. This further confirms the boundary holographic theory in 1D can indeed be viewed as a dimensional reduction of a 2D CFT.
It also noteworthy that the microcanonical entropy reduces to the Cardy formula for $\lambda = 1/ \mathcal E = 1/ (2 \mathcal E_0),$ which corresponds to the setup where the Dirichlet wall is located at the horizon. 

The negative heat capacity of the cosmic patch, and hence thermal instability, can be further seen from the microcanonical entropy (\ref{eq:SmicrocanonicaldS}). It is straightforward to verify the entropy associated to the cosmic horizon patch is a \emph{convex} function of energy $\mathcal{E}$, a common characteristic of an unstable thermal system. Convexity of the entropy coincides with the canonical partition function diverging in a steepest descent approximation (cf. \cite{Brown:1994su}). In contrast, entropy for the black hole horizon patch, or states dual to AdS$_{2}$ are \emph{concave} in the deformed energy, coinciding with thermal stability.

Finally, both  the canonical and microcanonical entropy of the deformed boundary theory match exactly with the  Bekenstein-Hawking entropy \eqref{BHentropy} for 2D dilaton gravity in the bulk,
\beq
S_{\text{BH}} =  \frac{\Phi_0}{4 G_2} + \frac{\Phi_h}{4 G_2  }\,,
\eeq
upon identifying $\Phi_0 / 4G_2$ with the constant term $S_0$ in the boundary entropy formulae \eqref{eq:ScanonicalAdS}-\eqref{eq:SmicrocanonicaldS} and $\Phi_h / 4 G_2$ with the second term in these formulae.  
This can be checked explicitly using the holographic dictionary for $\lambda$ and $c$, \eqref{eq:lambda bulk} and \eqref{dictionaryc} respectively, and inserting the expressions for the temperatures and energies in the previous sections.

We observe that the thermal entropy is independent of $\lambda$ in an ensemble of fixed $\mathcal{E}_{0}$, i.e.,  the number of states does not change if we fix $\mathcal E_0$. In the bulk, fixing $\mathcal{E}_{0}$ amounts to fixing the ADM mass. For any $\lambda$ the Bekenstein-Hawking entropy matches with the Cardy entropy $S= 2 \pi \sqrt{c \mathcal E_0 / 3}$, and is clearly independent of $\lambda$. Alternatively,  if $\mathcal E$ or $\tilde \beta$ are held fixed instead of $\mathcal E_0$, then the entropy clearly does depend on $\lambda.$ Note that $\mathcal E$ and $\tilde \beta$   are the thermodynamic variables of the deformed theory, whereas $\mathcal E_0$ is the energy of the undeformed theory. Thus,   whether the entropy is a function of the deformation parameter $\lambda$ depends on the ensemble, but for the ensemble that fixes the thermodynamic variables of the deformed theory itself (e.g., $\mathcal E$ or $\tilde 
\beta$) the entropy is indeed    a function of $\lambda.$

\section{\texorpdfstring{Interpolating geometries via $\text{T}\overline{\text{T}}$}{} deformations}\label{sec:centaur TTbar}

In the previous section we reviewed a general class of dilaton gravity theories with a timelike Dirichlet boundary that has a dual description in terms of a one-dimensional $\text{T}\overline{\text{T}}$ deformed quantum mechanical theory \cite{Gross:2019ach,Gross:2019uxi}. Evidence for this comes from the equivalence of the microscopic deformed energy spectrum and the quasi-local energy of the gravitational theory. Notably, we found the generic solution to the microscopic flow equation is capable of describing either the cosmological or black hole horizon patch in dS$_{2}$ JT gravity, giving a microscopic perspective of quasi-local energy in dS$_{2}$, and consequently, the entropy.

An interesting class of dilaton gravity models to consider are those which interpolate between  
dS$_{2}$ and AdS$_{2}$ spaces \cite{Anninos:2017hhn,Anninos:2018svg,Anninos:2020cwo,Anninos:2022hqo}. The simplest such set-up is dubbed the ``centaur geometry'' \cite{Anninos:2017hhn}, having a dS$_{2}$ interior but  AdS$_{2}$ asymptotics. Motivation for these dilaton models is that they provide a natural setting to develop dS$_2$ (static patch) holography, where the dS static patch is accessible to probe from the asymptotic AdS boundary.\footnote{There is a no-go theorem \cite{Freivogel:2005qh} which states dS$_{d+1}$ geometries glued to asymptotically AdS$_{d+1}$ space for $d>1$ will be hidden behind the black hole horizon of the asymptotically AdS$_{d+1}$. The $d=1$ interpolating geometries evade this theorem.} 
Here we apply the $\text{T}\overline{\text{T}}$ formalism of the previous section to explicitly develop a microscopic picture of the centaur model (such an analysis was alluded to in \cite{Gross:2019ach} but not developed).

\subsection{Centaur  (A)dS\texorpdfstring{$_2$}{} geometries}

There are two different centaur geometries: one where the interior dS$_2$ space includes the cosmological horizon, and another where dS$_2$ space with a black hole horizon, both glued to an AdS$_2$ exterior geometry. In the former, the dilaton decreases towards the boundary, and in the latter the dilaton increases towards the boundary. Like in the previous sections, we associate a positive value of the dilaton to the cosmological horizon, and a negative value  to the black hole horizon, following our convention in \cite{Svesko:2022txo}.\footnote{This is different from the convention in the original work \cite{Anninos:2017hhn}, where the dilaton is taken to be negative near the cosmological horizon, and positive near the black hole horizon.}

\begin{figure}[t!]
    \centering
    \begin{subfigure}[t]{0.45\textwidth}
    \includegraphics[width=\textwidth]{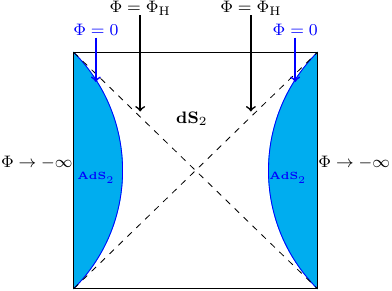}
        \caption{ }
    \end{subfigure}\hspace{1.5cm}\begin{subfigure}[t]{0.42\textwidth}
    \includegraphics[width=\textwidth]{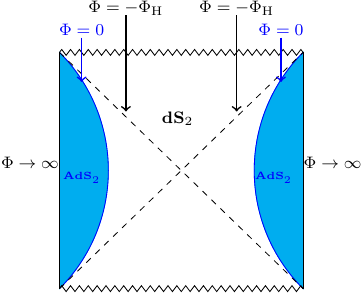}
       \caption{ }
    \end{subfigure}
    \caption{Penrose diagram of  centaur geometry, interpolating between the dS$_2$ space (white) and    AdS$_2$ space (blue). (a) Centaur geometry with   a cosmological horizon. In this case the dilaton is positive on the horizon and decreases towards the boundary, where it becomes minus infinity. (b) Centaur geometry with a black hole horizon. The dilaton is negative at the horizon and increases towards to the AdS boundary, where it diverges. 
    }
    \label{fig:Centaur}
\end{figure}

The centaur geometry can be obtained as a solution of a two-dimensional dilaton-gravity theory (\ref{eq:I centaur}) for an array of potentials $V(\Phi)$. For concreteness, we take
\begin{equation}\label{eq:pot (A)dS2}
    V(\Phi)=\mp\tfrac{2}{\ell^2}\abs{\Phi}~.
\end{equation}
For this dilaton potential the centaur geometries are described by the solution \eqref{eq:static patch sol}, where the blackening factor is, respectively, given by 
    \begin{equation}\label{eq:cases Centaur adS dS new}
\text{(a) centaur with cosmological   horizon:}\quad   N(r)=
    \begin{cases}
    \frac{r_{  H}^2-r^2}{\ell^2}~,\quad & r_{H}\geq r\geq 0\,,\\
    \frac{r_{  H}^2+r^2}{\ell^2},\quad &0 \geq r\geq -\infty ~,
    \end{cases}
    \end{equation}
    \begin{equation}\label{eq:cases Centaur adS dS}
 \text{(b) centaur with black hole horizon:} \quad     N(r)=
    \begin{cases}
    \frac{r_{  H}^2-r^2}{\ell^2}~,\quad &-r_{H}\leq r\leq 0\,,\\
    \frac{r_{  H}^2+r^2}{\ell^2},\quad &0 \leq r\leq\infty ~.
    \end{cases}
    \end{equation}
    The minus sign in the potential  \eqref{eq:pot (A)dS2} yields a centaur solution with a dS$_2$ cosmological horizon (at $r=r_H$), and the plus sign corresponds to a centaur geometry with a black hole horizon (at $r=- r_H$). 
In the former case, the region $r > 0$ describes dS$_2$ space, and the region $r < 0$ corresponds to a global AdS$_2$ patch. In the latter case, on the other hand, the region with $r< 0$ corresponds to dS$_2$, and the region with $r > 0$ to global  AdS$_2$. In both cases the interface between AdS$_2$ and dS$_2$ is located at $r=0.$ The Penrose diagrams of the two centaur geometries are displayed in Figure \ref{fig:Centaur}. 

Notice in the former case the dilaton $\Phi=\Phi_r r/\ell$ is positive at the cosmological horizon and decreases towards the AdS boundary, where it becomes minus infinity. In contrast, in the latter case the dilaton   is negative at the black hole horizon and increases monotonically towards the AdS boundary, where it blows up. Since $\Phi_0 + \Phi$ can be interpreted as the inverse of the gravitational coupling, $\Phi \to + \infty$ corresponds to a weak gravitational regime and $\Phi \to -\infty  $   to a strongly gravitating regime (since $\Phi_0 + \Phi \to 0$ in that case).  Hence, in the presence of a cosmological horizon the semiclassical description of the boundary of the centaur geometry breaks down. Therefore, it is not possible to define a seed (undeformed) dual quantum mechanical theory at the boundary of a centaur geometry with a cosmological horizon. Perhaps it is possible to remedy this issue by defining the seed theory at a cutoff surface away from the boundary. Putting this issue aside, below we will compare the energy spectra and heat capacities of the $\text{T}\overline{\text{T}}$ deformed theories dual the two centaur models.

Next, we  introduce a timelike Dirichlet boundary with a fixed  dilaton $\Phi_B \propto r_B$ and a fixed temperature, whose inverse value is  on shell given by the inverse Tolman temperature 
\begin{equation}
    \begin{aligned}
      \text{(a) centaur with cosmological horizon:} \quad    \beta_{\text{T}}&=\begin{cases}
        \frac{\beta_{\text{H}}}{\ell}\sqrt{r_{ H}^2-r_{  B}^2}\,,&\quad  r_H\geq r_B\geq 0~,\\
            \frac{\beta_{\text{H}}}{\ell}\sqrt{r_{  H}^2+r_{ B}^2}\,,&\quad 0 \geq r_B\geq -\infty ~,
        \end{cases}
    \end{aligned}
\end{equation}
\begin{equation}
    \begin{aligned}
      \text{(b) centaur with black hole horizon:} \quad    \beta_{\text{T}}&=\begin{cases}
        \frac{\beta_{\text{H}}}{\ell}\sqrt{r_{ H}^2-r_{  B}^2}\,,&\quad -r_H\leq r_B\leq 0~,\\
            \frac{\beta_{\text{H}}}{\ell}\sqrt{r_{  H}^2+r_{ B}^2}\,,&\quad 0 \leq r_B\leq \infty ~,
        \end{cases}
    \end{aligned}
\end{equation}
where $\beta_{\text{H}}= 2\pi \ell^2 /r_H$.
In both centaur geometries, the inverse temperatures range from $[0,2\pi \ell]$ in the dS region, and $[2 \pi \ell, \infty]$ in the AdS patch. 

Additionally, the quasi-local energy at an arbitrary Dirichlet wall is given on shell by 
\begin{equation}
    \begin{aligned}
        E_{\text{BY}}&= \begin{cases}
         \frac{1}{8\pi G_2 \ell} \left ( \Phi_{B} \pm\frac{|\Phi_B |\beta_{\text{T}}}{ \sqrt{4 \pi^2 \ell^2 - \beta_{\text{T}}^2 }} \right)\,, \quad & 0\le \beta_{\text{T}} \le 2 \pi \ell\,,\\
            \frac{ 1}{8\pi G_2 \ell} \left ( \Phi_B \pm\frac{|\Phi_B |\beta_{\text{T}}}{\sqrt{ \beta_{\text{T}}^2-4 \pi^2 \ell^2}} \right)  \,, \quad &2 \pi \ell \le\beta_{\text{T}}
       < \infty \,,\end{cases} 
       \label{centaurbyenergy}
    \end{aligned}
\end{equation}
where the plus sign corresponds to a centaur geometry with a cosmological horizon, and the minus sign corresponds to a geometry with a dS$_2$ black hole horizon. 
Note that for the `cosmological' centaur the Brown-York energy diverges to plus infinity at the interface where $ \beta_{\text{T}} = 2\pi \ell$, whereas for the `black hole' centaur the Brown-York energy diverges to minus infinity at the interface  and it goes to zero at the AdS boundary. Interestingly, upon multiplying the quasi-local energy with   $\sqrt{N(r_B)}$ and taking the limit $r_B \to \mp \infty$  (or $\beta_{\text{T}} \to \infty $) we find that the   ADM mass of the centaur geometry is negative in the presence of a     black hole horizon and positive in the presence of  a cosmological horizon 
\beq
M_{\text{ADM}}^{\text{cent}} = \pm \frac{\Phi_r r_H^2 }{16 \pi G_2 \ell^3}\,.\eeq

\subsection{Thermodynamics of the dual deformed  theory}

\begin{figure}
    \centering
    \begin{subfigure}[t]{0.49\textwidth}
        \includegraphics[width=\textwidth]{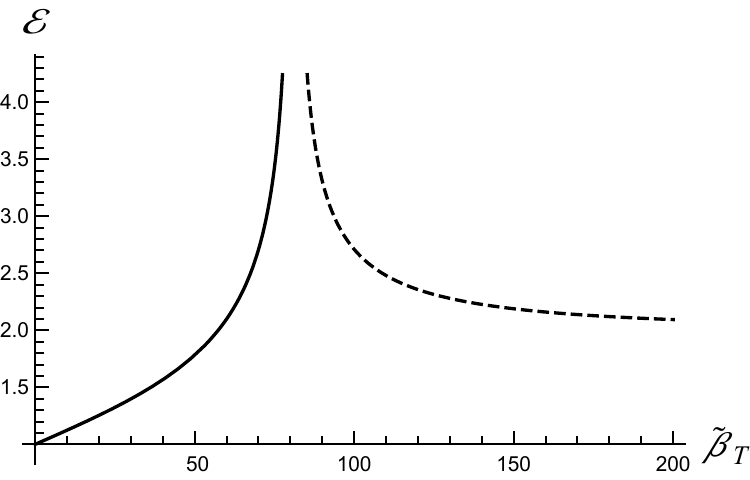}
    \end{subfigure}  \hfill  \begin{subfigure}[t]{0.49\textwidth}
        \includegraphics[width=\textwidth]{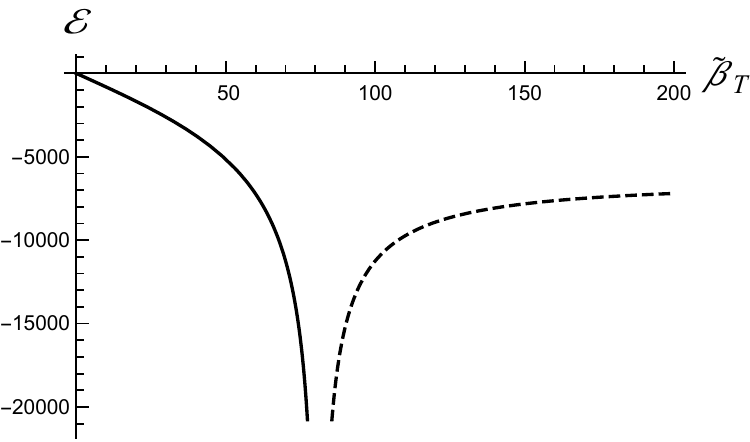}
    \end{subfigure}
    \caption{Energy spectrum (\ref{eq:Energy lambda beta_T centaur}) as a function of $\tilde{\beta}_{\text{T}}$  for $\lambda=1$ and $c=1000$ for the microscopic theory dual to the centaur model  in the presence of  (a) a   cosmological horizon, and (b) a  black hole horizon. The solid line represents the region $\tilde \beta_{\text{T}}^2 \le  \frac{2}{3}\pi^2 c \lambda$, and the dashed line for $\tilde \beta_{\text{T}}^2 \ge  \frac{2}{3}\pi^2 c \lambda$.}
    \label{fig:Energy_centaur_beta}
\end{figure}

\begin{figure}
    \centering
    \begin{subfigure}[t]{0.49\textwidth}
    \includegraphics[width=\textwidth]{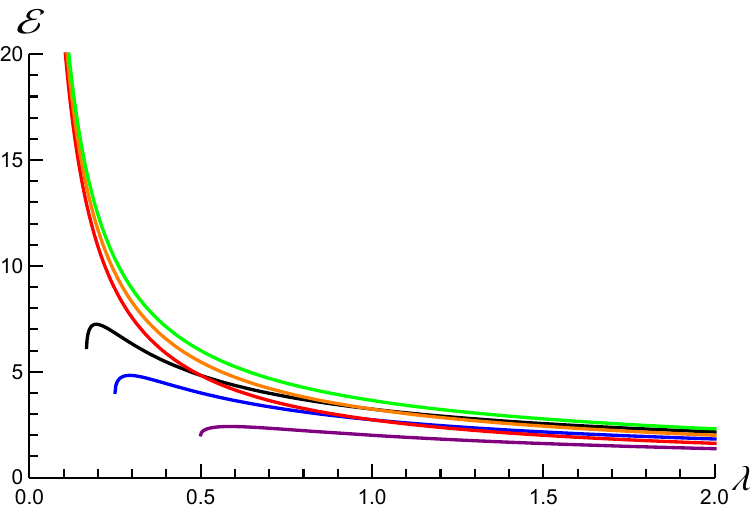}\caption{}
    \end{subfigure}\hfill\begin{subfigure}[t]{0.49\textwidth}
    \includegraphics[width=\textwidth]{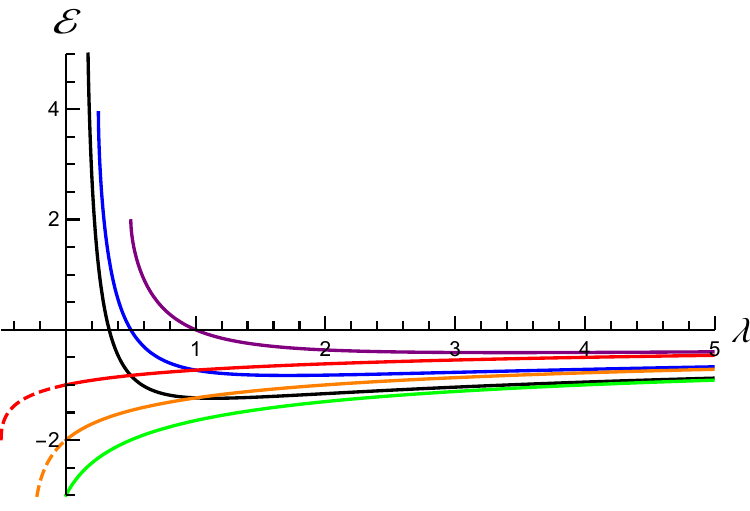}\caption{}
    \end{subfigure}
    \caption{The energy spectrum (\ref{eq:energies centaur}) of the centaur model for different representative values of $\mathcal{E}_H$ in the presence of (a) a dS$_2$ cosmological horizon, and (b) a dS$_2$ black hole horizon. Like in Figure \ref{fig:En lambda L}, we have allowed for an analytic continuation for the deformation parameter $\lambda=-\frac{1}{2\mathcal{E}_H}$ in the global AdS$_2$ region (dashed lines), corresponding to the lower line in (\ref{eq:energies centaur}); while the upper line corresponds to the dS$_2$ region.  Purple, blue and black solid lines indicate $\mathcal{E}_H=1,~2,~3$ respectively in the dS$_2$ region; while red, orange and green correspond to $\mathcal{E}_H=1,~2,~3$ in the AdS$_2$ region. 
    }
    \label{fig:Energies TTbar centaur}
\end{figure}

We want to find the energy spectrum of the deformed dual microscopic theory. 
We can apply the previous general solutions of the flow equation (\ref{eq:Flow eq Dilaton gravity})   to the particular centaur dilaton potential in (\ref{eq:pot (A)dS2}) to derive   for the energy spectrum. For the dS region of the centaur geometry the deformed energies are given by \eqref{eq:micendS2}. For the AdS region we  fix  the constant in \eqref{eq:enspecgendila} to be  $C_1  = 2 \mathcal E_H $.   The  deformed energy spectrum   for the centaur model thus takes the form, as a function of the dimensionless boundary temperature $\tilde \beta_{\text{T}}=\beta_{\text{T}}/r_B $,
\begin{equation}\label{eq:Energy lambda beta_T centaur}
    \begin{aligned}
        \mathcal E (
    \lambda, \tilde \beta_{\text{T}}
        )&= \begin{cases}
        \frac{1}{\lambda}\left(1\pm\frac{\tilde{\beta}_{\text{T}}}{\sqrt{\frac{2}{3}\pi^2 c \lambda -\tilde{\beta}^{2}_{\text{T}}}}\right)\,,\quad & 0 \le \tilde \beta_{\text{T}}^2 \le \frac{2}{3}\pi^2 c \lambda\,,\\
            \frac{1}{\lambda} \left ( 1 \pm \frac{\tilde \beta_{\text{T}}}{\sqrt{ \tilde \beta_{\text{T}}^2-\frac{2}{3}\pi^2 c \lambda}} \right) \,, \quad &\frac{2}{3}\pi^2 c \lambda \le \tilde \beta_{\text{T}}^2< \infty\,,
        \end{cases} 
    \end{aligned}
\end{equation}
or, as a function of  $\mathcal E_{  H}$,
\begin{equation}\label{eq:energies centaur}
    \mathcal{E}{(\lambda, \mathcal E_H)}=\begin{cases}
        \frac{1}{\lambda} \left (1\pm\sqrt{{2\lambda}\mathcal{E}_{H}-1}\right)~,& \frac{1}{2\mathcal{E}_H} \leq  \lambda \leq  \infty ~,\\
       \frac{1}{\lambda} \left(1\pm \sqrt{ {2\lambda}\mathcal{E}_{H}+1}\right)~,&0 \le \lambda \le \infty~.
    \end{cases} 
\end{equation}
The upper expression after the curly bracket corresponds to states dual to dS$_2$, whereas the lower expressions are for states dual to AdS$_2$. Moreover, the plus sign corresponds to a centaur geometry with a cosmological horizon, and the minus sign holds for states dual to a centaur geometry with a dS$_2$ black hole horizon.
This energy spectrum agrees with the Brown-York energy   \eqref{centaurbyenergy} if we identify   $\mathcal{E}  =|r_{B}|E_{\text{BY}}$ and $ \mathcal{E}_{H}\equiv \Phi_rr_H^2/(16\pi G_2 \ell^2) .$
The energy spectra for the deformed theories are shown in Figures~\ref{fig:Energy_centaur_beta} and~\ref{fig:Energies TTbar centaur}, for theories dual to cosmological centaur models and black hole centaur models.
Note at the interface between the dS and AdS regions, at fixed $\mathcal E_H$ the energy goes to zero as $\lambda \to \infty$, but at fixed $\lambda$ the energy diverges at the interface. 

By integrating the first law \eqref{firstlaw} at fixed $\lambda$  we can derive the thermal entropy for states dual to the centaur geometry (\ref{eq:Energy lambda beta_T centaur}). For the canonical ensemble, we find
\begin{equation}\label{eq:ScanonicalCentaur}
    \begin{aligned}
       S (\lambda, \tilde{\beta}_{\rm T})&= \begin{cases}
       S_0 \pm\frac{ \frac{2}{3} \pi^2 c}{\sqrt{ \frac{2}{3} \pi^2 c \lambda - \tilde{\beta}_{\rm T}^2}} \,,\quad & \tilde{\beta}_{\rm T}\leq\sqrt{\frac{3}{2\pi^2 c \lambda}}\,, \\
       S_0 \pm \frac{ \frac{2}{3} \pi^2 c}{\sqrt{\tilde{\beta}_{\rm T}^2-\frac{2}{3} \pi^2 c \lambda}}      \,,\quad &\tilde{\beta}_{\rm T}\geq\sqrt{\frac{3}{2\pi^2 c \lambda}}\,,
        \end{cases} 
    \end{aligned}
    \end{equation}
    where the upper expression corresponds to the case where the boundary (on which the microscopic theory lives) is located in the dS$_2$ region, and for the lower expression the Dirichlet wall is located in the (global) AdS$_2$ region of the centaur geometry.   We note that the upper  expression for the entropy agrees with \eqref{eq:ScanonicaldS}, but the lower expression differs from \eqref{eq:ScanonicalAdS} since the AdS$_2$ region in the centaur geometry corresponds to global AdS, whereas the  AdS$_2$ region considered  in \eqref{eq:ScanonicalAdS} corresponds to an eternal AdS$_2$ black hole (or Rindler-AdS). This is related to the fact that the temperature in the AdS$_2$ region here is bounded above (at fixed~$\lambda$), whereas in the case considered in   \eqref{eq:ScanonicalAdS} it is unbounded above.

The microcanonical entropy of thermal states dual to the centaur geometry is 
\begin{align}\label{eq:SmicrocanonicalCentaur}
    &\text{For  states dual to the dS$_2$  region:}~~S (\lambda, \mathcal E) =S_0  \pm 2 \pi \sqrt{\frac{c}{6}\left ( \mathcal E (\lambda\mathcal E  -2) + \frac{2}{\lambda} \right)}~,\\ &\text{For  states dual to the AdS$_2$  region:}~~S (\lambda, \mathcal E) =S_0 \pm  2 \pi \sqrt{\frac{c}{6} \mathcal E (\lambda \mathcal E-2)}~.
\end{align}
Note that the entropy diverges at the interface $\lambda = \infty$ ($r_B =0$)  between the AdS$_2$ and dS$_2$ patches. Again, the microcanonical entropy for thermal states dual to a (global) AdS$_2$ region differs slightly from \eqref{eq:SmicrocanonicalAdS}.

\begin{figure}[t!]
    \centering
    \begin{subfigure}[c]{0.49\textwidth}
        \includegraphics[width=\textwidth]{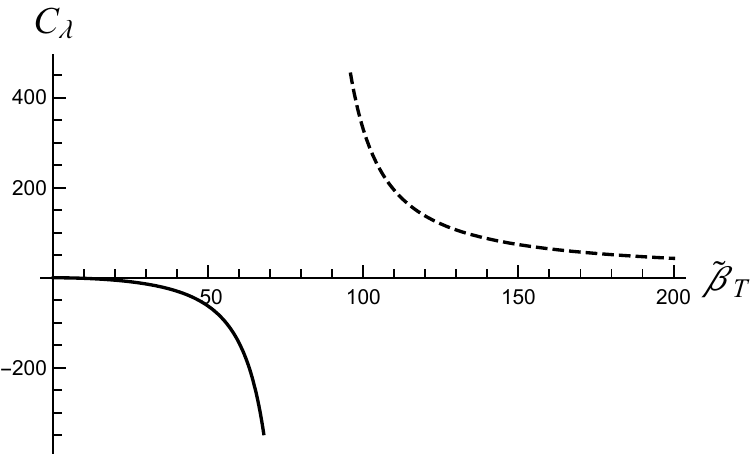}\caption{}
    \end{subfigure}\hfill\begin{subfigure}[c]{0.49\textwidth}
        \includegraphics[width=\textwidth]{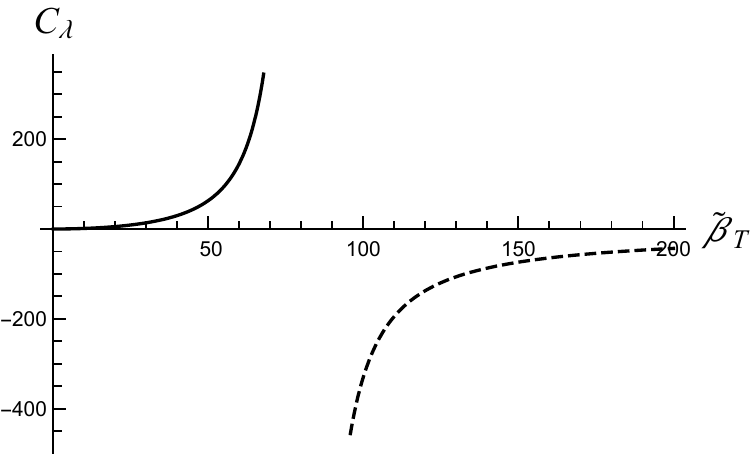}\caption{}
    \end{subfigure}
    \caption{Heat capacity of the centaur model  with (a) a cosmological horizon,  corresponding to the upper sign in (\ref{eq:heat capacity centaur}), and (b) a black hole horizon, i.e. the lower sign in (\ref{eq:heat capacity centaur}). The solid line represents the region $0\leq\tilde \beta_{\text{T}} \le \frac{2}{3}\pi^2 c\lambda$, and the dashed line for $\tilde \beta_{\text{T}}^2 \ge \frac{2}{3}\pi^2 c\lambda$. We have set $c=1000$ and $\lambda=1$ in the plots.}
    \label{fig:centaur_heat_capacity}
\end{figure}

Furthermore, the heat capacity  at fixed $\lambda$ can be easily computed from the energy spectrum, using the definition in \eqref{eq:heatcapmic},
\beq \label{eq:heat capacity centaur}
 C_\lambda = \begin{cases}
 \mp \frac{1  }{\lambda}\frac{\frac{2}{3} \pi^2c \lambda \tilde{\beta}_{\text{T}}^2}{(\frac{2}{3} \pi^2c \lambda  - \tilde{\beta}_{\text{T}}^{2})^{3/2}} \,,\quad & 0 \le \tilde \beta_{\text{T}}^2 \le \frac{2}{3} \pi^2c \lambda \,,  \\
  \pm \frac{1  }{\lambda}\frac{\frac{2}{3} \pi^2c \lambda \tilde{\beta}_{\text{T}}^2}{(  \tilde{\beta}_{\text{T}}^{2}-\frac{2}{3} \pi^2c \lambda )^{3/2}} \,, \quad & \tilde \beta_{\text{T}}^2\geq\frac{2}{3} \pi^2c \lambda \,,
 \end{cases}
 \eeq
 where the upper sign corresponds to the `cosmological' centaur model and the lower sign to the `black hole' centaur model.  The heat capacity at fixed $\lambda$ as a function of the $\tilde{\beta}_{\rm T}$ is shown in Figure \ref{fig:centaur_heat_capacity}.
 We emphasize that the heat capacity switches sign for both centaur models from the dS region to the AdS region. This is a distinctive feature of deformed microscopic theories dual to the centaur geometry, since for instance for  
 theories dual to AdS$_2$ JT gravity the heat capacity is  positive for all temperatures, see \eqref{eq:heatcapmic}. It is further interesting to note that the heat capacity also depends on the nature of the horizon, i.e. the signs in the heat capacity are precisely opposite for the `cosmological' and `black hole' centaur models. It is especially interesting that for microscopic theories that live on a Dirichlet wall inside the AdS region the heat capacity is positive in the presence of a cosmological horizon  and   negative in the presence of a dS$_2$ black hole horizon. This can again be contrasted to the heat capacity of an AdS$_2$ black hole horizon, which is always positive. Thus, the heat capacity is a useful observable to distinguish the microscopic dual theory of the centaur geometry from the dual theory of AdS$_2$ space.

\section{Dilaton gravity from {\texorpdfstring{$\text{T}\overline{\text{T}}+\Lambda_2$}{}}}\label{Sec:TTbar+Lambda2}

Above we provided a holographic interpretation of quasi-local thermodynamics of the centaur model via  $\text{T}\overline{\text{T}}$ deformations. Here we derive the two-dimensional analog of
$\text{T}\overline{\text{T}}+\Lambda_2$ deformations \cite{Gorbenko:2018oov,Lewkowycz:2019xse,Shyam:2021ciy,Coleman:2021nor,Batra:2024kjl},\footnote{See \cite{Silverstein:2024xnr} for recent developments on $T^2+\Lambda_3$ deformations.} originally used to connect patches of dS$_{3}$ to AdS$_{3}$ and provide a microstate counting of dS$_{3}$ entropy. This is accomplished by first generalizing $\text{T}\overline{\text{T}}+\Lambda_2$ flows to connect more general (A)dS$_{3}$ patches, namely BTZ and Schwarzschild-dS$_3$ of the same mass. Notably, the resulting deformed energy spectrum maps each original energy level (where $\lambda=0$) to different values of $\lambda$. That is, there is not a single theory, at a given value of~$\lambda,$ which has all of these energy levels as part of its spectrum; the flow we consider is tuned such that only one of the states looks like Schwarzschild-dS$_{3}$ from a particular value $\lambda_0$ onwards.  We then dimensionally reduce these more general flows, and show they can be used to holographically describe $\text{AdS}_{2}$ to a $\text{dS}_{2}$ region characterized by an appropriate dilaton-gravity theory. We will see the microscopic energy spectrum and quasi-local thermodynamics are compatible with a $\text{T}\overline{\text{T}}$-deformation of the quantum theory dual to the centaur geometry. 

\subsection{\texorpdfstring{Review of $\text{T}\overline{\text{T}}+\Lambda_2$}{} deformations and patchwise holography}

We begin with a brief review of solvable irrelevant $\text{T}\overline{\text{T}}+\Lambda_2$ flows and how they holographically correspond to connecting an  $\text{AdS}_{3}$ patch to a $\text{dS}_{3}$ patch. We extend this to more general (A)dS$_{3}$ patches in the subsequent section before deriving its two-dimensional analog via dimensional reduction. 

On the field theory side, to accommodate a $\text{dS}_{3}$ region, the idea is to slightly modify the 1-parameter family of deformed theories (\ref{eq:defQFTaction}) by including a two-dimensional `cosmological constant' $\Lambda_{2}$ \cite{Gorbenko:2018oov,Lewkowycz:2019xse}
\beq I_{\text{QFT}}=I_{\text{CFT}}+\mu\int \rmd^{2}x\sqrt{\gamma}\text{T}\overline{\text{T}}+\frac{(1-\eta)}{\mu}\int \rmd^{2}x\sqrt{\gamma}\;.\label{eq:TTbarL2act}\eeq
Here $\eta=\pm1$, where $\eta=+1$ returns the standard $\text{T}\overline{\text{T}}$ deformed theory. The associated flow equation is 
\beq \partial_{\mu}\log Z_{\text{QFT}}(\mu,\eta)=-\int\rmd^{2}x\sqrt{\gamma}\langle\text{T}\overline{\text{T}}\rangle+\frac{(1-\eta)}{\mu^{2}}\int\rmd^{2}x\sqrt{\gamma}\;,\label{eq:flowZL2}\eeq
together with the initial condition $Z_{\text{QFT}}(\mu=0,\eta=1)=Z_{\text{CFT}}$. The Burgers equation (\ref{eq:curly En}) defining the microscopic energy spectrum $\mathcal{E}_{n}(\lambda)$ becomes (see Appendix A of \cite{Gorbenko:2018oov})\footnote{To compare, note the dimensionless flow parameter $y$ in \cite{Gorbenko:2018oov} is related to ours via $2\pi^{2}y_{\text{there}}=\lambda_{\text{here}}$.}
\beq 4\pi\partial_{\lambda}\mathcal{E}_{n}-2\lambda\mathcal{E}_{n}\partial_{\lambda}\mathcal{E}_{n}-\mathcal{E}^{2}_{n}+P_{n}^{2}L^{2}+\left(\frac{2\pi}{\lambda}\right)^{\hspace{-1mm}2}(1-\eta)=0\;.\label{eq:BurgeqnEL2}\eeq
where $\lambda$ is defined in (\ref{eq:def lambda}). The general solution yields
\beq E_{n}(\lambda,\eta)=\frac{2\pi}{\lambda L}\left(1\pm\sqrt{\eta-\frac{4C_{1}\lambda}{2\pi^{2}}+\lambda^{2}J_{n}^{2}}\right)\;,\label{eq:gensolL2}\eeq
where recall $\mathcal{E}_{n}=E_{n}L$. How the integration constant $C_{1}$ and which $\pm$ branch are determined will be described momentarily.

A key insight of \cite{Gorbenko:2018oov} is that the flow equation (\ref{eq:flowZL2}) serves as a prescription that is to be applied piecewise. Specifically, the total flow trajectory is comprised of two legs: (i) start with a pure $\text{T}\overline{\text{T}}$ deformation (where $\eta=1$), flowing from $\lambda=0$ to some positive constant $\lambda=\lambda_{0}$, then (ii) at $\lambda=\lambda_{0}$, turn on $\eta=-1$, thus connecting the pure $\text{T}\overline{\text{T}}$ segment of the trajectory to a $\text{T}\overline{\text{T}}+\Lambda_{2}$ segment.\footnote{The two segments are connected by imposing the general boundary condition $Z_{\text{QFT}}(\lambda=\lambda_{0},~\eta=-1)=Z_{\text{QFT}}(\lambda=\lambda_{0},~\eta=1)$.} For special values of $\lambda_{0}$, the combined flow dualizes to sewing together gravitating patches of $\text{AdS}_{3}$ to $\text{dS}_{3}$ space \cite{Gorbenko:2018oov,Shyam:2021ciy,Coleman:2021nor}. 
To see how this works, consider three-dimensional Einstein gravity with cosmological constant $2\Lambda_{3}=-\frac{2\eta}{\ell^{2}_{3}}$, and a Gibbons-Hawking-York boundary term plus a local counterterm 
\beq
    I_{\text{EH}}=\frac{1}{16\pi G_{3}}\int_{\mathcal{M}}\rmd^3 x\sqrt{-g}\qty(R+\frac{2\eta}{\ell^{2}_{3}})+\frac{1}{8\pi G_{3}}\int_{\partial\mathcal{M}}\rmd^2 x\sqrt{h}\left(K-\frac{1}{\ell_{3}}\right)\;.
\label{eq:3D action}\eeq
Clearly, a flow between $\eta=+1$ to $\eta=-1$ corresponds to a transition from bulk spacetimes with $\Lambda_{3}<0$ to those with $\Lambda_{3}>0$, e.g., the  BTZ black hole and $\text{dS}_{3}$. The goal is to smoothly connect a patch of $\text{AdS}_{3}$ to a patch of $\text{dS}_{3}$, particularly one which contains the $\text{dS}_{3}$ cosmological horizon, whilst respecting appropriate boundary conditions. In the dual field theory language, this amounts to finding the match point in the dimensionless coupling, $\lambda_{0}$. The point $\lambda=\lambda_{0}$ can be inferred on the gravity side by seeing when the $\text{AdS}_{3}$ or BTZ black hole metric looks the same as $\text{dS}_{3}$, at least in some limit. 

One such limit is the near-horizon region of the BTZ black hole at its Hawking-Page phase transition point (where the black hole horizon $r_{h}=\ell_{3}$) and the near-horizon region of the $\text{dS}_{3}$ static patch (where the cosmological horizon is $r_{c}=\ell_{3})$ \cite{Coleman:2021nor}. Using the standard relation between the ADM mass and horizon radius of a (static) BTZ black hole, $8G_{3}M_{\text{ADM}}\ell^{2}_{3}=r_{h}^{2}$, and the holographic identification $\ell_{3} M_{\text{ADM}}=\Delta_{n}+\bar{\Delta}_{n}-\frac{c}{12}$, notice at the Hawking-Page phase transition point, where $M_{\text{ADM}}=(8G_{3})^{-1}$, the dual CFT satisfies $\Delta_{n}+\bar{\Delta}_{n}=\frac{c}{6}$ or $M_n = \frac{c}{12}$ at this transition.

To admit a smooth pure gravity description, the deformed energy spectrum (\ref{eq:gensolL2}) is required to be continuous at the matching point $\lambda=\lambda_{0}$, while maintaining the seed CFT boundary condition, $E_{n}(\lambda=0,\eta=1)=2\pi M_{n}/L$. The deformed energy spectrum (for $J_n =0$) is
\beq E_{n}(\lambda,\eta)=\frac{2\pi}{\lambda L}\left(1-\sqrt{\eta(1-2M_{n}\lambda)}\right)\;,\label{eq:micenergyHP}\eeq
where the integration constant $C_{1}$ 
and `$-$' are chosen to ensure the seed boundary condition is satisfied. 
Notice for $\eta=1$ and $M_n = \frac{c}{12}$,
the square root vanishes when $\lambda=\lambda_{0}=\frac{6}{c}$. Next, recall the quasi-local Brown-York energy associated with the $\text{dS}_{3}$ cosmological horizon patch (cf. Eq. (C.12) of \cite{Banihashemi:2022htw}, where here a  shift is owed to the local counterterm appearing in the bulk action (\ref{eq:3D action}) is included)
\beq E_{\text{BY}}^{\text{dS}_{3}}=\frac{1}{4G_{3}}\left(\frac{r_{B}}{\ell_{3}}+\sqrt{1-\frac{r_{B}^{2}}{\ell_{3}^{2}}}\right)\;.\eeq
 This energy matches the microscopic spectrum (\ref{eq:micenergyHP}) for $\eta=-1$, $\lambda=\frac{4G_{3}\ell_{3}}{r_{B}^{2}}$ when $M_{n}=\frac{\ell_{3}}{8G_{3}}=\frac{c}{12}$; the only difference is that the minus sign in front of the square root in (\ref{eq:micenergyHP}) should be replaced by a plus sign. Moreover, for  $M_{n}=\frac{c}{12}$  the square root again vanishes when $\lambda_{0}=\frac{6}{c}$. 

The combined trajectory is then comprised of: (i) pure $\text{T}\overline{\text{T}}$ deformation via $\lambda=0$ to $\lambda=\frac{6}{c}$, choosing the `$-$' branch, followed by (ii) a $\text{T}\overline{\text{T}}+\Lambda_{2}$ deformation from $\lambda=\frac{6}{c}$ to $\lambda\gg1$, choosing the `$+$' branch. The deformed energy spectrum is then piecewise, 
\begin{equation}\label{eq:defEL2}
    E_{n}(\lambda)=\begin{cases}
        \frac{2\pi}{\lambda L}\left(1-\sqrt{1-\frac{\lambda c}{6}}\right)~,&\lambda\in[0,\frac{6}{c}]\\
        \frac{2\pi}{\lambda L}\left(1+\sqrt{\frac{\lambda c}{6}-1}\right)~,&\lambda\in[\frac{6}{c},\infty)
    \end{cases}
\end{equation}
where the energy is continuous at $\lambda=\frac{c}{6}$. From the bulk perspective the flow is described as follows. First, (i) move the conformal boundary of the BTZ black hole (near its Hawking-Page transition point) inward until $r_{B}\simeq\ell_{3}$ ($\lambda=\frac{4G_{3}}{\ell_{3}}$). At this point the $\text{AdS}_{3}$ near horizon geometry is indistinguishable from the near horizon region of $\text{dS}_{3}$. Then, (ii) move the Dirichlet wall in dS$_3$ outward toward the pole ($r_{B}=0$) filling in the complementary region with the $\text{dS}_{3}$ static patch including the cosmological horizon. 

This patchwise prescription, arguably, allows for a microscopic accounting of the Gibbons-Hawking entropy. Namely, the microstates associated with the 
$M_n = \frac{c}{12}$
BTZ black hole are attributed to the $\text{dS}_{3}$ cosmic horizon patch, where the Gibbons-Hawking entropy is given by the Cardy entropy \cite{Shyam:2021ciy,Coleman:2021nor}. Indeed,
\beq S_{\text{Cardy}}=2\pi\sqrt{\frac{c}{3} M_n}=\frac{2\pi\ell_{3}}{4G_{3}}=S_{\text{GH}}\;.\eeq
In fact, this argument extends to include the first logarithmic correction $-3\log (S_{\text{GH}})$ (see \cite{Anninos:2020hfj} for a first principles derivation of the quantum-corrected Gibbons-Hawking entropy). This is due to the fact that the $1/c$ correction to the Cardy entropy is known to reproduce the logarithmic correction to the BTZ black hole entropy \cite{Carlip:2000nv}. Further, the $\text{dS}_{3}$ entropy is guaranteed to be finite on account of the deformed spectrum being finite. In \cite{Batra:2024kjl} the $\text{T}\overline{\text{T}}+\Lambda_{2}$ prescription was generalized to further account for bulk matter fields, going beyond the model-independent pure (semi-classical) gravity sector.

\subsection{\texorpdfstring{Generalizing  $\text{T}\overline{\text{T}}+\Lambda_2$ to arbitrary energies: from BTZ to conical dS$_3$}{}}

Above we briefly reviewed the $\text{T}\overline{\text{T}}+\Lambda_2$ flow that is dual to gluing a patch of the BTZ black hole at the Hawking-Page transition point, where $r_h = \ell_3$, to a patch of pure dS$_3$ geometry. We emphasize that this  flow is restricted to a certain energy level, namely where the undeformed CFT starts in a state with energy $M_n = c/12$. On the field theory side, however, the $\text{T}\overline{\text{T}}+\Lambda_2$ flow equation allows for a deformed energy spectrum that depends on an \emph{arbitrary} dimensionless energy parameter $M_n,$ given by \eqref{eq:micenergyHP} (for $J_n=0$) 
\beq E_{n}(\lambda,\eta, M_n)=\frac{2\pi}{\lambda L}\left(1\pm \sqrt{\eta(1-2M_{n}\lambda)}\right)\;.\label{eq:micenergyHPb}\eeq
This suggests there exists a $\text{T}\overline{\text{T}}+\Lambda_2$ flow that starts from an undeformed CFT state with arbitrary energy $M_n$ (instead of setting it equal to $  c/12$). In fact, we can combine two different trajectories as before: (i) pure $\text{T}\overline{\text{T}}$ deformation from $(\eta=1,\lambda =0)$ to $(\eta=1,\lambda =1/(2M_n))$   along the `$-$' branch, followed by (ii) a $\text{T}\overline{\text{T}}+\Lambda_2$ deformation from $(\eta=-1,\lambda =  1/(2 M_n) )$   to large $\lambda$ along the `$+$' branch.  

Here we describe the holographic dual of this more general  flow. We will argue it  corresponds to connecting a BTZ black hole with some arbitrary mass $M_{\text{ADM}}(= M_n / \ell_3)$ to  a locally dS$_3$ geometry with a conical defect and the same mass $M_{\text{ADM}}$. The $\text{T}\overline{\text{T}}+\Lambda_2$ 
 flow  reviewed in the previous subsection is a special case of this more general flow where $M_{\text{ADM}} = 1/(8 G_3)$. A version of this more general $\text{T}\overline{\text{T}}+\Lambda_2$ flow was considered in the original work  \cite{Coleman:2021nor}, specifically flows dual to gluing the BTZ geometry to  dS$_3$ with a conical \emph{excess}. Here we consider the case of a conical \emph{defect} --- the essential difference being the range of  $M_{\text{ADM}}$.  
 
 Further, the $\text{T}\overline{\text{T}}+\Lambda_2$ flow for arbitrary  $M_n$  is interesting for the purpose of this paper, since, upon a dimensional reduction, it yields a flow that is dual to gluing AdS$_2$ to dS$_2.$ We will discuss this in the next subsection, where we also  compare this flow to the $\text{T}\overline{\text{T}}$ deformation that is dual to moving from the exterior to the interior in the centaur geometry (considered in section \ref{sec:centaur TTbar}). The dimensional reduction of the arbitrary $\text{T}\overline{\text{T}}+\Lambda_2$ flow with arbitrary $M_n$ is more appropriate for comparison with the centaur flow, since in both cases the deformed energy spectrum depends on an arbitrary mass parameter, $M_n$ vs. $\mathcal E_H$, or, equivalently, an arbitrary temperature $\tilde T$. On the other hand, it is difficult to fully compare the $\text{T}\overline{\text{T}}+\Lambda_2$ deformed spectrum with $M_n = c/12$, Eq. \eqref{eq:defEL2}, to the deformed spectrum that is dual to the centaur model, since the former  holds only for a specific undeformed CFT energy whereas the latter depends on an arbitrary undeformed energy $\mathcal E_H$. 
 This generalization  of $\text{T}\overline{\text{T}}+\Lambda_2$
 deformations for arbitrary energies $M_n$ is also interesting in itself (see also \cite{Coleman:2021nor}). 

Let us now describe the three-dimensional bulk dual geometry for the  $\text{T}\overline{\text{T}}+\Lambda_2$ flow with arbitrary $M_n $. Both the static BTZ black hole  \cite{Banados:1992wn,Banados:1992gq} and the dS$_3$ geometry with a conical defect,  \cite{DESER1984405,Balasubramanian:2001nb} are described by the line element (in Euclidean signature)
\begin{equation}
    ds^2 =   N(r) d\tau^2 + \frac{dr^2}{N(r)} + r^2 d \phi^2\,,
\end{equation}
where the blackening factor takes the form
\begin{equation}
    N(r)= \eta \left ( \frac{r^2}{\ell_3^2}-8  G_3 M_{\text{ADM}}\right)\,,
    \end{equation}
with $\eta = +1$ and $M_{\text{ADM}}>0$ for the BTZ geometry,  and $\eta = -1$ and $0 < M_{\text{ADM}} < 1/(8 G_3)$ for conical dS$_3.$ 
Conical dS$_3$ is identical to three-dimensional Schwarzschild-de Sitter (SdS) space, but it does not contain a black hole. Rather, in the global extension of the conical dS$_3$ geometry there are two conical singularities, one at each pole of a two-sphere. The conical singularities arise due to the presence of point masses at the two poles of dS$_3.$ For $M_{\text{ADM}}=1/(8 G_3)$ there is no conical defect in dS$_3$, and for $M_{\text{ADM}}=0$ the deficit angle is $2\pi.$ For   arbitrary $M_{\text{ADM}}$ the deficit angle of the conical dS$_3$ geometry is: $\delta \phi = 2\pi (1- \sqrt{8 G_3 M_{\text{ADM}}})$. There is a conical excess in dS$_3$ for $M_{\text{ADM}} > 1/(8 G_3)$, but we discard this solution as unphysical   since it corresponds to a negative point mass that is unbounded below.\footnote{The point mass $m$ is related to the ADM mass via $4 G_3 m = 1 - \sqrt{8 G_3 M_{\text{ADM}}}$ \cite{Klemm:2002ir}. The negative point mass is unbounded for the following reason (see \cite{Banihashemi:2022htw,Banihashemi:2022jys}). The cosmological horizon solution is a local maximum rather than a minimum of the Euclidean Einstein gravity action as a function of the mass parameter in the Schwarzschild-de Sitter line element. When the mass parameter is allowed to be negative, the on-shell action is unbounded from below, leading to an arbitrarily large cosmological horizon, and correspondingly, refer to a thermal reservoir whose energy is likewise unbounded from below. We regard such traits as un-physical and therefore discard such solutions. Note, however, negative point mass solutions are in principle valid in the context of gravitational perturbation theory in the presence of timelike Dirichlet walls (see, e.g., the discussion of \cite{Andrade:2015gja}).} 

The BTZ geometry has a black hole horizon and the conical dS$_3$ geometry only has a cosmological horizon. In both cases the horizon radius is given by $r_h = \ell_3 \sqrt{8 G_3  M_{\text{ADM}}}$ and hence the horizon entropy equals
\begin{equation}
    S_{\text{BH}}= \frac{2\pi r_h}{4G_3} =  \frac{2\pi \ell_3}{4G_3} \sqrt{8 G_3  M_{\text{ADM}}}\,. 
\end{equation}
Further,  for Dirichlet boundaries at fixed $r_B$   the Tolman temperature is
\begin{equation}
    \beta_{\text{T}} = \frac{2\pi \ell_3}{\sqrt{8 G_3 M_{\text{ADM}}}} \sqrt{ \eta \left (  \frac{r_B^2  }{ \ell_3^2 }- 8 G_3 M_{\text{ADM}} \right)}\,,
\end{equation}
and the Brown-York quasi-local energy is given by (see also \cite{Gorbenko:2018oov,Banihashemi:2022htw})
\begin{equation}\label{quasilocalBTZdS}
    E_{\text{BY}}= \frac{r_{B}}{4G_{3}\ell_{3}}\left(1\pm \sqrt{\eta \left ( 1-\frac{8G_{3}M_{\text{ADM}}\ell_{3}^{2}}{r_{B}^{2}} \right)} \right)\,.
\end{equation}
The first term in the quasi-local energy arises due to a local counterterm in the gravitational action, which is chosen to be the same for both geometries. The sign in front of the square root is `$+$' for the quasi-local system $r \in [r_B, \infty)$ in the BTZ geometry and for $r\in[  r_B, r_h]$ in conical dS$_3$, and `$-$' for the system $r\in [r_h, r_B]$  in BTZ  and for $r\in [0, r_B]$ in conical dS$_3$. 

Crucially, the quasi-local energy \eqref{quasilocalBTZdS} precisely matches with the deformed energy \eqref{eq:micenergyHP} due to the holographic dictionary  $M_n =  M_{\text{ADM}} \ell_3$ and $\lambda =4G_{3}\ell_{3}/r_B^2$ (and recall $L =2 \pi r_B$).
Thus, the piecewise $\text{T}\overline{\text{T}}+\Lambda_2$ flow for arbitrary $M_n$ described below \eqref{eq:micenergyHPb} corresponds in the bulk  to connecting a patch of the BTZ black hole to a patch of conical dS$_3$, both with mass $M_{\text{ADM}}$. In particular, the flow  consists of two bulk trajectories: (i) moving the conformal boundary of BTZ $(\eta =1)$ until the black hole horizon is reached at $r_B = r_h$ $(\lambda = 1/ (2 M_{\text{ADM}} \ell_3))$, at which point the near horizon geometry of BTZ is indistinguishable from that of conical dS$_3$; followed by (ii) moving a Dirichlet boundary inside the static patch of conical dS$_3$ $(\eta = -1)$ toward the pole at $r_B=0$ $(\lambda = \infty)$. We emphasize this piecewise construction is done for each value of $M_{n}$. The (dimensionless) deformed energy spectrum along the state-dependent flow   thus consists of two pieces and, in terms of the undeformed energy $M_n$, is given by (recall $\mathcal E(\lambda)  = E_n (\lambda) r_B$)
\begin{equation}\label{eq:defEL2bb}
    \mathcal E(\lambda, M_n)=\begin{cases}
        \frac{1}{\lambda }\left(1 -  \sqrt{ 1-2M_{n}\lambda }\right)\,,&\lambda\in[0,\frac{1}{2M_n}]\,,\\
        \frac{1}{\lambda }\left(1 +  \sqrt{  2M_{n}\lambda -1}\right)\,,&\lambda\in[\frac{1}{2 M_n},\infty)\,.
    \end{cases}
\end{equation}
Note that the deformed energy spectrum has each original energy level $M_n$ at $\lambda=0$ being mapped to different values of $\lambda$. In other words, there is not a single theory, at a given $\lambda$, which has all of these energy levels as part of its spectrum, unlike standard $\text{T}\overline{\text{T}}$ deformations where the deformed energy spectrum is computed along the flow at a given value of $\lambda$. Thus, the resulting spectrum of conical defect states in dS$_{3}$  do not live in the same theory.\footnote{We thank Edgar Shaghoulian for emphasizing this point.}

 In terms of the dimensionless (inverse) temperature, the spectrum is
\begin{equation}\label{eq:defEL2tempppbb}
    \mathcal E(\lambda, \tilde \beta_{\text{T}})=\begin{cases}
        \frac{1}{\lambda }\left(1 -   \frac{\tilde \beta_{\text{T}}}{ \sqrt{\frac{2}{3}\pi^2 c \lambda +\tilde \beta_{\text{T}}^2 }}\right)\,,&\tilde \beta_{\text{T}}\in\qty[0,\frac{2\pi}{\sqrt{12 M_n/c}}]\,,\\
        \frac{1}{\lambda }\left(1 +  \frac{\tilde \beta_{\text{T}}}{ \sqrt{\frac{2}{3}\pi^2 c \lambda -\tilde \beta_{\text{T}}^2 }}\right)\,,&\tilde \beta_{\text{T}}\in[0,\infty)\,.
    \end{cases}
\end{equation}
Note that the transition point in the flow is given by $\tilde \beta_{\text{T}}=0$ or $\lambda=\lambda_{0}= 1/(2 M_n)$, a state dependent value (reminiscent of Eq. (5.2) \cite{Coleman:2021nor}). The upper bound on the inverse dimensionless temperature in the upper line arises from the limit $r_B \to \infty$   (or $\lambda \to 0$),   given by the inverse Hawking temperature divided by $\ell_3$, $\lim_{r_B\to \infty}\tilde \beta_{\text{T}} = \beta_{\text{H}}/ \ell_3 = \frac{2\pi }{\sqrt{8 G_3 M_{\text{ADM}}}}$, which equals $\frac{2\pi}{\sqrt{12 M_n/c}}$ in the boundary theory.

\subsection{\texorpdfstring{Dimensional reduction of $\text{T}\overline{\text{T}}+\Lambda_2$}{}}

Let us now study  a spherical dimensional reduction of the 2D $\text{T}\overline{\text{T}}+\Lambda_2$ flow for arbitrary energies to a flow for a 1D quantum theory. It turns out that the deformed energy spectrum stays the same under this dimensional reduction, just as was the case for the dimensional reduction of the 2D $\text{T}\overline{\text{T}}$ flow to one dimension  lower. We show this by dimensionally reducing the 3D bulk action (\ref{eq:3D action}) to a 2D bulk action of the form (\ref{eq:I centaur}), specifically we find\footnote{See Appendix A of \cite{Svesko:2022txo} for a more general derivation of this type of action.}
\begin{equation}\label{eq:dim reduction to Dilaton gravity}
    I_E=-\frac{1}{16\pi G_N}\int_{\mathcal{M}} \rmd^2x\sqrt{g}(R\Phi+V(\Phi))-\frac{1}{8\pi G_N}\int_{\partial\mathcal{M}}\rmd \tau\sqrt{h}\,\Phi (K-\tfrac{1}{\ell})~,
\end{equation}
with a dilaton potential 
\begin{equation}\label{eq:eta potential 2d}
        V(\Phi)=\frac{2\eta}{\ell^2}{\Phi}~,
\end{equation} 
where we identified $\ell_3 = \ell.$
Notice there is no topological term in the 3D to 2D reduction.

The solutions to the dilaton equation  of motion  for the   potential (\ref{eq:eta potential 2d}) take the form   (\ref{eq:static patch sol}) with  
\begin{align}\label{eq:blackening}
    N(r)&=\eta\frac{r^2-r_H^2}{\ell^2}\,,
\end{align}
the blackening factor corresponding to (i) an AdS$_{2}$ black hole ($\eta=+1$), or (ii)  the half-reduction of dS$_3$ space with a conical deficit when $r_H\neq\ell$ ($\eta=-1$). Physically, we are implementing a generalization of the (s-wave reduced) situation studied in \cite{Coleman:2021nor} where the BTZ black hole has an arbitrary horizon radius, and the (S)dS$_3$ has a conical deficit. This allows us to carry out a more in-depth thermodynamical analysis for these geometries and the associated dimensional reduction of the $\text{T}\overline{\text{T}}+\Lambda_2$ flow.
The  geometric interpretation of this flow 
differs from the interpolating model \cite{Anninos:2017hhn,Anninos:2018svg} in that   AdS$_2$ space is glued  to a patch of dS$_2$ space   at their respective event horizons instead of the geometries being patched together from the start in a centaur model.

The deformed energy spectrum now follows from inserting the dilaton potential (\ref{eq:eta potential 2d}) into the general formula (\ref{eq:Flow eq Dilaton gravity}) 
\begin{equation}\label{eq:E spectrum TTbar+Lambda2}
    \mathcal{E}(\lambda,\mathcal E_0)=\tfrac{1}{\lambda}\left(1\pm\sqrt{\eta(1-{2\lambda}\mathcal{E}_0)} \right)~.
\end{equation}
Here we assume the states in the seed theory can have arbitrary energies $\mathcal{E}_0$. Importantly, this expression is equivalent to (\ref{eq:micenergyHPb}) by setting $\mathcal{E}_0=M_n$. The spectrum with the `$-$' sign   represents the solution of the flow equation that is smoothly connected with the seed theory.

\begin{figure}
    \centering
    \includegraphics[width=0.49\textwidth]{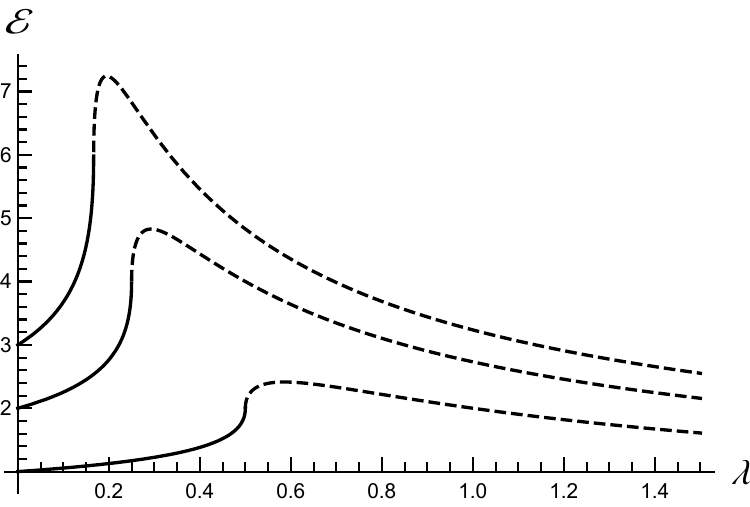}\hfill\includegraphics[width=0.49\textwidth]{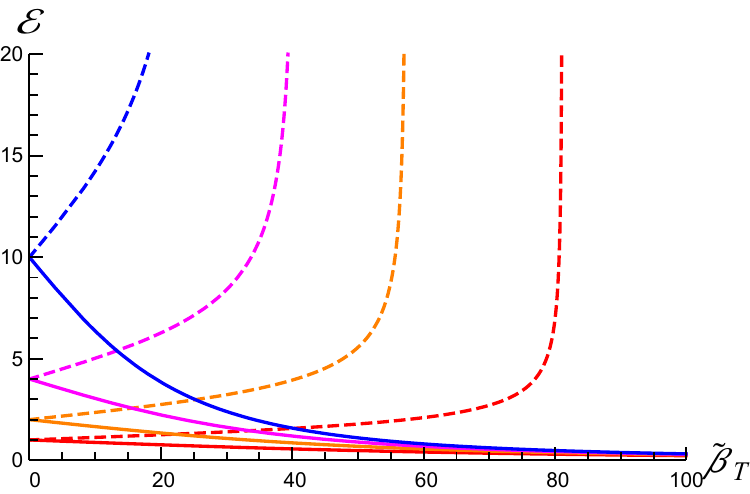}
    \caption{Energy spectrum of the thermal state dual to the s-wave reduction of the $\text{T}\overline{\text{T}}+\Lambda_2$ flow, \textbf{Left}: fixing $\mathcal{E}_0=1,~2,~3$ from bottom to top respectively 
    and varying $\lambda$ (\ref{eq:E spectrum TTbar+Lambda2}), with a relative $-$ sign for the AdS region, and $+$ for dS), and \textbf{Right}: fixing $\lambda$ and varying $\beta_{\tilde{\text{T}}}$ (\ref{eq:E TTbar+Lambda1}). In both cases, we set $c=1000$, and in the later case we take (from right to left) $\lambda=1$, $.5$, $.25$, $.1$; where the solid colored lines indicate $\eta=+1$, and the dashed lines $\eta=-1$.}
    \label{fig:E_TTL2}
\end{figure}

The Tolman temperature corresponding to (\ref{eq:blackening}) takes the form
\begin{equation}
    \beta_T=\frac{\beta_H}{\ell_3}\sqrt{\eta\frac{r_B^2-r_H^2}{\ell_3^2}}~.
\end{equation}
The energy levels (\ref{eq:E spectrum TTbar+Lambda2}) can be then expressed as
\beq
\mathcal{E}(\lambda,\tilde{\beta}_{\rm T})= \frac{1}{\lambda} \left ( 1 \pm \frac{\tilde{\beta}_{\text{T}}}{\sqrt{\frac{2}{3}\pi^2 c \lambda+ \eta \tilde{\beta}_{\text{T}}^{2}}}\right)\;,\label{eq:E TTbar+Lambda1}\eeq
where we allow, in principle, the relative $\pm$ sign solutions to the flow equation. However, for the purposes of representing a dimensional reduction of the $\text{T}\overline{\text{T}}+\Lambda_2$ flow, we are interested in analyzing the case with a relative `$-$' sign in front of the second term for the AdS$_2$ black hole patch $(\eta = 1)$, and a  `$+$' sign for the dS$_2$ patch $(\eta = -1)$, like in \eqref{eq:defEL2bb} and \eqref{eq:defEL2tempppbb}. Figure \ref{fig:E_TTL2} shows a comparison between the energy expressions \eqref{eq:defEL2bb} and \eqref{eq:defEL2tempppbb}.

\begin{figure}
    \centering
    \includegraphics[width=0.5\textwidth]{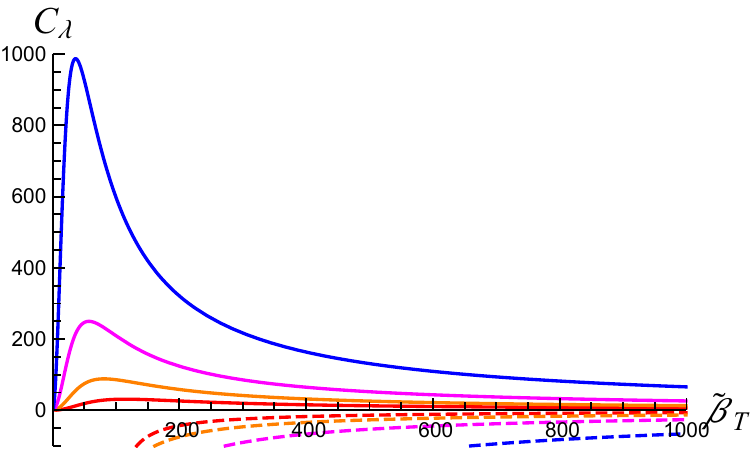}
    \caption{Heat capacity for the thermal state dual to the s-wave reduction of the $\text{T}\overline{\text{T}}+\Lambda_2$ flow as a function of the dimensionless inverse boundary temperature $\tilde \beta_{{\text{T}}}$. We are using the same parameters as on the right-hand side of Figure \ref{fig:E_TTL2}.}
    \label{fig:C_TTL2}
\end{figure}

The heat capacity \eqref{eq:heatcapmic} at fixed $\lambda$,
\beq 
 C_\lambda   =\mp \frac{1}{\lambda}\frac{\frac{2}{3} \pi^2c \lambda \tilde{\beta}_{\text{T}}^2}{(\frac{2}{3} \pi^2 c \lambda  +\eta\tilde{\beta}_{\text{T}}^2)^{3/2}} \,,
 \label{eq:heatcap TTbar+Lambda1}\eeq
is illustrated in Figure \ref{fig:C_TTL2}.
As in the centaur model, we find the boundary dual to the AdS$_2$ black hole that is smoothly connected to the seed theory is  thermodynamically stable, while the boundary dual of the cosmic patch dS$_2$ solution  is thermodynamically unstable. 

We complete our discussion of the spherical reduction of the $\text{T}\overline{\text{T}}+\Lambda_2$ flow with computing the thermal entropy using the first law $d \mathcal E = \tilde T dS$ at fixed $\lambda$. In the  canonical ensemble the entropy is
\begin{equation}
    S (\lambda, \tilde{\beta}_{\rm T}) = S_0\mp \frac{ \frac{2}{3} \pi^2 c \eta }{\sqrt{ \frac{2}{3} \pi^2 c \lambda +\eta \tilde{\beta}_{\rm T}^2}}~,
\end{equation}
while  in the microcanonical ensemble the thermodynamic entropy is given by 
\begin{align} \label{entropyformulattbarflow}
S (\lambda, \mathcal E) =S_0 \mp  2 \pi \eta \sqrt{\frac{c}{6} \left ( \eta\mathcal E (2-\lambda \mathcal E)+\frac{1-\eta}{\lambda} \right) }~.
\end{align}
Notice there is a lower bound on $\lambda \tilde T^2$, given by  $1/(\frac{2}{3}\pi^2 c )$, along the s-wave reduced $\text{T}\overline{\text{T}}+\Lambda_2$ flow for $\eta=-1$.
These entropy formulae agree with previous expressions derived in subsection~\ref{subsec:cardy} for the  entropies of thermal states dual to an (A)dS$_2$ patch. That is, by setting $\eta =1$ and choosing the lower sign in front of the second term in the expressions above we recover Eqs. \eqref{eq:ScanonicalAdS} and \eqref{eq:SmicrocanonicalAdS}, while setting $\eta=1$ and choosing the upper sign yields Eqs. \eqref{eq:ScanonicaldS} and \eqref{eq:SmicrocanonicaldS}. 
At the interface between  the AdS$_2$ and dS$_2$ patches, located at the event horizon of both patches, we have $\lambda = \frac{1}{2\mathcal E_0},$ $\tilde \beta_{\text{T}}=0$ and $ \mathcal E =\frac{1}{\lambda}$, which follows from \eqref{eq:defEL2bb} and~\eqref{eq:defEL2tempppbb}. We notice in this case  the AdS$_2$ and dS$_2$ entropies coincide and the canonical and microcanonical entropies \eqref{eq:ScanonicalAdS}-\eqref{eq:SmicrocanonicaldS}  reduce  to the Cardy formula \eqref{eq:EntropymicroAdS2}.

\section{Discussion and Outlook}\label{Sec:Discussion}

In this article we revisited the $\text{T}\overline{\text{T}}$ deformations of theories dual to two-dimensional dilaton theories of gravity \cite{Gross:2019uxi,Gross:2019ach}. Notably, we carried out a thorough investigation into the  microphysical interpretation of the quasi-local thermodynamics of (A)dS$_{2}$ JT gravity and the centaur model, which, in our view, had been lacking. In so doing, we uncovered many new formulae for the microscopic theory, including the energy spectrum, heat capacity, and found the entropy of the dual 1D theory obeys a deformed Cardy formula.

 We further considered a spherical dimensional reduction of a generalized $\text{T}\overline{\text{T}}+\Lambda_2$ flow connecting a BTZ black hole of arbitrary mass to a locally conical dS$_{3}$ spacetime.  This led us to a new type of flow connecting (A)dS$_{2}$ patches. Geometrically, the flow connects AdS$_{2}$ and dS$_{2}$ along their respective horizons, in contrast from the centaur model (where dS$_{2}$ is glued to a global AdS$_{2}$ region). Consequently, the thermal entropy of states dual to the AdS$_{2}$ region differs from the thermal entropy of states dual to the AdS$_{2}$ region in the centaur model, though both have a Cardy-like form.  At the boundary of AdS$_2$ and at the event horizon 
the thermal entropy reduces to the standard Cardy formula. 
Circumstantially, this  suggests the deformed boundary theory can  be viewed near the boundary \emph{and} the horizon as a dimensional reduction of a 2D CFT, but further work has to be done to examine if this is the case. 
 
 The heat capacity of the interpolating system, however, is qualitatively similar to the centaur model. In particular, 
the boundary dual to the AdS$_2$ black hole that is smoothly connected to the seed theory is  thermodynamically stable, while the boundary dual of the cosmic patch dS$_2$ solution  is  unstable. Thus, a signature of the change from AdS to dS is that the heat capacity at a fixed deformation parameter of the Dirichlet wall  changes sign.

\vspace{2mm}

\noindent There are multiple interesting directions to take this work, some of which we list below.

\vspace{2mm}

\noindent \textbf{Schwarzian and SYK description.} In this article we primarily focused on providing a holographic description of  bulk dilaton gravity theories in terms of $\text{T}\overline{\text{T}}$-deformed theories. Famously, AdS JT gravity can be understood in terms of a Schwarzian on the AdS$_{2}$ boundary, and, at low-energies, is effectively described by a Sachdev-Ye-Kitaev (SYK) model of fermions \cite{Maldacena:2016upp,Maldacena:2016hyu} (see also the review \cite{Sarosi:2017ykf}). The partition function of AdS JT gravity with a finite Dirichlet cutoff was evaluated in \cite{Iliesiu:2020zld} (see also \cite{Griguolo:2021wgy}), and found to precisely match a Schwarzian theory deformed by an operator analogous $\text{T}\overline{\text{T}}$-deformations. It would be interesting to directly compute the partition function of the deformed theories considered here, and see if there exists a similar Schwarzian description, expanding on \cite{Gross:2019uxi}. Moreover,  $\text{T}\overline{\text{T}}$-like deformations may be directly applied in SYK-type model \cite{Anninos:2022qgy}. Notably, it was found that a pair of uncoupled SYK models with complex coupling constants reproduces the thermodynamics of the centaur models. It would be interesting to develop an SYK-like description of the (A)dS$_{2}$ geometry found from dimensional reduction of $\text{T}\overline{\text{T}}+\Lambda_{2}$-deformed theories.

\vspace{2mm}

\noindent \textbf{de Sitter JT and DSSYK.} The half-reduction model of dS$_{2}$ JT gravity arguably has a holographic description in terms of double-scaled SYK (DSSYK) \cite{Susskind:2021esx,Susskind:2022bia,Susskind:2022dfz,Rahman:2022jsf,Narovlansky:2023lfz}. An analysis of 1D T$\overline{\text{T}}$ deformations in the DSSYK model
are addressed in \cite{Aguilar-Gutierrez:2024oea,A}.
Notably, the asymptotic boundary conditions and the holographic dictionary relating the deformation parameter and the radial bulk cutoff scale need to be generalized, allowing for complex bulk geometries. Interestingly, phase transitions between thermodynamically stable and unstable configurations can occur depending on the temperature, an aspect not present in our analysis.

\vspace{2mm}

 \noindent \textbf{Complex eigenvalues and Cauchy slice holography.} The appearance of complex eigenvalues in the energy spectrum of $T^2$ deformations in AdS/CFT (as well as in cosmology \cite{Araujo-Regado:2022jpj}) is a crucial component of Cauchy slice holography \cite{Araujo-Regado:2022gvw,Araujo-Regado:2022jpj}, a proposed duality between bulk gravity and a `boundary' theory that lives on Cauchy slices of the Lorentzian spacetime. Despite the presence of complex energies, bulk unitarity emerges nonetheless, 
 and there exist new types of covariant entropy bounds \cite{Soni:2024aop}. 
Here we explicitly excluded complex energy eigenvalues of the deformed theory. It would be worth adapting this formalism to lower-dimensional models of two-dimensional dilaton gravity, and, as in higher-dimensions reformulate the holographic principle in the language of Wheeler-DeWitt canonical quantization.

\vspace{2mm}

\noindent \textbf{Beyond finite Dirichlet walls.}  In this work we focused on gravity theories with a finite timelike boundary obeying Dirichlet boundary conditions. It is now known, however, such boundary conditions in four-dimensional general relativity do not admit a well-posed initial boundary value problem in Euclidean \cite{Anderson_2008,Witten:2018lgb} or Lorentzian signature \cite{Friedrich:1998xt,Fournodavlos:2020wde,Fournodavlos:2021eye,An:2021fcq,Anninos:2022ujl}.\footnote{Specifically, generic real-valued Dirichlet data do not satisfy the Einstein constraint equations at a finite boundary, nor does the Dirichlet problem have a unique solution (for certain choices of Dirichlet data).} Instead, it has been conjectured that `conformal boundary conditions' (CBCs) --- where the trace of the extrinsic curvature and conformal class of the induced metric of the finite boundary are held fixed --- supply a well-posed initial boundary value problem \cite{An:2021fcq} (there is in fact a one-parameter family of such boundary conditions \cite{Liu:2024ymn}). Such CBCs appeared previously in the fluid-gravity paradigm \cite{Bredberg:2011xw,Anninos:2011zn}, and lead to (conformal) quasi-local thermodynamics  \cite{Anninos:2023epi,Anninos:2024wpy,Anninos:2024xhc} (see also \cite{Banihashemi:2024yye}). In particular, for finite walls obeying CBCs in asymptotically de Sitter backgrounds, the dS static patch is thermally stable (for sufficiently large trace of extrinsic curvature) \cite{Anninos:2024wpy}, in stark contrast with the Dirichlet wall case. While the arguments leading to the issues of well-posedness for finite Dirichlet walls do not apply in three-dimensional general relativity or two-dimensional dilaton theories, it would be interesting to develop the analog of  $\text{T}\overline{\text{T}}$-deformations for theories dual to gravity with finite walls obeying CBCs. Such an investigation in the case of three-dimensional gravity was considered in \cite{Coleman:2020jte}. It would be natural to dimensionally reduce such flows, and develop a holographic description of the conformal quasi-local thermodynamics of the effective two-dimensional dilaton theories \cite{inprep}.

\section*{Acknowledgements}
We thank Dio Anninos, Eyoab Bahiru, Stefano Baiguera, Dami\'an Galante, Silvia Georgescu, Eleanor Harris, Norihiro Iizuka, Sam van Leuven, Mehrdad Mirbabayi, Andrew Rolph, Sunil Kumar Sake, Edgar Shaghoulian, Vasudev Shyam, Eva Silverstein, Aron Wall, and Nicolò Zenoni for illuminating discussions. We also thank Ricardo Espíndola for initial collaboration on this work. SEAG thanks the University of Amsterdam, the Delta Institute for Theoretical Physics, the International Centre for Theoretical Physics; and the School of Physics of the University of El Salvador for their hospitality and support during several phases of the project, and the Research Foundation - Flanders (FWO) for also providing mobility support (Grant No. K250423N). SEAG and MV also thank the organizers of the XIX Modave summer school, where part of our work was developed. The work of SEAG was partially supported by the FWO Research Project G0H9318N and the inter-university project iBOF/21/084, as well as from the Okinawa Institute of Science and Technology Graduate University. This article was made possible through the support of the ID\#62312 grant from the John Templeton Foundation, as part of the ‘The Quantum Information Structure of Spacetime’ Project (QISS), as well as Grant ID\# 62423 from the John Templeton Foundation. The opinions expressed in this article are those of the author(s) and do not necessarily reflect the views of the John Templeton Foundation.  AS is supported by
STFC grant ST/X000753/1.  MRV is supported by SNF Postdoc Mobility grant P500PT-206877 and the Spinoza Grant of the Dutch Science Organisation (NWO).

\appendix

\section{Flow equations dual to general 2D dilaton gravity} \label{app:TTdefsgen2D}

Here, following \cite{Gross:2019ach}, we provide additional details to the derivation of the flow equation for the effective one-dimensional quantum theory assumed to be dual to a general dilaton theory of gravity. The essential prescription, as presented for holographic CFTs in arbitrary dimensions  \cite{Hartman:2018tkw},\footnote{See also \cite{Taylor:2018xcy,Morone:2024ffm,Tsolakidis:2024wut}
for $\text{T}\overline{\text{T}}$ deformations in arbitrary dimensions.} is that the standard $\text{AdS}_{d+1}/\text{CFT}_{d}$ dictionary relating the bulk gravity and CFT partition functions is taken to hold at finite bulk radial cutoff $r_{B}$
\beq Z_{\text{EFT}}[r_{B};\gamma_{ij},J]=Z_{\text{grav}}[h_{ij}^{B}=r_{B}^{2}\gamma_{ij},\psi_{B}=r_{B}^{\Delta-d}J]\;.\label{eq:Dirchdic}\eeq
The left-hand side is the generating function for the (assumed holographic) effective field theory, which need not be a CFT itself, $\gamma_{ij}$ is the metric describing the field theory geometry, and $J$ is, for simplicity, taken to be a source for a scalar operator $\mathcal{O}$ of dimension $\Delta$. On the right-hand side is the (on-shell) gravitational partition function in an asymptotically AdS background with (Euclidean) metric $ds^{2}_{d+1}=g_{\mu\nu}dx^{\mu}dx^{\nu}=N(r)dr^{2}+r^{2}\gamma_{ij}dx^{i}dx^{j}$,\footnote{Here $N(r)\to1/r^{2}$ near the conformal boundary of AdS.} where the bulk metric and bulk scalar field $\psi$ are taken to obey Dirichlet boundary conditions, i.e. fixing the boundary induced metric $h_{ij}(r_{B},x)\equiv h^{B}_{ij}(x)$ and bulk fields $\psi(r_{B},x)\equiv\psi_{B}(x)$. The standard dictionary is recovered in the limit $r_{B}\to\infty$. While the status of the proposed Dirichlet dictionary (\ref{eq:Dirchdic}) is questionable for bulk gravity theories in dimensions $d\geq3$, there appears to be no hurdle for $d=2$ or, pertinently, $d=1$. 

In a semi-classical limit, the bulk partition function is given by the on-shell Euclidean action, $Z_{\text{grav}}=e^{-I_{\text{E}}}$. The flow of $I_{\text{E}}[r_{B}]$ is governed by the Hamilton-Jacobi equation with Hamiltonian $\mathcal{H}$ associated with bulk radial evolution.\footnote{The Hamilton-Jacobi equation describing the bulk radial flow is $\partial_{r_{B}}I_{\text{E}}[r_{B};\psi_{B}]=-\mathcal{H}[\psi_{B},\frac{\delta I_{\text{E}}}{\delta\psi_{B}}]$.} Working in this limit and employing the Hamilton-Jacobi equation, the dictionary (\ref{eq:Dirchdic}) implies the flow equation \cite{Hartman:2018tkw}
\beq \partial_{r_{B}}\hat{I}_{\text{EFT}}(r_{B},J)=-\mathcal{H}[r_{B}^{\Delta-d}J,-r_{B}^{d-\Delta}\sqrt{\gamma}\mathcal{O}]+\frac{d-\Delta}{r_{B}}\int \rmd x^{d}\sqrt{\gamma}J\mathcal{O}\;,\eeq
for effective (Euclidean) field theory action $\hat{I}_{\text{EFT}}=I_{\text{EFT}}-\int \rmd x^{d}\sqrt{\gamma}J\mathcal{O}$. Here $\mathcal{O}\equiv\frac{1}{\sqrt{\gamma}}\frac{\delta \hat{I}_{\text{EFT}}}{\delta J}$ at each step along the flow.

\subsection*{Flow equation for 2D dilaton gravity} 

\noindent \textbf{Bulk flow equation.} Following suit, to determine the flow equation for the one-dimensional effective field theory whose energy spectrum is guaranteed to coincide with the quasi-local Brown-York energy of the two-dimensional  gravity model in question, one first computes the Hamiltonian constraint of the gravity theory \cite{Davis:2004xi,Grumiller:2007ju} (see also Appendix C of \cite{Carrasco:2023fcj})
\beq 0=\mathcal{H}=16\pi G_{2}\pi_{\Phi}\pi^{\tau\tau}-\frac{1}{16\pi G_{2}}V(\Phi)\;,\label{eq:Hamconst}\eeq
where $\pi^{ab}=-\frac{\sqrt{h}}{16\pi G_{2}}h^{ab}n^{c}\nabla_{c}\Phi$ and $\pi_{\Phi}=-\frac{\sqrt{h}K}{8\pi G_{2}}$ are the canonical momenta conjugate to $h_{ab}$ and $\Phi$, respectively, in an ADM split, where $n^{c}=\sqrt{N}\partial^{c}_{r}$ is the unit normal to the timelike boundary. Since it will prove useful momentarily, note that the conjugate momenta $\pi^{ab}$ features in the (renormalized) quasi-local Brown-York stress-tensor \cite{Brown:1992br} (see, e.g., Appendix C of \cite{Pedraza:2021cvx})
\beq\label{eq:tilde T tau tau}
\tau^{ab}\equiv\frac{2}{\sqrt{h}}\frac{\delta I^{\text{on-shell}}_{E}}{\delta h_{ab}}=-\frac{1}{8\pi G_{2}}\left(n^{c}\nabla_{c}\Phi-\frac{\Phi}{\ell}\right)h^{ab}=\frac{2}{\sqrt{h}}\left(\pi^{ab}+\frac{\sqrt{h}}{16\pi G_{2}}\frac{\Phi}{\ell}h^{ab}\right)\;.\eeq
The Brown-York quasi-local energy is
\beq \label{eq:E BY}
E_{\text{BY}}=u_{a}u_{b}\tau^{ab}=\tau^{\tau}_{\;\tau}=\frac{\Phi_{r}r_{B}}{8\pi G_{2}\ell^{2}}\left(1-\frac{\ell}{r_{B}}\sqrt{N(r_{B})}\right)\;,\eeq
for unit normal $u^{a}=N^{-1/2}\partial^{a}_{\tau}$, matching  Eq. (\ref{eq:BYenbh}).\footnote{To recover the quasi-local energy for the de Sitter JT gravity (\ref{eq:EBY Nariai}), one notes the unit normal $n^{c}=-\sqrt{N}\partial^{c}_{r}$ for the `cosmological system'.} The second equality follows from $u_{\tau}u_{\tau}\tau^{ab}=h_{\tau\tau}\tau^{\tau\tau}=\tau^{\tau}_{\tau}$. Moreover, the momenta $\pi_{\Phi}$ appears in the scalar operator $\mathcal{O}_{\Phi}$ conjugate to $\Phi$
\beq \label{eq:eq:tilde O}
\mathcal{O}_\Phi\equiv\frac{1}{\sqrt{h}}\frac{\delta I^{\text{on-shell}}_{E}}{\delta\Phi}=-\frac{K}{8\pi G_{2}}+\frac{1}{8\pi G_{2}\ell}=\frac{1}{\sqrt{h}}\left(\pi_{\Phi}+\frac{\sqrt{h}}{8\pi G_{2}\ell}\right)\;.\eeq
This is the bulk analog of the scalar operator associated with source $J$ in the dictionary (\ref{eq:Dirchdic}). Thus, in terms of these variables, the Hamiltonian constraint (\ref{eq:Hamconst}) may be recast as
\beq
\begin{split}
 0=\mathcal{H}
 &=8\pi G_{2}\mathcal{O}_{\Phi}\tau^{\tau}_{\;\tau}-\frac{\mathcal{O}_{\Phi}\Phi}{\ell}-\frac{\tau^{\tau}_{\;\tau}}{\ell}+\frac{1}{16\pi G_{2}}\left[\frac{2\Phi}{\ell^{2}}-V(\Phi)\right]\;,
\end{split}
\label{eq:Hamconstdila}\eeq
where we used $\sqrt{h}=\sqrt{h_{\tau\tau}}$.
Observe the last term vanishes for $\text{AdS}_{2}$ JT gravity, i.e.,  $V(\Phi)=2\Phi/\ell^{2}$. 

The Hamilton-Jacobi equation governing the bulk radial flow is 
\beq \partial_{r_{B}}I_{\text{E}}^{\text{on-shell}}[r_{B};\Phi_{B}]=-\mathcal{H}\left[\Phi_{B},\frac{\delta I_{\text{E}}^{\text{on-shell}}}{\delta\Phi_{B}}\right]\;,\label{eq:HamJacbulk}\eeq
where $\Phi_{B}\equiv\Phi(r=r_{B})$. Let us analyze the left-hand side. A general variation of the dilaton theory (\ref{eq:I centaur}) is
\begin{equation}\label{eq:I variation}
\begin{aligned}
    \delta I_{\text{E}}=&-\frac{1}{2}\int\rmd^2 x\sqrt{g}\qty[{\mathscr E}^{\mu\nu}\delta g_{\mu\nu}+{\mathscr E}_\Phi\delta\Phi]+\int\rmd \tau\qty[\pi^{ij}\delta h_{ij}+\pi_\Phi\delta\Phi]+\frac{1}{8\pi G_2}\int \rmd \tau\sqrt{h}\qty[\frac{\delta\Phi}{\ell}+\frac{\Phi}{2\ell h}\delta h]\,,
\end{aligned}
\end{equation}
where ${\mathscr E}^{\mu\nu}$ and ${\mathscr E}_\Phi$ denote the equations of motion (\ref{eq:EOM gravity}) and (\ref{eq:EOM dilaton}), respectively. Thus, on-shell, 
\begin{align} 
\partial_{r_{B}}I_{\text{E}}[r_{B};\Phi_{B}]&=\int \rmd\tau\sqrt{h} \biggr[\frac{1}{\sqrt{h}}\left(\pi^{\tau\tau}+\frac{\sqrt{h}\Phi}{16\pi G_{2}\ell}h^{\tau\tau}\right)\partial_{r_{B}}h_{\tau\tau}+\frac{1}{\sqrt{h}}\left(\pi_{\Phi}+\frac{\sqrt{h}}{8\pi G_{2}\ell}\right)\partial_{r_{B}}\Phi\biggr] \nonumber \\
&=\int \rmd\tau\sqrt{h}\left[\frac{1}{2}\tau^{\tau\tau}(\partial_{r_{B}}h_{\tau\tau})+\mathcal{O}_{\Phi}(\partial_{r_{B}}\Phi)\right]\;.
\label{eq:partrBIE}\end{align}

\vspace{2mm}

\noindent \textbf{Boundary flow equation.} Now we determine the flow equation for the effective quantum mechanical on the finite boundary by assuming the 2D/1D analog of the dictionary (\ref{eq:Dirchdic}), together with 
\beq h^{B}_{\tau\tau}=(r_{B}/\ell)^{2}\gamma_{\tau\tau}\;,\quad \tau_{\tau\tau}=r_{B}\ell^{-1}T_{\tau\tau}\;,\quad \mathcal{O}_{\Phi}=r^{-2}_{B}\ell^{2}O\;,\label{eq:dict2}\eeq
where $\gamma_{ij}$ is the background metric of the 1D quantum mechanical theory, $T_{ij}$ is its stress-energy tensor and $O$ the dual operator of $\Phi$. It follows then
\beq 
\begin{split} \partial_{r_{B}}I_{\text{EFT}}=\partial_{r_{B}}I_{\text{E}}&=\int \rmd\tau\sqrt{\gamma}\left(\frac{\tau^{\tau}_{\;\tau}}{\ell}+\frac{\mathcal{O}_{\Phi}\Phi_{B}}{\ell}\right)\\
&=\int \rmd\tau\sqrt{\gamma}\left(8\pi G_{2}\mathcal{O}_{\Phi}\tau^{\tau}_{\;\tau}+\frac{1}{16\pi G_{2}}\left[\frac{2\Phi_{B}}{\ell^{2}}-V(\Phi_{B})\right]\right)
\end{split}
\eeq
where we substituted (\ref{eq:dict2}) into the bulk relation (\ref{eq:partrBIE}) in the first line, and used the Hamiltonian constraint (\ref{eq:Hamconstdila}) to get to the second line. Using the Hamiltonian constraint to replace $\mathcal{O}_{\Phi}$ via
\beq \mathcal{O}_{\Phi}=\frac{\left[\frac{\tau^{\tau}_{\;\tau}}{\ell}-\frac{1}{16\pi G_{2}}\left(\frac{2\Phi}{\ell^{2}}-V(\Phi)\right)\right]}{\left(8\pi G_{2}\tau^{\tau}_{\;\tau}-\frac{\Phi}{\ell}\right)}\;,\eeq
gives
\beq 
\begin{split}
\partial_{r_{B}}I_{\text{EFT}}=\int \rmd\tau\sqrt{\gamma}\biggr[\frac{\frac{8\pi G_{2}(\tau^{\tau}_{\;\tau})^{2}}{\ell}-\frac{\Phi_{B}}{16\pi G_{2}\ell}\left[\frac{2\Phi_{B}}{\ell^{2}}-V(\Phi_{B})\right]}{\left(8\pi G_{2}\tau^{\tau}_{\;\tau}-\frac{\Phi_{B}}{\ell}\right)}\biggr]\;.
\end{split}
\eeq
Implementing the dictionary (\ref{eq:dict2}), one finds after a little algebra
\beq r_{B}\partial_{r_{B}}I_{\text{EFT}}=\int\rmd\tau\sqrt{\gamma}\biggr[\frac{\lambda^{2}(T^{\tau}_{\;\tau})^{2}-\frac{1}{\ell^{2}}\left(1-\frac{\ell^{2}V(\Phi_{B})}{2\Phi_{B}}\right)}{(\lambda^{2}T^{\tau}_{\;\tau}-\lambda\ell^{-1})}\biggr]\;.\eeq
Using $r_{B}\partial_{r_{B}}=-2\lambda\partial_{\lambda}$ with $\lambda=8\pi G_{2}\ell^{2}/(\Phi_{r}r_{B}^{2})$, we recover 
the flow equation (\ref{eq:flowIFTgendil}). Note that this flow equation essentially arises from dimensional reduction of a massive gravity action description for a general family of stress-tensor deformations (see Section 3.4 of \cite{Tsolakidis:2024wut}).

Writing $T^{\tau}_{\;\tau}=E(\lambda)$, the flow equation for the deformed energy spectrum is 
\beq \partial_{\lambda}E=\frac{E^{2}-(\ell\lambda)^{-2}\left(1-\frac{\ell^{2}}{2\Phi_{r}}\sqrt{\frac{\lambda c}{6}}V\qty({\Phi_{r}}\sqrt{\frac{6}{\lambda c}})\right)}{2\ell^{-1}-2\lambda E}\;, \eeq
with generic solution
\beq E(\lambda)=\frac{1}{\ell \lambda}\left(1\pm\sqrt{1+\lambda \ell^{2} C_{1}+\frac{\lambda}{2}\int_{1}^{\lambda}\frac{d\tilde{\lambda}}{\tilde{\lambda}^{2}}\left[2-\frac{\ell^{2}}{\Phi_{r}}\sqrt{\frac{c\tilde{\lambda}}{6}}V\qty({\Phi_{r}}\sqrt{\frac{6}{\tilde{\lambda} c}})\right]}\right)\;.
\eeq
for constant $C_{1}$. Introducing a change of integration variable $\tilde{\lambda}$ to $y$ via  $y\equiv{\Phi_{r}}\sqrt{\frac{6}{c\lambda}}$, such that  the integral evaluates to 
\beq 
\begin{split}
\frac{\lambda}{2}\int_{1}^{\lambda}\frac{d\tilde{\lambda}}{\tilde{\lambda}^{2}}\left[2-\frac{\ell^{2}}{\Phi_{r}}\sqrt{\frac{\tilde{\lambda} c}{6}}V\qty({\Phi_{r}}\sqrt{\frac{6}{\tilde{\lambda} c}})\right]&=-\frac{\lambda c}{6\Phi_{r}^{2}}\int^{y}_{\Phi_{r}\sqrt{\frac{6}{c}}}dy[2y-\ell^{2}V(y)]\\
&=\lambda-1+\lambda\frac{c}{6}\frac{\ell^{2}}{\Phi_{r}^{2}}\int^{y}_{\Phi_{r}\sqrt{\frac{6}{c}}}dy V(y)\;.
\end{split}
\eeq
Adjusting constant $C_{1}$, we recover the generic solution (\ref{eq:enspecgendila}).

\bibliographystyle{JHEP}
\bibliography{references.bib}

\providecommand{\href}[2]{#2}\begingroup\raggedright\begin{thebibliography}{100}

\bibitem{Gibbons:1977mu}
G.W.~Gibbons and S.W.~Hawking, \emph{{Cosmological Event Horizons,
  Thermodynamics, and Particle Creation}},
  \href{https://doi.org/10.1103/PhysRevD.15.2738}{\emph{Phys. Rev. D}
  {\bfseries 15} (1977) 2738}.

\bibitem{Gibbons:1976ue}
G.W.~Gibbons and S.W.~Hawking, \emph{{Action Integrals and Partition Functions
  in Quantum Gravity}},
  \href{https://doi.org/10.1103/PhysRevD.15.2752}{\emph{Phys. Rev. D}
  {\bfseries 15} (1977) 2752}.

\bibitem{Banks:2000fe}
T.~Banks, \emph{{Cosmological breaking of supersymmetry?}},
  \href{https://doi.org/10.1142/S0217751X01003998}{\emph{Int. J. Mod. Phys. A}
  {\bfseries 16} (2001) 910}
  [\href{https://arxiv.org/abs/hep-th/0007146}{{\ttfamily hep-th/0007146}}].

\bibitem{Witten:2001kn}
E.~Witten, \emph{{Quantum gravity in de Sitter space}},  in \emph{{Strings
  2001: International Conference}}, 6, 2001
  [\href{https://arxiv.org/abs/hep-th/0106109}{{\ttfamily hep-th/0106109}}].

\bibitem{Goheer:2002vf}
N.~Goheer, M.~Kleban and L.~Susskind, \emph{{The Trouble with de Sitter
  space}}, \href{https://doi.org/10.1088/1126-6708/2003/07/056}{\emph{JHEP}
  {\bfseries 07} (2003) 056}
  [\href{https://arxiv.org/abs/hep-th/0212209}{{\ttfamily hep-th/0212209}}].

\bibitem{Sen:1995in}
A.~Sen, \emph{{Extremal black holes and elementary string states}},
  \href{https://doi.org/10.1142/S0217732395002234}{\emph{Mod. Phys. Lett. A}
  {\bfseries 10} (1995) 2081}
  [\href{https://arxiv.org/abs/hep-th/9504147}{{\ttfamily hep-th/9504147}}].

\bibitem{Strominger:1996sh}
A.~Strominger and C.~Vafa, \emph{{Microscopic origin of the Bekenstein-Hawking
  entropy}}, \href{https://doi.org/10.1016/0370-2693(96)00345-0}{\emph{Phys.
  Lett. B} {\bfseries 379} (1996) 99}
  [\href{https://arxiv.org/abs/hep-th/9601029}{{\ttfamily hep-th/9601029}}].

\bibitem{York:1986it}
J.W.~York, Jr., \emph{{Black hole thermodynamics and the Euclidean Einstein
  action}}, \href{https://doi.org/10.1103/PhysRevD.33.2092}{\emph{Phys. Rev. D}
  {\bfseries 33} (1986) 2092}.

\bibitem{Whiting:1988qr}
B.F.~Whiting and J.W.~York, Jr., \emph{{Action Principle and Partition Function
  for the Gravitational Field in Black Hole Topologies}},
  \href{https://doi.org/10.1103/PhysRevLett.61.1336}{\emph{Phys. Rev. Lett.}
  {\bfseries 61} (1988) 1336}.

\bibitem{Braden:1990hw}
H.W.~Braden, J.D.~Brown, B.F.~Whiting and J.W.~York, Jr., \emph{{Charged black
  hole in a grand canonical ensemble}},
  \href{https://doi.org/10.1103/PhysRevD.42.3376}{\emph{Phys. Rev. D}
  {\bfseries 42} (1990) 3376}.

\bibitem{Brown:1992br}
J.D.~Brown and J.W.~York, Jr., \emph{{Quasilocal energy and conserved charges
  derived from the gravitational action}},
  \href{https://doi.org/10.1103/PhysRevD.47.1407}{\emph{Phys. Rev. D}
  {\bfseries 47} (1993) 1407}
  [\href{https://arxiv.org/abs/gr-qc/9209012}{{\ttfamily gr-qc/9209012}}].

\bibitem{Brown:1992bq}
J.D.~Brown and J.W.~York, Jr., \emph{{The Microcanonical functional integral.
  1. The Gravitational field}},
  \href{https://doi.org/10.1103/PhysRevD.47.1420}{\emph{Phys. Rev. D}
  {\bfseries 47} (1993) 1420}
  [\href{https://arxiv.org/abs/gr-qc/9209014}{{\ttfamily gr-qc/9209014}}].

\bibitem{Brown:1994gs}
J.D.~Brown, J.~Creighton and R.B.~Mann, \emph{{Temperature, energy and heat
  capacity of asymptotically anti-de Sitter black holes}},
  \href{https://doi.org/10.1103/PhysRevD.50.6394}{\emph{Phys. Rev. D}
  {\bfseries 50} (1994) 6394}
  [\href{https://arxiv.org/abs/gr-qc/9405007}{{\ttfamily gr-qc/9405007}}].

\bibitem{Banihashemi:2022jys}
B.~Banihashemi and T.~Jacobson, \emph{{Thermodynamic ensembles with
  cosmological horizons}},
  \href{https://doi.org/10.1007/JHEP07(2022)042}{\emph{JHEP} {\bfseries 07}
  (2022) 042} [\href{https://arxiv.org/abs/2204.05324}{{\ttfamily
  2204.05324}}].

\bibitem{Carlip:2003ne}
S.~Carlip and S.~Vaidya, \emph{{Phase transitions and critical behavior for
  charged black holes}},
  \href{https://doi.org/10.1088/0264-9381/20/16/319}{\emph{Class. Quant. Grav.}
  {\bfseries 20} (2003) 3827}
  [\href{https://arxiv.org/abs/gr-qc/0306054}{{\ttfamily gr-qc/0306054}}].

\bibitem{Hayward:1990zm}
G.~Hayward, \emph{{Euclidean action and the thermodynamics of manifolds without
  boundary}}, \href{https://doi.org/10.1103/PhysRevD.41.3248}{\emph{Phys. Rev.
  D} {\bfseries 41} (1990) 3248}.

\bibitem{Svesko:2022txo}
A.~Svesko, E.~Verheijden, E.P.~Verlinde and M.R.~Visser, \emph{{Quasi-local
  energy and microcanonical entropy in two-dimensional nearly de Sitter
  gravity}}, \href{https://doi.org/10.1007/JHEP08(2022)075}{\emph{JHEP}
  {\bfseries 08} (2022) 075}
  [\href{https://arxiv.org/abs/2203.00700}{{\ttfamily 2203.00700}}].

\bibitem{Draper:2022ofa}
P.~Draper and S.~Farkas, \emph{{Euclidean de Sitter black holes and
  microcanonical equilibrium}},
  \href{https://doi.org/10.1103/PhysRevD.105.126021}{\emph{Phys. Rev. D}
  {\bfseries 105} (2022) 126021}
  [\href{https://arxiv.org/abs/2203.01871}{{\ttfamily 2203.01871}}].

\bibitem{Anninos:2022hqo}
D.~Anninos and E.~Harris, \emph{{Interpolating geometries and the stretched
  dS$_{2}$ horizon}},
  \href{https://doi.org/10.1007/JHEP11(2022)166}{\emph{JHEP} {\bfseries 11}
  (2022) 166} [\href{https://arxiv.org/abs/2209.06144}{{\ttfamily
  2209.06144}}].

\bibitem{Banihashemi:2022htw}
B.~Banihashemi, T.~Jacobson, A.~Svesko and M.~Visser, \emph{{The minus sign in
  the first law of de Sitter horizons}},
  \href{https://doi.org/10.1007/JHEP01(2023)054}{\emph{JHEP} {\bfseries 01}
  (2023) 054} [\href{https://arxiv.org/abs/2208.11706}{{\ttfamily
  2208.11706}}].

\bibitem{Anninos:2011af}
D.~Anninos, S.A.~Hartnoll and D.M.~Hofman, \emph{{Static Patch Solipsism:
  Conformal Symmetry of the de Sitter Worldline}},
  \href{https://doi.org/10.1088/0264-9381/29/7/075002}{\emph{Class. Quant.
  Grav.} {\bfseries 29} (2012) 075002}
  [\href{https://arxiv.org/abs/1109.4942}{{\ttfamily 1109.4942}}].

\bibitem{Anninos:2017hhn}
D.~Anninos and D.M.~Hofman, \emph{{Infrared Realization of dS$_2$ in AdS$_2$}},
  \href{https://doi.org/10.1088/1361-6382/aab143}{\emph{Class. Quant. Grav.}
  {\bfseries 35} (2018) 085003}
  [\href{https://arxiv.org/abs/1703.04622}{{\ttfamily 1703.04622}}].

\bibitem{Susskind:2021omt}
L.~Susskind, \emph{{De Sitter Holography: Fluctuations, Anomalous Symmetry, and
  Wormholes}}, \href{https://doi.org/10.3390/universe7120464}{\emph{Universe}
  {\bfseries 7} (2021) 464} [\href{https://arxiv.org/abs/2106.03964}{{\ttfamily
  2106.03964}}].

\bibitem{Leuven:2018ejp}
S.~Leuven, E.~Verlinde and M.~Visser, \emph{{Towards non-AdS Holography via the
  Long String Phenomenon}},
  \href{https://doi.org/10.1007/JHEP06(2018)097}{\emph{JHEP} {\bfseries 06}
  (2018) 097} [\href{https://arxiv.org/abs/1801.02589}{{\ttfamily
  1801.02589}}].

\bibitem{Shyam:2021ciy}
V.~Shyam, \emph{{$ \mathrm{T}\overline{\mathrm{T}} $ +
  \ensuremath{\Lambda}$_{2}$ deformed CFT on the stretched dS$_{3}$ horizon}},
  \href{https://doi.org/10.1007/JHEP04(2022)052}{\emph{JHEP} {\bfseries 04}
  (2022) 052} [\href{https://arxiv.org/abs/2106.10227}{{\ttfamily
  2106.10227}}].

\bibitem{Coleman:2021nor}
E.~Coleman, E.A.~Mazenc, V.~Shyam, E.~Silverstein, R.M.~Soni, G.~Torroba
  et~al., \emph{{De Sitter microstates from T$ \overline{T} $ +
  \ensuremath{\Lambda}$_{2}$ and the Hawking-Page transition}},
  \href{https://doi.org/10.1007/JHEP07(2022)140}{\emph{JHEP} {\bfseries 07}
  (2022) 140} [\href{https://arxiv.org/abs/2110.14670}{{\ttfamily
  2110.14670}}].

\bibitem{Silverstein:2022dfj}
E.~Silverstein, \emph{{Black hole to cosmic horizon microstates in string/M
  theory: timelike boundaries and internal averaging}},
  \href{https://arxiv.org/abs/2212.00588}{{\ttfamily 2212.00588}}.

\bibitem{Batra:2024kjl}
G.~Batra, G.B.~De~Luca, E.~Silverstein, G.~Torroba and S.~Yang,
  \emph{{Bulk-local dS$_3$ holography: the Matter with $T\bar T+\Lambda_2$}},
  \href{https://arxiv.org/abs/2403.01040}{{\ttfamily 2403.01040}}.

\bibitem{Zamolodchikov:2004ce}
A.B.~Zamolodchikov, \emph{{Expectation value of composite field T anti-T in
  two-dimensional quantum field theory}},
  \href{https://arxiv.org/abs/hep-th/0401146}{{\ttfamily hep-th/0401146}}.

\bibitem{Smirnov:2016lqw}
F.A.~Smirnov and A.B.~Zamolodchikov, \emph{{On space of integrable quantum
  field theories}},
  \href{https://doi.org/10.1016/j.nuclphysb.2016.12.014}{\emph{Nucl. Phys. B}
  {\bfseries 915} (2017) 363}
  [\href{https://arxiv.org/abs/1608.05499}{{\ttfamily 1608.05499}}].

\bibitem{Dubovsky:2012wk}
S.~Dubovsky, R.~Flauger and V.~Gorbenko, \emph{{Solving the Simplest Theory of
  Quantum Gravity}}, \href{https://doi.org/10.1007/JHEP09(2012)133}{\emph{JHEP}
  {\bfseries 09} (2012) 133} [\href{https://arxiv.org/abs/1205.6805}{{\ttfamily
  1205.6805}}].

\bibitem{Cavaglia:2016oda}
A.~Cavagli\`a, S.~Negro, I.M.~Sz\'ecs\'enyi and R.~Tateo, \emph{{$T
  \bar{T}$-deformed 2D Quantum Field Theories}},
  \href{https://doi.org/10.1007/JHEP10(2016)112}{\emph{JHEP} {\bfseries 10}
  (2016) 112} [\href{https://arxiv.org/abs/1608.05534}{{\ttfamily
  1608.05534}}].

\bibitem{McGough:2016lol}
L.~McGough, M.~Mezei and H.~Verlinde, \emph{{Moving the CFT into the bulk with
  $ T\overline{T} $}},
  \href{https://doi.org/10.1007/JHEP04(2018)010}{\emph{JHEP} {\bfseries 04}
  (2018) 010} [\href{https://arxiv.org/abs/1611.03470}{{\ttfamily
  1611.03470}}].

\bibitem{Kraus:2018xrn}
P.~Kraus, J.~Liu and D.~Marolf, \emph{{Cutoff AdS$_{3}$ versus the $
  T\overline{T} $ deformation}},
  \href{https://doi.org/10.1007/JHEP07(2018)027}{\emph{JHEP} {\bfseries 07}
  (2018) 027} [\href{https://arxiv.org/abs/1801.02714}{{\ttfamily
  1801.02714}}].

\bibitem{Gorbenko:2018oov}
V.~Gorbenko, E.~Silverstein and G.~Torroba, \emph{{dS/dS and $ T\overline{T}
  $}}, \href{https://doi.org/10.1007/JHEP03(2019)085}{\emph{JHEP} {\bfseries
  03} (2019) 085} [\href{https://arxiv.org/abs/1811.07965}{{\ttfamily
  1811.07965}}].

\bibitem{Strominger:1997eq}
A.~Strominger, \emph{{Black hole entropy from near horizon microstates}},
  \href{https://doi.org/10.1088/1126-6708/1998/02/009}{\emph{JHEP} {\bfseries
  02} (1998) 009} [\href{https://arxiv.org/abs/hep-th/9712251}{{\ttfamily
  hep-th/9712251}}].

\bibitem{Carlip:2000nv}
S.~Carlip, \emph{{Logarithmic corrections to black hole entropy from the Cardy
  formula}}, \href{https://doi.org/10.1088/0264-9381/17/20/302}{\emph{Class.
  Quant. Grav.} {\bfseries 17} (2000) 4175}
  [\href{https://arxiv.org/abs/gr-qc/0005017}{{\ttfamily gr-qc/0005017}}].

\bibitem{Bousso:2001mw}
R.~Bousso, A.~Maloney and A.~Strominger, \emph{{Conformal vacua and entropy in
  de Sitter space}},
  \href{https://doi.org/10.1103/PhysRevD.65.104039}{\emph{Phys. Rev. D}
  {\bfseries 65} (2002) 104039}
  [\href{https://arxiv.org/abs/hep-th/0112218}{{\ttfamily hep-th/0112218}}].

\bibitem{Cardy:1986ie}
J.L.~Cardy, \emph{{Operator Content of Two-Dimensional Conformally Invariant
  Theories}}, \href{https://doi.org/10.1016/0550-3213(86)90552-3}{\emph{Nucl.
  Phys. B} {\bfseries 270} (1986) 186}.

\bibitem{Gross:2019ach}
D.J.~Gross, J.~Kruthoff, A.~Rolph and E.~Shaghoulian, \emph{{$T\overline{T}$ in
  AdS$_2$ and Quantum Mechanics}},
  \href{https://doi.org/10.1103/PhysRevD.101.026011}{\emph{Phys. Rev. D}
  {\bfseries 101} (2020) 026011}
  [\href{https://arxiv.org/abs/1907.04873}{{\ttfamily 1907.04873}}].

\bibitem{Jackiw:1984je}
R.~Jackiw, \emph{{Lower Dimensional Gravity}},
  \href{https://doi.org/10.1016/0550-3213(85)90448-1}{\emph{Nucl. Phys. B}
  {\bfseries 252} (1985) 343}.

\bibitem{Teitelboim:1983ux}
C.~Teitelboim, \emph{{Gravitation and Hamiltonian Structure in Two Space-Time
  Dimensions}}, \href{https://doi.org/10.1016/0370-2693(83)90012-6}{\emph{Phys.
  Lett. B} {\bfseries 126} (1983) 41}.

\bibitem{Gross:2019uxi}
D.J.~Gross, J.~Kruthoff, A.~Rolph and E.~Shaghoulian, \emph{{Hamiltonian
  deformations in quantum mechanics, $T\bar T$, and the SYK model}},
  \href{https://doi.org/10.1103/PhysRevD.102.046019}{\emph{Phys. Rev. D}
  {\bfseries 102} (2020) 046019}
  [\href{https://arxiv.org/abs/1912.06132}{{\ttfamily 1912.06132}}].

\bibitem{Maldacena:2019cbz}
J.~Maldacena, G.J.~Turiaci and Z.~Yang, \emph{{Two dimensional Nearly de Sitter
  gravity}}, \href{https://doi.org/10.1007/JHEP01(2021)139}{\emph{JHEP}
  {\bfseries 01} (2021) 139}
  [\href{https://arxiv.org/abs/1904.01911}{{\ttfamily 1904.01911}}].

\bibitem{Cavaglia:1998xj}
M.~Cavaglia, \emph{{Geometrodynamical formulation of two-dimensional dilaton
  gravity}}, \href{https://doi.org/10.1103/PhysRevD.59.084011}{\emph{Phys. Rev.
  D} {\bfseries 59} (1999) 084011}
  [\href{https://arxiv.org/abs/hep-th/9811059}{{\ttfamily hep-th/9811059}}].

\bibitem{Lemos:1996bq}
J.P.S.~Lemos, \emph{{Thermodynamics of the two-dimensional black hole in the
  Teitelboim-Jackiw theory}},
  \href{https://doi.org/10.1103/PhysRevD.54.6206}{\emph{Phys. Rev. D}
  {\bfseries 54} (1996) 6206}
  [\href{https://arxiv.org/abs/gr-qc/9608016}{{\ttfamily gr-qc/9608016}}].

\bibitem{Creighton:1995uj}
J.D.E.~Creighton and R.B.~Mann, \emph{{Quasilocal thermodynamics of
  two-dimensional black holes}},
  \href{https://doi.org/10.1103/PhysRevD.54.7476}{\emph{Phys. Rev. D}
  {\bfseries 54} (1996) 7476}.

\bibitem{Grumiller:2007ju}
D.~Grumiller and R.~McNees, \emph{{Thermodynamics of black holes in two (and
  higher) dimensions}},
  \href{https://doi.org/10.1088/1126-6708/2007/04/074}{\emph{JHEP} {\bfseries
  04} (2007) 074} [\href{https://arxiv.org/abs/hep-th/0703230}{{\ttfamily
  hep-th/0703230}}].

\bibitem{Moitra:2019xoj}
U.~Moitra, S.K.~Sake, S.P.~Trivedi and V.~Vishal, \emph{{Jackiw-Teitelboim
  Model Coupled to Conformal Matter in the Semi-Classical Limit}},
  \href{https://doi.org/10.1007/JHEP04(2020)199}{\emph{JHEP} {\bfseries 04}
  (2020) 199} [\href{https://arxiv.org/abs/1908.08523}{{\ttfamily
  1908.08523}}].

\bibitem{Pedraza:2021cvx}
J.F.~Pedraza, A.~Svesko, W.~Sybesma and M.R.~Visser, \emph{{Semi-classical
  thermodynamics of quantum extremal surfaces in Jackiw-Teitelboim gravity}},
  \href{https://doi.org/10.1007/JHEP12(2021)134}{\emph{JHEP} {\bfseries 12}
  (2021) 134} [\href{https://arxiv.org/abs/2107.10358}{{\ttfamily
  2107.10358}}].

\bibitem{Witten:1998zw}
E.~Witten, \emph{{Anti-de Sitter space, thermal phase transition, and
  confinement in gauge theories}},
  \href{https://doi.org/10.4310/ATMP.1998.v2.n3.a3}{\emph{Adv. Theor. Math.
  Phys.} {\bfseries 2} (1998) 505}
  [\href{https://arxiv.org/abs/hep-th/9803131}{{\ttfamily hep-th/9803131}}].

\bibitem{Savonije:2001nd}
I.~Savonije and E.P.~Verlinde, \emph{{CFT and entropy on the brane}},
  \href{https://doi.org/10.1016/S0370-2693(01)00467-1}{\emph{Phys. Lett. B}
  {\bfseries 507} (2001) 305}
  [\href{https://arxiv.org/abs/hep-th/0102042}{{\ttfamily hep-th/0102042}}].

\bibitem{Visser:2021eqk}
M.R.~Visser, \emph{{Holographic thermodynamics requires a chemical potential
  for color}}, \href{https://doi.org/10.1103/PhysRevD.105.106014}{\emph{Phys.
  Rev. D} {\bfseries 105} (2022) 106014}
  [\href{https://arxiv.org/abs/2101.04145}{{\ttfamily 2101.04145}}].

\bibitem{Cotler:2019nbi}
J.~Cotler, K.~Jensen and A.~Maloney, \emph{{Low-dimensional de Sitter quantum
  gravity}}, \href{https://doi.org/10.1007/JHEP06(2020)048}{\emph{JHEP}
  {\bfseries 06} (2020) 048}
  [\href{https://arxiv.org/abs/1905.03780}{{\ttfamily 1905.03780}}].

\bibitem{Castro:2022cuo}
A.~Castro, F.~Mariani and C.~Toldo, \emph{{Near-extremal limits of de Sitter
  black holes}}, \href{https://doi.org/10.1007/JHEP07(2023)131}{\emph{JHEP}
  {\bfseries 07} (2023) 131}
  [\href{https://arxiv.org/abs/2212.14356}{{\ttfamily 2212.14356}}].

\bibitem{Sybesma:2020fxg}
W.~Sybesma, \emph{{Pure de Sitter space and the island moving back in time}},
  \href{https://doi.org/10.1088/1361-6382/abff9a}{\emph{Class. Quant. Grav.}
  {\bfseries 38} (2021) 145012}
  [\href{https://arxiv.org/abs/2008.07994}{{\ttfamily 2008.07994}}].

\bibitem{Kames-King:2021etp}
J.~Kames-King, E.M.H.~Verheijden and E.P.~Verlinde, \emph{{No Page curves for
  the de Sitter horizon}},
  \href{https://doi.org/10.1007/JHEP03(2022)040}{\emph{JHEP} {\bfseries 03}
  (2022) 040} [\href{https://arxiv.org/abs/2108.09318}{{\ttfamily
  2108.09318}}].

\bibitem{Balasubramanian:2001nb}
V.~Balasubramanian, J.~de~Boer and D.~Minic, \emph{{Mass, entropy and
  holography in asymptotically de Sitter spaces}},
  \href{https://doi.org/10.1103/PhysRevD.65.123508}{\emph{Phys. Rev. D}
  {\bfseries 65} (2002) 123508}
  [\href{https://arxiv.org/abs/hep-th/0110108}{{\ttfamily hep-th/0110108}}].

\bibitem{Banados:1992wn}
M.~Banados, C.~Teitelboim and J.~Zanelli, \emph{{The Black hole in
  three-dimensional space-time}},
  \href{https://doi.org/10.1103/PhysRevLett.69.1849}{\emph{Phys. Rev. Lett.}
  {\bfseries 69} (1992) 1849}
  [\href{https://arxiv.org/abs/hep-th/9204099}{{\ttfamily hep-th/9204099}}].

\bibitem{Banados:1992gq}
M.~Banados, M.~Henneaux, C.~Teitelboim and J.~Zanelli, \emph{{Geometry of the
  (2+1) black hole}},
  \href{https://doi.org/10.1103/PhysRevD.48.1506}{\emph{Phys. Rev. D}
  {\bfseries 48} (1993) 1506}
  [\href{https://arxiv.org/abs/gr-qc/9302012}{{\ttfamily gr-qc/9302012}}].

\bibitem{Tsolakidis:2024wut}
E.~Tsolakidis, \emph{{Massive gravity generalization of $ T\overline{T} $
  deformations}}, \href{https://doi.org/10.1007/JHEP09(2024)167}{\emph{JHEP}
  {\bfseries 09} (2024) 167}
  [\href{https://arxiv.org/abs/2405.07967}{{\ttfamily 2405.07967}}].

\bibitem{Hartman:2018tkw}
T.~Hartman, J.~Kruthoff, E.~Shaghoulian and A.~Tajdini, \emph{{Holography at
  finite cutoff with a $T^2$ deformation}},
  \href{https://doi.org/10.1007/JHEP03(2019)004}{\emph{JHEP} {\bfseries 03}
  (2019) 004} [\href{https://arxiv.org/abs/1807.11401}{{\ttfamily
  1807.11401}}].

\bibitem{Taylor:2018xcy}
M.~Taylor, \emph{{TT deformations in general dimensions}},
  \href{https://arxiv.org/abs/1805.10287}{{\ttfamily 1805.10287}}.

\bibitem{Brown:1994su}
J.D.~Brown and J.W.~York, Jr., \emph{{The Path integral formulation of
  gravitational thermodynamics}},  in \emph{{The Black Hole 25 Years After}},
  pp.~1--24, 1, 1994 [\href{https://arxiv.org/abs/gr-qc/9405024}{{\ttfamily
  gr-qc/9405024}}].

\bibitem{Bloete:1986qm}
H.W.J.~Bloete, J.L.~Cardy and M.P.~Nightingale, \emph{{Conformal Invariance,
  the Central Charge, and Universal Finite Size Amplitudes at Criticality}},
  \href{https://doi.org/10.1103/PhysRevLett.56.742}{\emph{Phys. Rev. Lett.}
  {\bfseries 56} (1986) 742}.

\bibitem{Anninos:2018svg}
D.~Anninos, D.A.~Galante and D.M.~Hofman, \emph{{De Sitter horizons \&
  holographic liquids}},
  \href{https://doi.org/10.1007/JHEP07(2019)038}{\emph{JHEP} {\bfseries 07}
  (2019) 038} [\href{https://arxiv.org/abs/1811.08153}{{\ttfamily
  1811.08153}}].

\bibitem{Anninos:2020cwo}
D.~Anninos and D.A.~Galante, \emph{{Constructing AdS$_{2}$ flow geometries}},
  \href{https://doi.org/10.1007/JHEP02(2021)045}{\emph{JHEP} {\bfseries 02}
  (2021) 045} [\href{https://arxiv.org/abs/2011.01944}{{\ttfamily
  2011.01944}}].

\bibitem{Freivogel:2005qh}
B.~Freivogel, V.E.~Hubeny, A.~Maloney, R.C.~Myers, M.~Rangamani and S.~Shenker,
  \emph{{Inflation in AdS/CFT}},
  \href{https://doi.org/10.1088/1126-6708/2006/03/007}{\emph{JHEP} {\bfseries
  03} (2006) 007} [\href{https://arxiv.org/abs/hep-th/0510046}{{\ttfamily
  hep-th/0510046}}].

\bibitem{Lewkowycz:2019xse}
A.~Lewkowycz, J.~Liu, E.~Silverstein and G.~Torroba, \emph{{$ T\overline{T} $
  and EE, with implications for (A)dS subregion encodings}},
  \href{https://doi.org/10.1007/JHEP04(2020)152}{\emph{JHEP} {\bfseries 04}
  (2020) 152} [\href{https://arxiv.org/abs/1909.13808}{{\ttfamily
  1909.13808}}].

\bibitem{Silverstein:2024xnr}
E.~Silverstein and G.~Torroba, \emph{{Timelike-bounded $dS_4$ holography from a
  solvable sector of the $T^2$ deformation}},
  \href{https://arxiv.org/abs/2409.08709}{{\ttfamily 2409.08709}}.

\bibitem{Anninos:2020hfj}
D.~Anninos, F.~Denef, Y.T.A.~Law and Z.~Sun, \emph{{Quantum de Sitter horizon
  entropy from quasicanonical bulk, edge, sphere and topological string
  partition functions}},
  \href{https://doi.org/10.1007/JHEP01(2022)088}{\emph{JHEP} {\bfseries 01}
  (2022) 088} [\href{https://arxiv.org/abs/2009.12464}{{\ttfamily
  2009.12464}}].

\bibitem{DESER1984405}
S.~Deser and R.~Jackiw, \emph{Three-dimensional cosmological gravity: Dynamics
  of constant curvature},
  \href{https://doi.org/https://doi.org/10.1016/0003-4916(84)90025-3}{\emph{Annals
  of Physics} {\bfseries 153} (1984) 405}.

\bibitem{Klemm:2002ir}
D.~Klemm and L.~Vanzo, \emph{{De Sitter gravity and Liouville theory}},
  \href{https://doi.org/10.1088/1126-6708/2002/04/030}{\emph{JHEP} {\bfseries
  04} (2002) 030} [\href{https://arxiv.org/abs/hep-th/0203268}{{\ttfamily
  hep-th/0203268}}].

\bibitem{Andrade:2015gja}
T.~Andrade, W.R.~Kelly, D.~Marolf and J.E.~Santos, \emph{{On the stability of
  gravity with Dirichlet walls}},
  \href{https://doi.org/10.1088/0264-9381/32/23/235006}{\emph{Class. Quant.
  Grav.} {\bfseries 32} (2015) 235006}
  [\href{https://arxiv.org/abs/1504.07580}{{\ttfamily 1504.07580}}].

\bibitem{Maldacena:2016upp}
J.~Maldacena, D.~Stanford and Z.~Yang, \emph{{Conformal symmetry and its
  breaking in two dimensional Nearly Anti-de-Sitter space}},
  \href{https://doi.org/10.1093/ptep/ptw124}{\emph{PTEP} {\bfseries 2016}
  (2016) 12C104} [\href{https://arxiv.org/abs/1606.01857}{{\ttfamily
  1606.01857}}].

\bibitem{Maldacena:2016hyu}
J.~Maldacena and D.~Stanford, \emph{{Remarks on the Sachdev-Ye-Kitaev model}},
  \href{https://doi.org/10.1103/PhysRevD.94.106002}{\emph{Phys. Rev. D}
  {\bfseries 94} (2016) 106002}
  [\href{https://arxiv.org/abs/1604.07818}{{\ttfamily 1604.07818}}].

\bibitem{Sarosi:2017ykf}
G.~S\'arosi, \emph{{AdS$_{2}$ holography and the SYK model}},
  \href{https://doi.org/10.22323/1.323.0001}{\emph{PoS} {\bfseries Modave2017}
  (2018) 001} [\href{https://arxiv.org/abs/1711.08482}{{\ttfamily
  1711.08482}}].

\bibitem{Iliesiu:2020zld}
L.V.~Iliesiu, J.~Kruthoff, G.J.~Turiaci and H.~Verlinde, \emph{{JT gravity at
  finite cutoff}},
  \href{https://doi.org/10.21468/SciPostPhys.9.2.023}{\emph{SciPost Phys.}
  {\bfseries 9} (2020) 023} [\href{https://arxiv.org/abs/2004.07242}{{\ttfamily
  2004.07242}}].

\bibitem{Griguolo:2021wgy}
L.~Griguolo, R.~Panerai, J.~Papalini and D.~Seminara, \emph{{Nonperturbative
  effects and resurgence in Jackiw-Teitelboim gravity at finite cutoff}},
  \href{https://doi.org/10.1103/PhysRevD.105.046015}{\emph{Phys. Rev. D}
  {\bfseries 105} (2022) 046015}
  [\href{https://arxiv.org/abs/2106.01375}{{\ttfamily 2106.01375}}].

\bibitem{Anninos:2022qgy}
D.~Anninos, D.A.~Galante and S.U.~Sheorey, \emph{{Renormalisation Group Flows
  of the SYK Model}},  \href{https://arxiv.org/abs/2212.04944}{{\ttfamily
  2212.04944}}.

\bibitem{Susskind:2021esx}
L.~Susskind, \emph{{Entanglement and Chaos in De Sitter Space Holography: An
  SYK Example}}, \href{https://doi.org/10.22128/jhap.2021.455.1005}{\emph{JHAP}
  {\bfseries 1} (2021) 1} [\href{https://arxiv.org/abs/2109.14104}{{\ttfamily
  2109.14104}}].

\bibitem{Susskind:2022bia}
L.~Susskind, \emph{{De Sitter Space, Double-Scaled SYK, and the Separation of
  Scales in the Semiclassical Limit}},
  \href{https://arxiv.org/abs/2209.09999}{{\ttfamily 2209.09999}}.

\bibitem{Susskind:2022dfz}
L.~Susskind, \emph{{Scrambling in Double-Scaled SYK and De Sitter Space}},
  \href{https://arxiv.org/abs/2205.00315}{{\ttfamily 2205.00315}}.

\bibitem{Rahman:2022jsf}
A.A.~Rahman, \emph{{dS JT Gravity and Double-Scaled SYK}},
  \href{https://arxiv.org/abs/2209.09997}{{\ttfamily 2209.09997}}.

\bibitem{Narovlansky:2023lfz}
V.~Narovlansky and H.~Verlinde, \emph{{Double-scaled SYK and de Sitter
  Holography}},  \href{https://arxiv.org/abs/2310.16994}{{\ttfamily
  2310.16994}}.

\bibitem{Aguilar-Gutierrez:2024oea}
S.E.~Aguilar-Gutierrez, \emph{{$T^2$ deformations in the double-scaled SYK
  model: Stretched horizon thermodynamics}},
  \href{https://arxiv.org/abs/2410.18303}{{\ttfamily 2410.18303}}.

\bibitem{A}
S.E.~Aguilar-Gutierrez, \emph{Moving boundaries with the double-scaled syk
  model: $t^2$ deformations, thermodynamics, and krylov complexity},  To
  appear.

\bibitem{Araujo-Regado:2022jpj}
G.~Araujo-Regado, \emph{{Holographic Cosmology on Closed Slices in 2+1
  Dimensions}},  \href{https://arxiv.org/abs/2212.03219}{{\ttfamily
  2212.03219}}.

\bibitem{Araujo-Regado:2022gvw}
G.~Araujo-Regado, R.~Khan and A.C.~Wall, \emph{{Cauchy slice holography: a new
  AdS/CFT dictionary}},
  \href{https://doi.org/10.1007/JHEP03(2023)026}{\emph{JHEP} {\bfseries 03}
  (2023) 026} [\href{https://arxiv.org/abs/2204.00591}{{\ttfamily
  2204.00591}}].

\bibitem{Soni:2024aop}
R.M.~Soni and A.C.~Wall, \emph{{A New Covariant Entropy Bound from Cauchy Slice
  Holography}},  \href{https://arxiv.org/abs/2407.16769}{{\ttfamily
  2407.16769}}.

\bibitem{Anderson_2008}
M.T.~Anderson, \emph{On boundary value problems for einstein metrics},
  \href{https://doi.org/10.2140/gt.2008.12.2009}{\emph{Geometry \&amp;
  Topology} {\bfseries 12} (2008) 2009–2045}.

\bibitem{Witten:2018lgb}
E.~Witten, \emph{{A note on boundary conditions in Euclidean gravity}},
  \href{https://doi.org/10.1142/S0129055X21400043}{\emph{Rev. Math. Phys.}
  {\bfseries 33} (2021) 2140004}
  [\href{https://arxiv.org/abs/1805.11559}{{\ttfamily 1805.11559}}].

\bibitem{Friedrich:1998xt}
H.~Friedrich and G.~Nagy, \emph{{The Initial boundary value problem for
  Einstein's vacuum field equations}},
  \href{https://doi.org/10.1007/s002200050571}{\emph{Commun. Math. Phys.}
  {\bfseries 201} (1999) 619}.

\bibitem{Fournodavlos:2020wde}
G.~Fournodavlos and J.~Smulevici, \emph{{The Initial Boundary Value Problem for
  the Einstein Equations with Totally Geodesic Timelike Boundary}},
  \href{https://doi.org/10.1007/s00220-021-04141-8}{\emph{Commun. Math. Phys.}
  {\bfseries 385} (2021) 1615}
  [\href{https://arxiv.org/abs/2006.01498}{{\ttfamily 2006.01498}}].

\bibitem{Fournodavlos:2021eye}
G.~Fournodavlos and J.~Smulevici, \emph{{The Initial Boundary Value Problem in
  General Relativity: The Umbilic Case}},
  \href{https://doi.org/10.1093/imrn/rnab359}{\emph{Int. Math. Res. Not.}
  {\bfseries 2023} (2023) 3790}
  [\href{https://arxiv.org/abs/2104.08851}{{\ttfamily 2104.08851}}].

\bibitem{An:2021fcq}
Z.~An and M.T.~Anderson, \emph{{The initial boundary value problem and
  quasi-local Hamiltonians in General Relativity}},
  \href{https://arxiv.org/abs/2103.15673}{{\ttfamily 2103.15673}}.

\bibitem{Anninos:2022ujl}
D.~Anninos, D.A.~Galante and B.~M\"uhlmann, \emph{{Finite features of quantum
  de Sitter space}},
  \href{https://doi.org/10.1088/1361-6382/acaba5}{\emph{Class. Quant. Grav.}
  {\bfseries 40} (2023) 025009}
  [\href{https://arxiv.org/abs/2206.14146}{{\ttfamily 2206.14146}}].

\bibitem{Liu:2024ymn}
X.~Liu, J.E.~Santos and T.~Wiseman, \emph{{New Well-Posed boundary conditions
  for semi-classical Euclidean gravity}},
  \href{https://doi.org/10.1007/JHEP06(2024)044}{\emph{JHEP} {\bfseries 06}
  (2024) 044} [\href{https://arxiv.org/abs/2402.04308}{{\ttfamily
  2402.04308}}].

\bibitem{Bredberg:2011xw}
I.~Bredberg and A.~Strominger, \emph{{Black Holes as Incompressible Fluids on
  the Sphere}}, \href{https://doi.org/10.1007/JHEP05(2012)043}{\emph{JHEP}
  {\bfseries 05} (2012) 043} [\href{https://arxiv.org/abs/1106.3084}{{\ttfamily
  1106.3084}}].

\bibitem{Anninos:2011zn}
D.~Anninos, T.~Anous, I.~Bredberg and G.S.~Ng, \emph{{Incompressible Fluids of
  the de Sitter Horizon and Beyond}},
  \href{https://doi.org/10.1007/JHEP05(2012)107}{\emph{JHEP} {\bfseries 05}
  (2012) 107} [\href{https://arxiv.org/abs/1110.3792}{{\ttfamily 1110.3792}}].

\bibitem{Anninos:2023epi}
D.~Anninos, D.A.~Galante and C.~Maneerat, \emph{{Gravitational observatories}},
  \href{https://doi.org/10.1007/JHEP12(2023)024}{\emph{JHEP} {\bfseries 12}
  (2023) 024} [\href{https://arxiv.org/abs/2310.08648}{{\ttfamily
  2310.08648}}].

\bibitem{Anninos:2024wpy}
D.~Anninos, D.A.~Galante and C.~Maneerat, \emph{{Cosmological observatories}},
  \href{https://doi.org/10.1088/1361-6382/ad5824}{\emph{Class. Quant. Grav.}
  {\bfseries 41} (2024) 165009}
  [\href{https://arxiv.org/abs/2402.04305}{{\ttfamily 2402.04305}}].

\bibitem{Anninos:2024xhc}
D.~Anninos, R.~Arias, D.A.~Galante and C.~Maneerat, \emph{{Gravitational
  Observatories in AdS$_4$}},
  \href{https://arxiv.org/abs/2412.16305}{{\ttfamily 2412.16305}}.

\bibitem{Banihashemi:2024yye}
B.~Banihashemi, E.~Shaghoulian and S.~Shashi, \emph{{Flat space gravity at
  finite cutoff}},  \href{https://arxiv.org/abs/2409.07643}{{\ttfamily
  2409.07643}}.

\bibitem{Coleman:2020jte}
E.~Coleman and V.~Shyam, \emph{{Conformal boundary conditions from cutoff
  AdS$_{3}$}}, \href{https://doi.org/10.1007/JHEP09(2021)079}{\emph{JHEP}
  {\bfseries 09} (2021) 079}
  [\href{https://arxiv.org/abs/2010.08504}{{\ttfamily 2010.08504}}].

\bibitem{inprep}
D.~Galante, C.~Maneerat and A.~Svesko., \emph{{In preparation}}.

\bibitem{Morone:2024ffm}
T.~Morone, S.~Negro and R.~Tateo, \emph{{Gravity and $T\bar{T}$ flows in higher
  dimensions}},  \href{https://arxiv.org/abs/2401.16400}{{\ttfamily
  2401.16400}}.

\bibitem{Davis:2004xi}
J.L.~Davis and R.~McNees, \emph{{Boundary counterterms and the thermodynamics
  of 2-D black holes}},
  \href{https://doi.org/10.1088/1126-6708/2005/09/072}{\emph{JHEP} {\bfseries
  09} (2005) 072} [\href{https://arxiv.org/abs/hep-th/0411121}{{\ttfamily
  hep-th/0411121}}].

\bibitem{Carrasco:2023fcj}
R.~Carrasco, J.F.~Pedraza, A.~Svesko and Z.~Weller-Davies, \emph{{Gravitation
  from optimized computation: Einstein and beyond}},
  \href{https://doi.org/10.1007/JHEP09(2023)167}{\emph{JHEP} {\bfseries 09}
  (2023) 167} [\href{https://arxiv.org/abs/2306.08503}{{\ttfamily
  2306.08503}}].

\end{thebibliography}\endgroup
\end{document}